\documentclass[twocolumn,aps,showpacs,multicol,amsmath,amssymb]{revtex4-1}
\usepackage{graphicx}
\usepackage{epstopdf}
\usepackage{subfigure}
\usepackage{sidecap}
\usepackage{amssymb}
\usepackage{amsmath}
\usepackage{bm}% bold math
\graphicspath{{Figures/}}

\newcommand{\si}{\sigma}

\newcommand{\lam}{\lambda}

\newcommand{\de}{\delta}
\newcommand{\De}{\Delta}
\newcommand{\Ga}{\Gamma}
\newcommand{\epla}{\epsilon^\lambda_{0\tau}}
\newcommand{\eplax}{\epsilon^\lambda_{01}}
\newcommand{\eplay}{\epsilon^\lambda_{02}}

\newcommand{\hrho}{\hat{\rho}}

\newcommand{\be}{\begin{equation}}
\newcommand{\ee}{\end{equation}}

\newcommand{\bea}{\begin{eqnarray}}
\newcommand{\eea}{\end{eqnarray}}
\newcommand{\bd}{\begin{displaymath}}
\newcommand{\ed}{\end{displaymath}}
\newcommand{\ba}{\begin{array}}
\newcommand{\ea}{\end{array}}
\newcommand{\bi}{\begin{itemize}}
\newcommand{\ei}{\end{itemize}}
\newcommand{\bc}{\begin{center}}
\newcommand{\ec}{\end{center}}
\newcommand{\bfl}{\begin{flushleft}}
\newcommand{\efl}{\end{flushleft}}
\newcommand{\bfr}{\begin{flushright}}
\newcommand{\efr}{\end{flushright}}
\newcommand{\non}{\nonumber}

\newcommand{\bl}{\begin{aligned}}
\newcommand{\el}{\end{aligned}}

\newcommand{\hh}{\hat{h}}
\newcommand{\hG}{\hat{G}}

\newcommand{\pJ}{J^\perp}

\newcommand{\bJ}{\bar{J}}

\newcommand{\bav}{\bar{V}}
\newcommand{\bartau}{\bar{\tau}}

\newcommand{\tg}{\tilde{g}}
\newcommand{\tilH}{\tilde{H}}

\newcommand{\fs}{\frac{1}{2}}
\newcommand{\fN}{\frac{1}{N_s}}

\newcommand{\expa}{\bigl(\frac{\bav^2}{D_c} +\frac{\Delta^2}{D_c}\bigr)}
\newcommand{\expb}{\bigl[\bigl(\frac{\bav^2}{D^2_c}\bigr)^2 +\frac{\Delta_0^2}{D^2_c}
+2\frac{\De_0\de_s}{D_c^3}\bigr]^\fs}

\newcommand{\om}{i\omega_n}

\newcommand{\ra}{\rangle}
\newcommand{\la}{\langle}

\def\ket#1{\left\vert #1 \right\rangle}
\def\dg{^{\dagger}}

\def\bR{{\bf R}} 
\def\bk{{\bf k}}

\def\bQ{{\bf Q}}  
  
 \def\bd{{\bf d}}  \def\bJ{{\bf J}}

\def\bs{{\bf s}} \def\bJ{{\bf J}}

\def\da{\downarrow} \def\ua{\uparrow}
 
\def\dg{\dagger}

\def\ket{\rangle}
\def\={\!\!\!&=&\!\!\!}
\def\+{\!\!\!&&\!\!\!+~}
\def\-{\!\!\!&&\!\!\!-~}

\usepackage{color}
\usepackage[dvipsnames]{xcolor}
\usepackage{xcolor}
\usepackage[colorlinks=true,citecolor=blue]{hyperref}

\usepackage{hyperref,cleveref}
\begin{document}

\title{Heavy quasiparticle bands in the underscreened quasiquartet Kondo lattice}

\author {Peter Thalmeier$^{1}$ and Alireza Akbari$^{2,3,4}$}
\affiliation{ 
$^{1}$Max Planck Institute for the  Chemical Physics of Solids, D-01187
Dresden, Germany}
\affiliation{$^2$Asia Pacific Center for Theoretical Physics, Pohang, Gyeongbuk 790-784, Korea}
\affiliation{$^3$Department of Physics, POSTECH, Pohang, Gyeongbuk 790-784, Korea}
\affiliation{$^4$Max Planck POSTECH Center for Complex Phase Materials, POSTECH, Pohang 790-784, Korea}

\begin{abstract}
We study the quasiparticle spectrum in an underscreened Kondo-lattice (KL) model that involves a single spin
degenerate conduction band and two crystalline-electric-field (CEF) split Kramers doublets coupled by both
orbital-diagonal and non-diagonal exchange interactions. We find the three quasiparticle bands of the model
 using a constrained fermionic mean field approach. While two bands are similar to the one-orbital model 
a new genuinely heavy band inside the main hybridization gap appears in the quasiquartet model. Its dispersion is due to effective hybridization with conduction states but the bandwidth is controled by the size of the CEF splitting. Furthermore several new indirect and direct hybridztion gaps may be identified.
By solving the selfconsistency equation  we calculate the CEF- splitting and exchange dependence of 
effective Kondo low energy scale, hybridization gaps and band widths. We also derive the quasiparticle spectral
densities and their partial orbital contributions. We suggest that the two-orbital KL model can exhibit mixed CEF/Kondo excitonic magnetism.
\end{abstract}

%\pacs{ }

\maketitle
%%%%%%%%%%%%%%%%%%%%%%%%%%%%%%%%%%%%%%%%%%%%%%%%%%%%%%

\section{Introduction}

The Anderson lattice and Kondo lattice (KL) models provide the basic understanding for strongly correlated
f-electron systems like heavy fermion metals, superconductors and Kondo insulators 
\cite{newns:87,hewson:93,tsunetsugu:97,kuramoto:00,thalmeier:05}. In the KL model the charge fluctuations between conduction and f-electrons are already eliminated leading to conduction electrons that interact 
through an effective Schrieffer-Wolff exchange term with a lattice of localized moments resulting from the total angular momentum \bJ~ of the f-electron shell. Their $(2J+1)$-fold state degeneracy is reduced by the action of the crystalline electric field leaving  (for noninteger J) only twofold degenerate (e.g. for common tetragonal symmetry D$_{4h}$) Kramers doublets or at the most 
fourfold degenerate quartets (in cubic symmery $O_h$) as in Ce- \cite{ohkawa:85,shiina:97} or Sm -hexaboride \cite{sundermann:18}  Kondo compounds. 

In Kondo lattice studies it is frequently assumed that the degeneracy of conduction states is the same as that of localized states, leading to a SU(N) internal symmetry of the KL Hamiltonian. This is reasonable for the $N=2$ degeneracy of a doublet ground state because of the Kramers degeneracy of conduction electrons when inversion and time reversal symmetry is preserved. However, fourfold degeneracy  of conduction states may only appear along symmetry lines or possibly symmetry planes, but generally not throughout the whole  Brillouin zone. Therefore for $N> 2$ this model is rather artificial. It is nevertheless useful because it is accessible
to a simple constrained mean field approach which becomes exact in the large N degeneracy limit \cite{lacroix:79,newns:87}. 
This leads very naturally to hybridized itinerant bands with partly heavy f-character that are described by the simple dispersions
$E_{1,2\bk}=\fs(\epsilon_\bk+\lam)\pm\fs\sqrt{(\epsilon_\bk-\lam)^2+4\bav^2}$ where $\lambda, \bav$ are selfconsistently determined effective f-level position and hybridization, respectively, the former defining the low energy Kondo scale of the system. In the original Anderson model one generally may also have a $\bk$-dependent hybridization $\bav_\bk$ with nodes leading to pseudogap behaviour \cite{ikeda:96,hanzawa:02}.
Most of qualitative understanding of heavy band and hybridization gap formation  and its physical consequences is based on this simple result for the equal degeneracy or 'fully screened' model. This designation stems from the corresponding impurity model where the local f-moment will be fully sreened at low temperature by the exchange with conduction electrons leading to just an enhanced Pauli susceptibility. 
However, even in the fully screened case $(N=2)$ many Kondo compounds become magnetically ordered \cite{irkhin:17}, which is commonly eplained as a rigid heavy band polarization \cite{,zhang:00,li:10,liu:13,li:15} in the lattice model. The type of magnetic order then depends on the filling of conduction band and strength of Kondo coupling. A more advanced DMFT approach to KL  magnetism beyond the rigid band model is used in Refs.  \onlinecite{beach:08,hoshino:13}.

For application to realistic Kondo systems, in particular Ce and Yb compounds the SU(2) model seems oversimplified. In the case of cubic compounds the ground state may be a quartet, then more than one hybridization gap may appear, e.g. in cubic $YbB_{12}$ \cite{akbari:09} or more complicated order than magnetic is observed as e.g. in cubic $CeB_6$\cite{akbari:12}. Even in tetragonal systems, in particular in Yb - compounds, the projection to the SU(2) model for the lowest Kramers doublet is too restrictive because of a closeby first excited CEF Kramers state , forming a quasi-quartet with the ground state. This case is e.g. realized in YbRu$_2$Ge$_2$  \cite{jeevan:06,takimoto:08,jeevan:11} and other Yb- and Ce- compounds \cite{huesges:18,hafner:18} with $\Gamma_6-\Gamma_7$ low lying quasi-quartet. This opens a new possibility, namely induced or excitonic magnetic or even multipolar order due to orbitally non-diagonal exchange couplings \cite{takimoto:08}.

This extension should also have profound consequences for the Kondo 
physics. Firstly the inclusion of the excited state implies that we have an underscreening situation with total number of $2N$  f-states and $N=2$ for the conduction band degeneracy which can drastically change the spectral properties of hybridized bands \cite{saso:03} . Secondly the CEF splitting should strongly influence the Kondo energy scale which becomes dependent on the splitting size, as is well known in the impurity models \cite{cornut:72} and also on the difference in diagonal and the additional off-diagonal exchange couplings. A most interesting aspect is the influence of (partial) Kondo screening on the possible excitonic order in the two-orbital quasi- quartet KL. This requires first a thorough understandig how the localized split CEF states turn into  quasiparticle bands due to the Kondo effect. Sofar the underscreened KL has been less extensively investigated. Existing work \cite{perkins:07,thomas:11,thomas:14} discusses possible magnetic phases and the ground state phase diagram without emphasis on CEF effects. In this work we perform a detailed study of a two-orbital underscreened KL model with quasi- quartet CEF states, in particular in view of its quasiparticle dispersion, hybridisation gaps , CEF-splitting dependent Kondo energy scale and spectral properties. This is a prerequisite for discussing physical applications like induced (excitonic) magnetism or multipolar order and possible spin exciton modes in the paramagnetic phase as well as the broken symmetry phases for such a more realistic Kondo lattice model.

The paper is organized as follows: In Sec.~\ref{sec:model} we introduce the quasi-quartet KL model and its fermionic representation. The treatment within constrained mean field theory for the strongly correlated f- electron limit is presented in Sec.~\ref{sec:meanfield}. Then in Sec .~\ref{sec:quasi} the quasiparticle bands are derived and their properties like band widths, effective mass and hybridization gaps are discussed. Sec.~\ref{sec:Green} introduces the Green's functions of the model that give the basis for formulating the selfconsistency requirements and constraints in Sec.~\ref{sec:constraint}. The numerical solutions, in particular spectral properties are discussed in Sec.~\ref{sec:numerical} and finally Sec.~\ref{sec:conclusion} gives the conclusions and outlook on further applications of the model.

\section{Model of the quasi-quartet Kondo lattice}
\label{sec:model}

\begin{figure}
%\vspace{0.5cm}
\includegraphics[width=0.95\columnwidth]{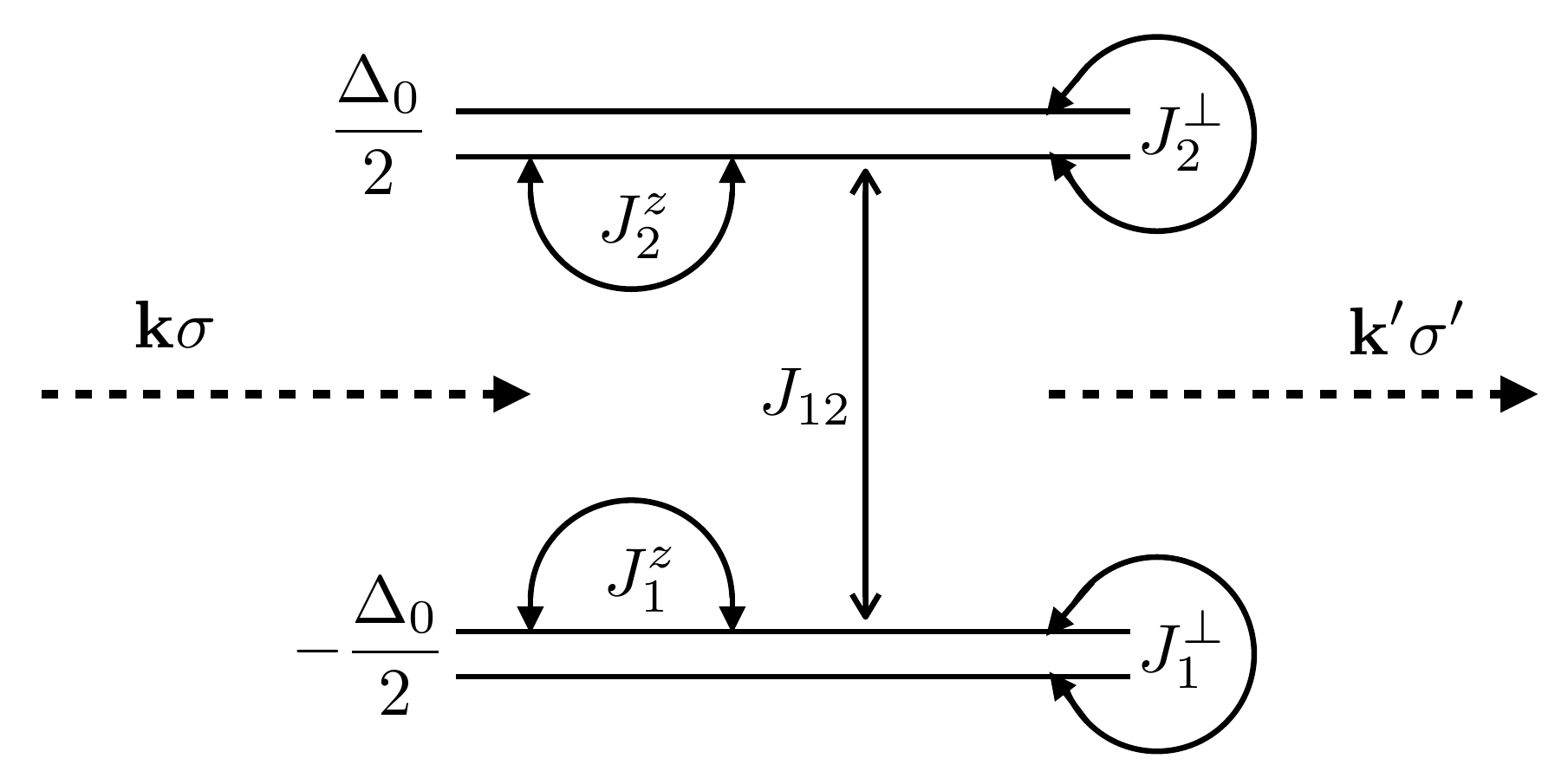}
\caption{Sketch of the quasi-quartet CEF level scheme consisting of two Kramers doublets ($\tau=1,2$ CEF orbital index, e.g. for $\Ga_6$, $\Ga_7$)  with splitting $\Delta_0$.  The various diagonal ($\pJ_\tau, J^z_\tau$) and off-diagonal ($J_{12}$) interactions are indicated which scatter conduction electrons from state $(\bk\si)$ to $(\bk'\si')$. }
\label{fig:CEFscheme}
\end{figure}

We investigate the Kondo-lattice (KL) model for a quasi-quartet system of 4f-CEF states, having in mind $Yb^{3+}(4f^{13})$ or 
 $Ce^{3+}(4f^{1})$  Kondo ions with one f-hole or electron, respectively. The doublet constituents of this model CEF-scheme split by an energy $\Delta_0=2\Delta$ are treated as Kramers $S=\frac{1}{2}$ pseudo-spins. It is sketched in Fig.\ref{fig:CEFscheme} and its exchange interactions with conduction (c-) electrons are indicated. A more detailed discussion based on Ref.~\onlinecite{takimoto:08} is given in  Appendix \ref{sec:app1}.  The effect of $c$-$f$ hybridization and $f$-$f$ Coulomb interaction is described by the Anderson lattice model \cite{hewson:93}. The Coulomb interaction is the largest energy scale and may be eliminated by a Schrieffer-Wolff transformation \cite{fazekas:99}. This leads to a Kondo-type Hamiltonian with effective antiferromagnetic exchange of strength $(g_J-1)I_{ex}$ between the conduction electron spins and the localized 4f- moments which (partly) screen them at low temperature. We consider a model with only {\it one} conduction band but {\it two} pseudo-spins representing the two lowest $4f$ Kramers doublets of the $(2J+1)$  CEF scheme. Therefore the degeneracy of conduction states is $N=2$ whereas there are $2N=4$ localized quasi-quartet states. In the impurity case with just one $f$-site such model is termed 'underscreened' \cite{nozieres:80,perkins:07} because in this case (for $\Delta_0=0$) the local 4f- moment cannot be fully screened to form a singlet ground state so that a residual spin $S^*=\frac{1}{2}$ survives, leading to a Curie type susceptibility contribution at low temperature. Nevertheless the Kondo  fixed point and associated local Fermi liquid is stable in the underscreened case because the residual FM coupling of $S^*$ to the renormalized conduction states scales logarithmically to zero \cite{nozieres:80,coleman:15}. This is in contrast to the overscreened case (more than N conduction channels) where the Kondo fixed point is unstable leading to non-Fermi liquid behaviour.
While the impurity case is understood  there are few treatments for the underscreened quasi-quartet Kondo lattice model, which is however, of some practical importance in tetragonal ($D_{4h}$ symmetry) Ce- and Yb- compounds. The model is defined by the Hamiltonian
 \be
 H=H_0+(g_J-1)I_{ex}\sum_i\bs_i\cdot\bJ_i ,
 \label{eqn:cfHam}
 \ee
 where $g_J$ is the Land\'e factor of the lowest total angular momentum $(\bJ)$ state of $Ce^{3+}(4f^1, J=\frac{5}{2})$ electron or 
 $Yb^{3+}(4f^{13}, J=\frac{7}{2})$ hole and $\bs$ the conduction electron spin. The first part of the  Hamiltonian describes noninteracting conduction electrons and CEF states. When we restrict the latter to the two lowest Kramers doublet states in Fig.~\ref{fig:CEFscheme} one may transform it to a fermionic representation as described in Appendix \ref{sec:app1} according to
 \begin{eqnarray}
 \bl
 H=\;
 & 
 H_0 +H_{cf} + H_{cf}^{(12)},
 \non\\
 H_0=\;
 &\sum_{\bk\si}\epsilon_\bk c^\dag_{\bk\si}c_{\bk\si}+
 \frac{1}{2}\Delta_0\sum_{i\tau\si}(-1)^\tau f^\dag_{i\tau\si}f_{i\tau\si},
 \\
 H_{cf}=\;
 &\frac{1}{4}\sum_{i\tau}J^z_\tau(f^\dag_{i\tau\ua}f_{i\tau\ua}-f^\dag_{i\tau\da}f_{i\tau\da})
 (c^\dag_{i\tau\ua}c_{i\tau\ua}-c^\dag_{i\tau\da}c_{i\tau\da})
 \\
 &+\fs\sum_{i\tau}J^\perp_\tau\bigl( f^\dag_{i\tau\ua}f_{i\tau\da}c^\da_{i\da}c_{i\ua}+
 f^\dag_{i\tau\da}f_{i\tau\ua}c^\dag_{i\ua}c_{i\da}\bigr),
  \\
 H^{(12)}_{cf}=\;
 &\frac{1}{2}J_{12}\sum_{i\tau}(f^\dag_{i\tau\da}f_{i\bar{\tau}\ua}c^\da_{i\ua}c_{i\da}
 +f^\dag_{i\tau\ua}f_{i\bar{\tau}\da}c^\da_{i\da}c_{i\ua}).
 \label{eqn:modelHam}
 \el
 \\
 \end{eqnarray}
Here $c^\dag_{\bk\si}$ create the single-band conduction states with dispersion denoted by $\epsilon_\bk$ and spin degeneracy by $\si=\ua,\da$, respectively. The second term in $H_0$ describes the 4f- quasi quartet of Fig.~\ref{fig:CEFscheme} with the two constituent Kramers doublets at lattice sites $\bR_i$ and with pseudo-spin index $\si=\ua,\da$. The doublets are denoted by the orbital index $\tau=1,2$ (lower and upper doublet, respectively).  The second term $H_{cf}$ corresponds to {\it elastic} exchange scattering of c-electrons from each doublet while the third term $H^{(12)}_{cf}$ is associated with the {\it inelastic} (off-diagonal) scattering between the orbitally different doublets. Due to tetragonal symmetry the former is described by sets of constants $J^z_\tau, J^\perp_\tau$ ($\tau=1,2$) for exchange parallel and perpendicular to the tetragonal z-axis. For the inelastic term $J^z_{12}=0$  and we define $J_{12}\equiv J^\perp_{12}$ (for details see appendix \ref{sec:app1}). In this work we restrict to the case where all effective couplings are  antiferromagnetic, although more general cases are possible (Appendix \ref{sec:app1}).
Furthermore we investigate only the paramagnetic phase within the constrained mean field approach. Then only the set of three transverse exchange parameters $J_\tau^\perp,J_{12}$ contribute to the ground state energy and quasiparticle energies within mean-field decoupling scheme \cite{lacroix:79,newns:87} of $H$ carried out in the following section.

\section{Constrained mean field theory}
\label{sec:meanfield}

For this purpose we introduce as (non-magnetic) order parameter, the effective homogeneous hybridization field generalized from Refs.\cite{lacroix:79,newns:87} or Refs.~\cite{zhang:00,li:10,liu:13,li:15} and defined by
\bea
\bl
&
V_\tau=\la \hat{V}_{i\tau}\ra =\la f^\dag_{i\tau\da}c_{i\da}+c^\dag_{i\ua}f_{i\tau\ua}\ra,
\\
&
V_{\tau\bartau}=\la \hat{V}_{i\tau\bartau}\ra 
=\la f^\dag_{i\tau\da}c_{i\da}+c^\dag_{i\ua}f_{i\bartau\ua}^{}\ra.
\label{eqn:OP}
\el
\eea
We set the gauge to $V_\tau=V^*_\tau$ and $V_{\tau\bartau}=V^*_{\tau\bartau}$ and restrict to the symmetric case $V_{\tau\bartau}=V_{\bartau\tau}$. It is easy to show that $V_{12}=V_{21}=\fs(V_1+V_2)$. Here $V_\tau$ describes the amplitude of each f-doublet to form a singlet state with a conduction electron at the same site.

The effect of large $f$-$f$ Coulomb interaction is to exclude doubly occupied states for electrons (holes) in the f-electron Hilbert space. This may be achieved by adding a Lagrange term according to 
\be
H^\lam=H+\sum_i\lam_i(\sum_{\tau} \hat{n}^f_{i\tau}-1); \;\;\;
\hat{n}^f_{i\tau}=\sum_\si f^\dag_{i\tau\si}f_{i\tau\si}^{}.
\label{eqn:constraint}
\ee
The constraint of single f-occupancy is enforced only globally in the mean field approach. Using Eq~(\ref{eqn:OP}) the  decoupling leads to 
\be
\bl
H^\lam_{mf}
=\;&
E^\lam_0+\tilH^\lam_{mf}
\\
=\;&
E^\lam_0+\sum_{\bk\si}\epsilon_\bk c^\dag_{\bk\si}c_{\bk\si}
+ \sum_{\bk\tau\si}\epla f^\dag_{\bk\tau\si}f_{\bk\tau\si}
\\
&
+\sum_{\bk\tau\si}\bav_\tau(f^\dag_{\bk\tau\si}c_{\bk\si}+c^\dag_{\bk\si}f_{\bk\tau\si}) .
\label{eqn:mfHam}
\el
\ee
Here we defined
\be
\bl
E^\lam_0
=\;
&\fs\sum_\tau J^\perp_\tau V^2_\tau+J_{12}V_{12}^2-\lam,
 \\
\epla
=\;
&\lam+(-1)^\tau\frac{\Delta_0}{2},
\el
\ee
furthermore we introduced the mixed hybridization amplitudes given by
\be
\bl
\left(
 \begin{array}{c}
\bav_1\\
\bav_2
\end{array}
\right)
=
-\fs
\left(
 \begin{array}{cc}
J^\perp_1+\fs J_{12}& \fs J_{12} \\
\fs J_{12}& J^\perp_2+\fs J_{12} \\
\end{array}
\right)
\left(
 \begin{array}{c}
V_1 \\
V_2 
\end{array}
\right).
\label{eqn:hybtrans}
\el
\ee
This transformation expresses the influence of nondiagonal (inelastic) exchange terms in the singlet formation.

%
% %%%%%%%%%%%%%%%%%%%%% figure %%%%%%%%%%%%%%%%%%%%%%%%%%%%
%%%%%%%%%%%%%%%%%%%%%%fig%%%%%%%%%%%%%%%%%%%%%%%%%%%%%%%
%
The total ground state energy (per site, $N_s$= number of sites) is then given by
\bea
\bl
E^\lam_{gs}/N_s=&\la H_{mf}^\lam \ra/N_s
\\=
&
\fN\sum_{\bk\si}\epsilon_\bk n^c_{\bk\si}
+\fs\De_0(n^f_2-n^f_1)
\\
&+\lam(n^f_1+n^f_2-1)
+\fs(J^\perp_1V_1^2+J ^\perp_2V_2^2)+J_{12}V_{12}^2 
\\\non
&-\fs\sum_\tau( J^\perp_\tau  V_\tau+J_{12}V_{12})\frac{2}{N_s}\sum_{\bk\si}\la f^\dag_{\bk\tau\si}c_{\bk\si}\ra,
\label{eqn:gsenergy}
\el
\\
\eea
where we defined 
\bea
 n^c_{\bk\si}=\la c^\dag_{\bk\si}c_{\bk\si}\ra;  \;\;\;
 n^f_\tau=\sum_\si \la  f^\dag_{i\tau\si}f_{i\tau\si} \ra,
\eea
as the c- and f-electron occupations, respectively. For noninteracting conduction electrons we have  $n^c_{\bk\si}=\Theta_H(\mu-\epsilon_{\bk\si})$ where $\mu$ is the chemical potential and $\Theta_H$ the  Heaviside function.

Minimization of the ground state energy with respect to $\lam$ and $V_\tau$ leads to 
\bea
n_f&=&n^f_1+n^f_2=1,\non\\
V_\tau&=&\fN\sum_{\bk\si}\la f^\dag_{\bk\tau\si}c_{\bk\si} \ra
,
\label{eqn:selfcons}
\eea
which express single occupancy constraint and hybridization selfconsistency respectively. Furthermore minimization with 
respect to $V_{12}$ only gives $V_{12}=\fs(V_1+V_2)$ consistent with the definitions in Eq.~(\ref{eqn:OP}). Then, introducing the 
spinors $\Psi^\dag_{\bk\si}=(c^\dag_{\bk\si},f^\dag_{1\bk\si},f^\dag_{2\bk\si})$ we obtain the bilinear mean-field Hamiltonian given by
\be
\tilde{H}_{mf}^\lam=\sum_{\bk m}\Psi^\dg_{\bk m} \hh_\bk \Psi_{\bk m}; \;\;\;
\hh_{\bk}=
\left(
 \begin{array}{ccc}
\epsilon_\bk& \bav_1 & \bav_2 \\
\bav_1& \epsilon_{01}^\lam& 0 \\
\bav_2& 0& \epsilon_{02}^\lam
\end{array}
\right).
%\nonumber
\label{eqn:bilHam}
\ee
We abbreviate $\De=\fs\De_0$ so that the effective f-level energies are $\epsilon^\lam_{01}=\lam-\De;\;\; \epsilon^\lam_{02}=\lam+\De$.  Furthermore $\epsilon^\lam_\bk =\epsilon_\bk-\lam$ will be used. 
%
% %%%%%%%%%%%%%%%%%%%%% figure %%%%%%%%%%%%%%%%%%%%%%%%%%%%
\begin{figure}
%\vspace{0.5cm}
%\includegraphics[width=0.4\columnwidth]{BZdisp.pdf}
\includegraphics[width=0.92\columnwidth]{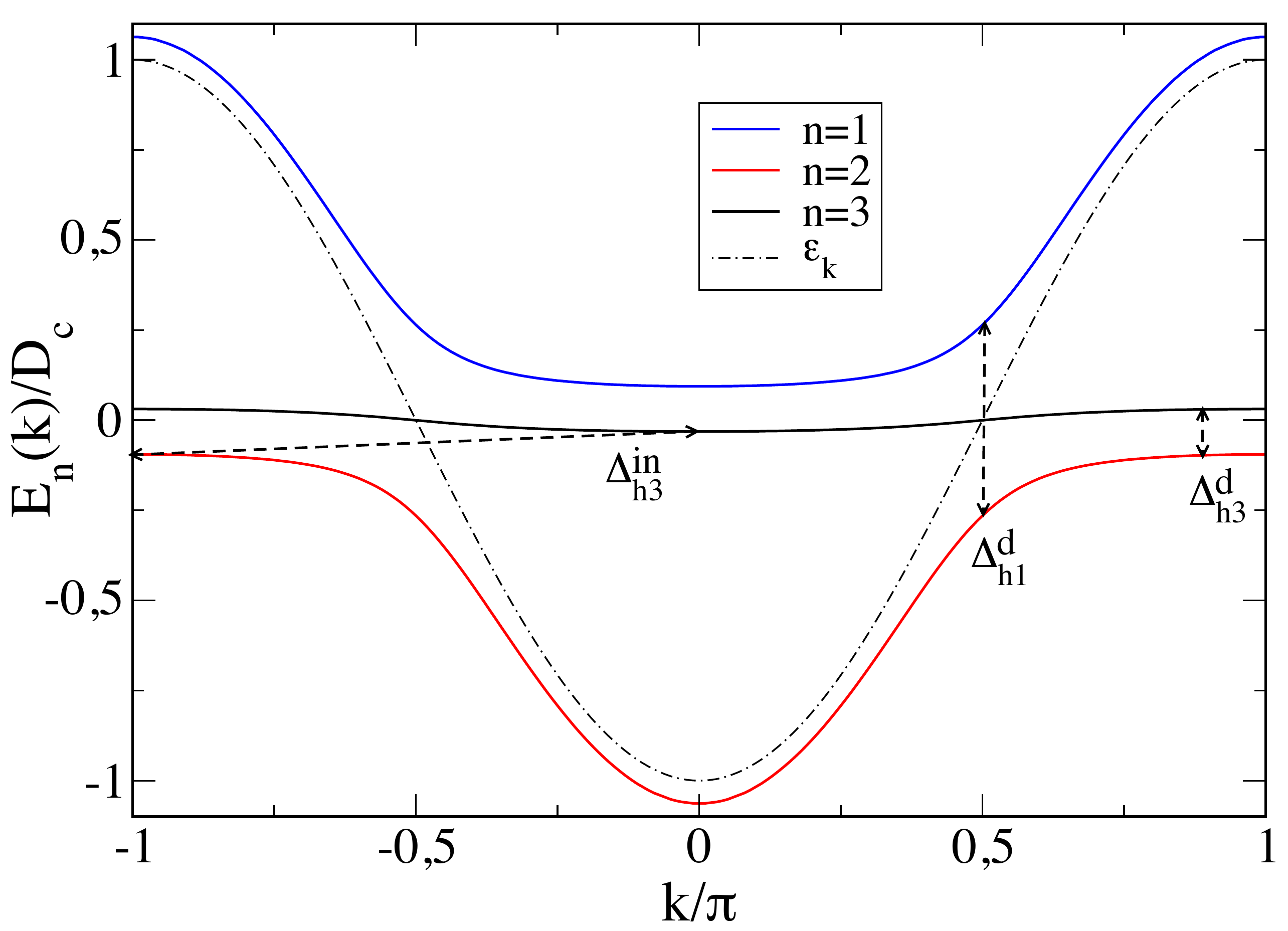}
\caption{Hybridized quasiparticle bands $E_{n\bk}$ (a) along straight BZ path $M(-\pi,-\pi)-\Gamma (0,0)-M(\pi,\pi)$ with 
some direct and indirect hybridization gaps indicated (Sec.~\ref{sec:widthgap}). There are {\it two} different direct hybridization gaps in the two-orbital KL model. The larger $(\De^d_{h1})$ is determined by the effective hybridzation scale $\bav$, the smaller $(\De^d_{h3})$  by the Kondo scale $T^*$.  The latter is of the same size as the indirect gap $\De^{in}_{h3}$. Parameters are  $\pJ_1=0.470, \pJ_2=0.767, J_{12}=0.2,  \De=0.056$. Then the particle-hole symmetric case is realized with selfconsistently determined $\bav_1=\bav_2=0.183$ and setting $\mu=-0.096$ (top of band $n=2)$. All energies in this and  subsequent figure are given in units of the half-conduction band with $D_c$.}
\label{fig:dispersion}
\end{figure}
%%%%%%%%%%%%%%%%%%%%%%fig%%%%%%%%%%%%%%%%%%%%%%%%%%%%%%%
%
\section{Quasiparticle bands and states}
\label{sec:quasi}

In this section we discuss the basic properties like dispersions, wave functions, hybridization gaps and effective masses of the quasiparticle  bands which may be found in closed form from diagonalizing the above bilinear mean field Hamiltonian.

\subsection{Hybridized dispersions and wave functions}
\label{sec:dispersion}

\begin{figure*}
%\vspace{0.5cm}
\includegraphics[width=0.95\columnwidth]{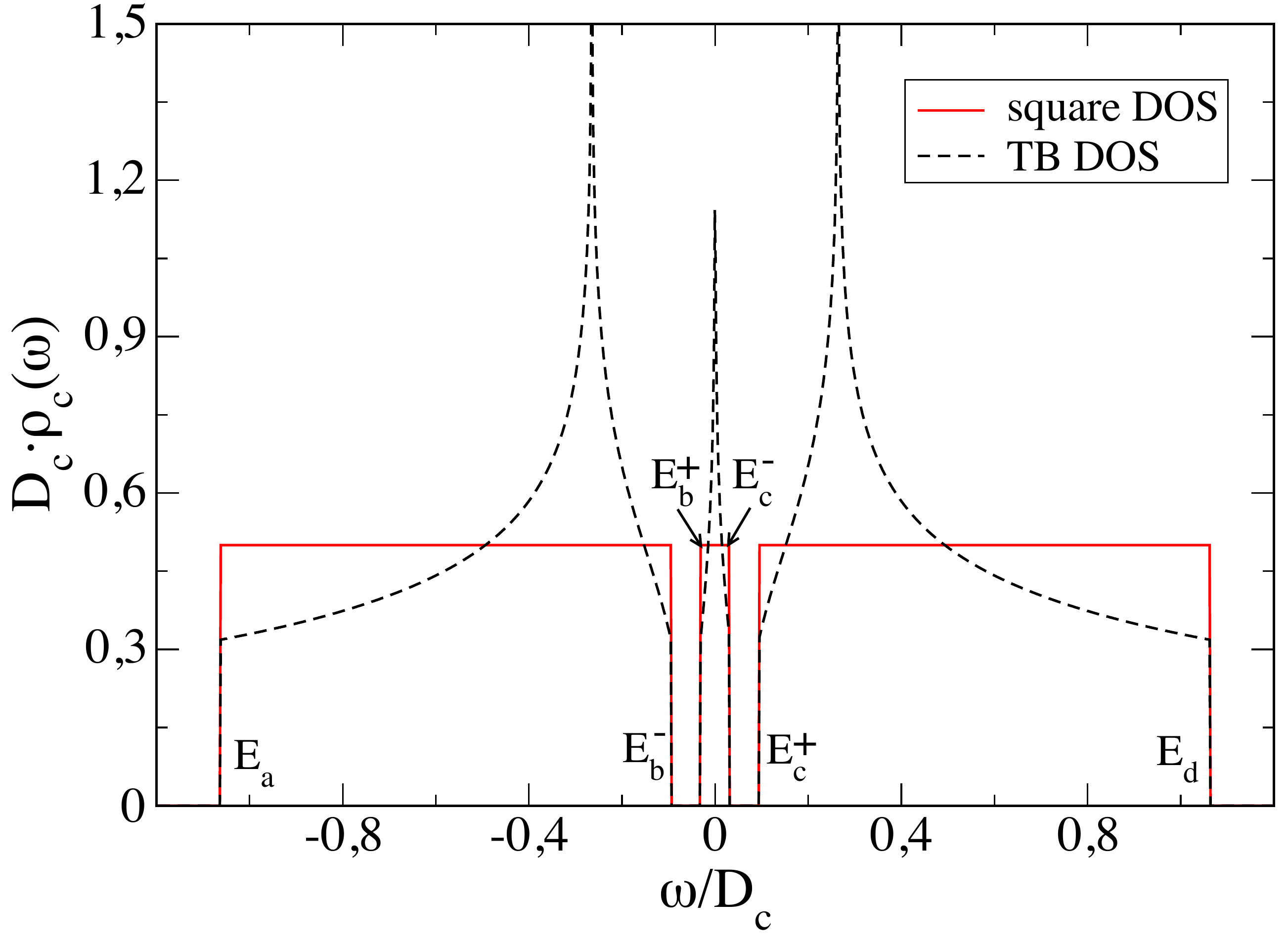}\hfill
\includegraphics[width=0.95\columnwidth]{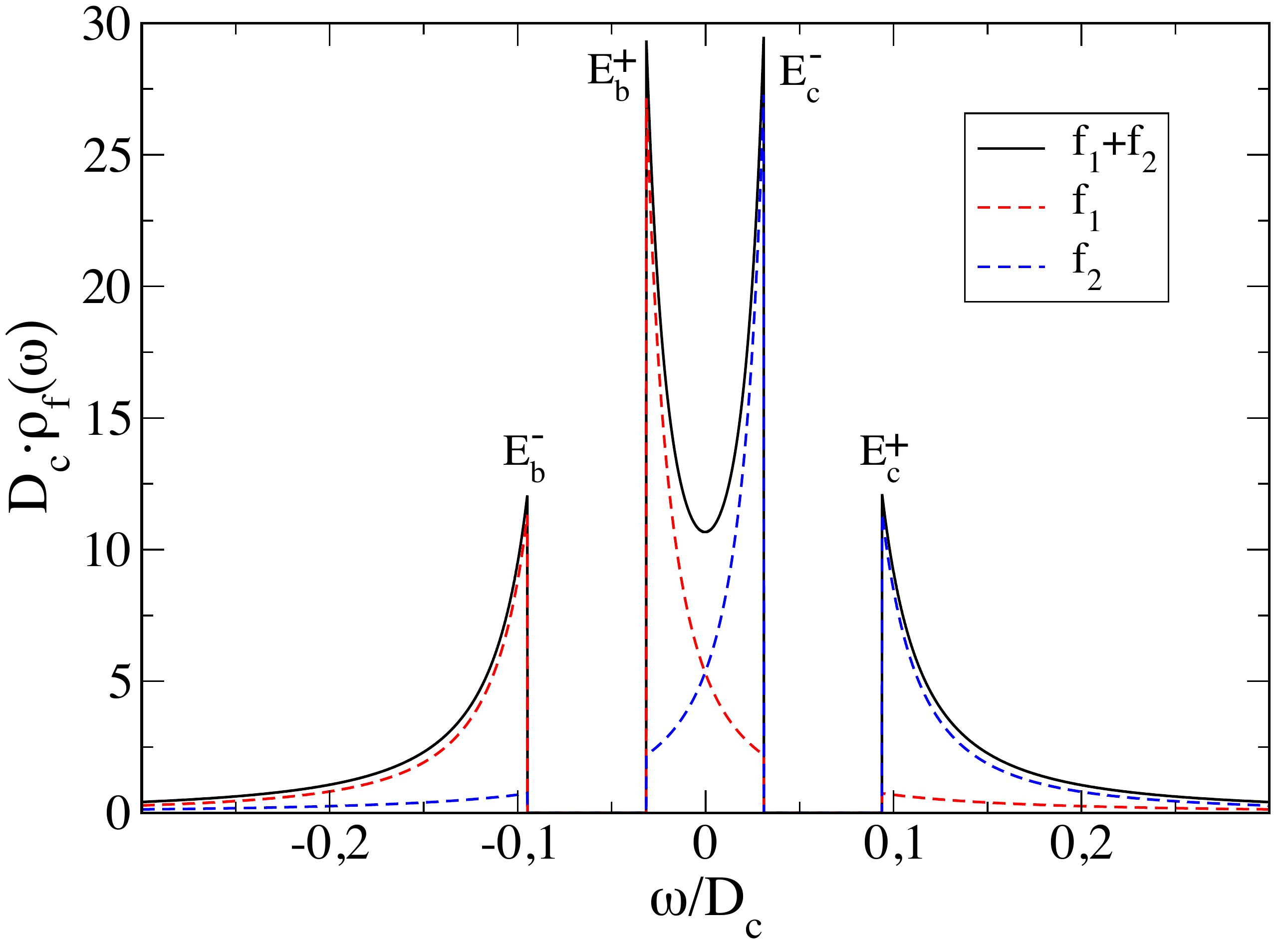}\\
\includegraphics[width=0.95\columnwidth]{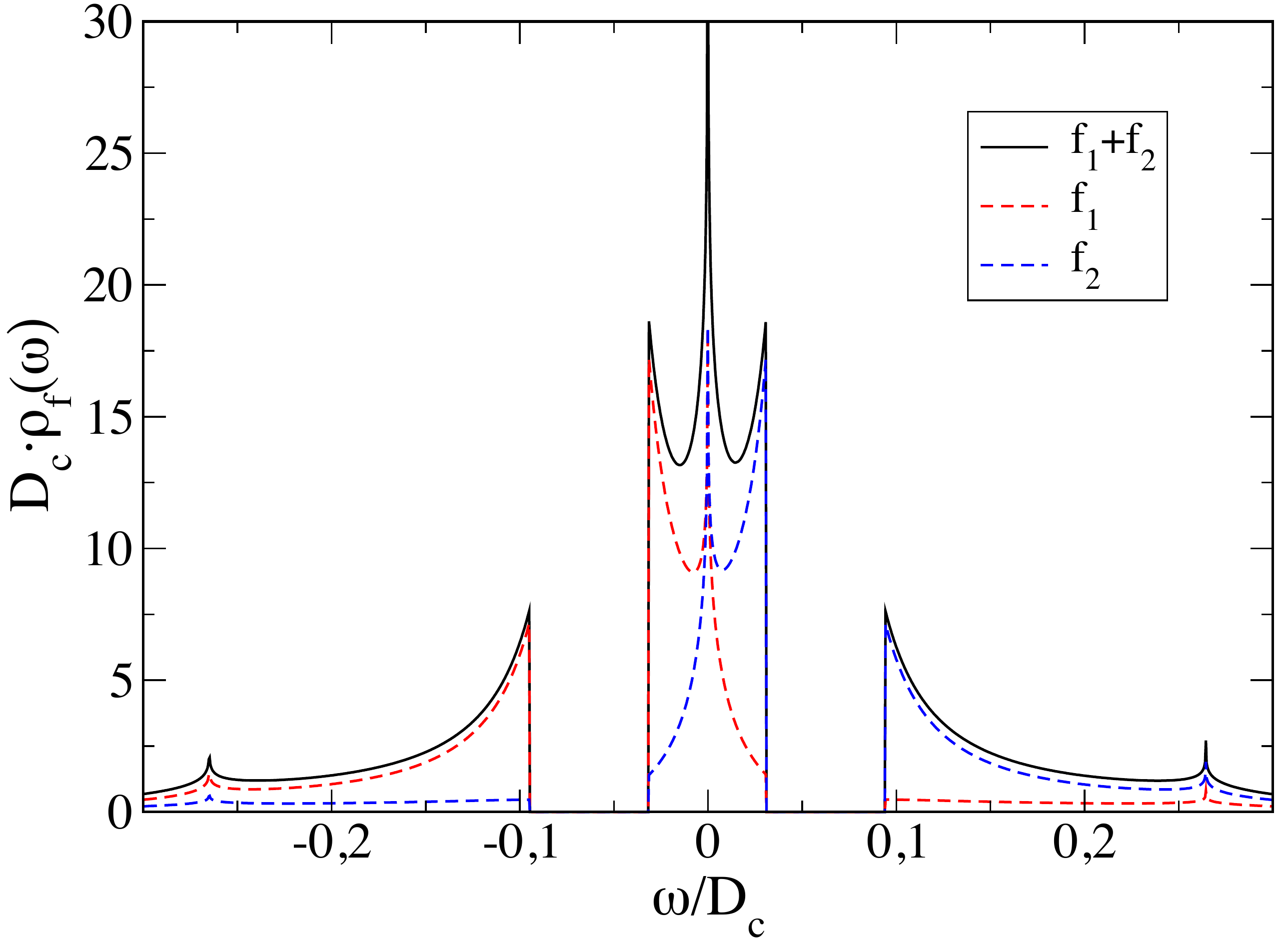}
\caption{Renormalized c- and f- DOS for parameters $\pJ_1=0.470, \pJ_2=0.767, J_{12}=0.2;\; 
\De=0.056$, same as in Fig.~\ref{fig:dispersion}. Then the particle-hole symmetric case is realized with selfconsistently determined $\bav_1=\bav_2=0.183$ and setting $\mu=-0.096$, corresponding to $\lambda =0$. (a) renormalized conduction electron DOS (Eq.~(\ref{eqn:DOSrenorm})) for square  and TB DOS models (Eq.~(\ref{eqn:DOS}). (b) Renormalized partial and total f-DOS based on square DOS model of $\rho^0_c(\omega)$ (c) same quantities based on TB DOS model. (In all figures the notation a,b,c is from left to right and top to bottom).}
\label{fig:spec_sym}
\end{figure*}

The hybridized quasiparticle dispersions are obtaind by finding the zeroes of the characteristic polynomial of $\hh_{\bk}$ given by
$d_\bk(\om)=det(\om-\hh)$ ($\omega_n$ is a Matsubara frequency). Its evaluation leads to
\be
\bl
d_\bk(\om)=&(\om-\epsilon_\bk)(\om-\eplax)(\om-\eplay)\\
&-\bav_1^2(\om-\eplay)-\bav_2^2(\om-\eplax)\\
=&(\om-E_{1\bk})(\om-E_{2\bk})(\om-E_{3\bk}).
\label{eqn:determinant}
\el
\ee
The three hybridized quasiparticle bands $E_{n\bk}$ $(n=1,2,3)$ are obtained from solving $d_\bk(\om)=0$ as
\bea
\bl
E_{1\bk}&=\lam +\frac{1}{3}\epsilon^\lam_\bk+2\sqrt[3]{r_\bk}\cos
(\frac{\phi_\bk}{3}),
\\
E_{2\bk}&=\lam +\frac{1}{3}\epsilon^\lam_\bk+2\sqrt[3]{r_\bk}\cos(\frac{\phi_\bk}{3}+\frac{2\pi}{3}),
\\
E_{3\bk}&=\lam +\frac{1}{3}\epsilon^\lam_\bk+2\sqrt[3]{r_\bk}\cos(\frac{\phi_\bk}{3}+\frac{4\pi}{3}).
\label{eqn:qpbands}
\el
\eea
%
% %%%%%%%%%%%%%%%%%%%%% figure %%%%%%%%%%%%%%%%%%%%%%%%%%%%
\begin{figure}
%\vspace{0.5cm}
\includegraphics[width=0.95\columnwidth]{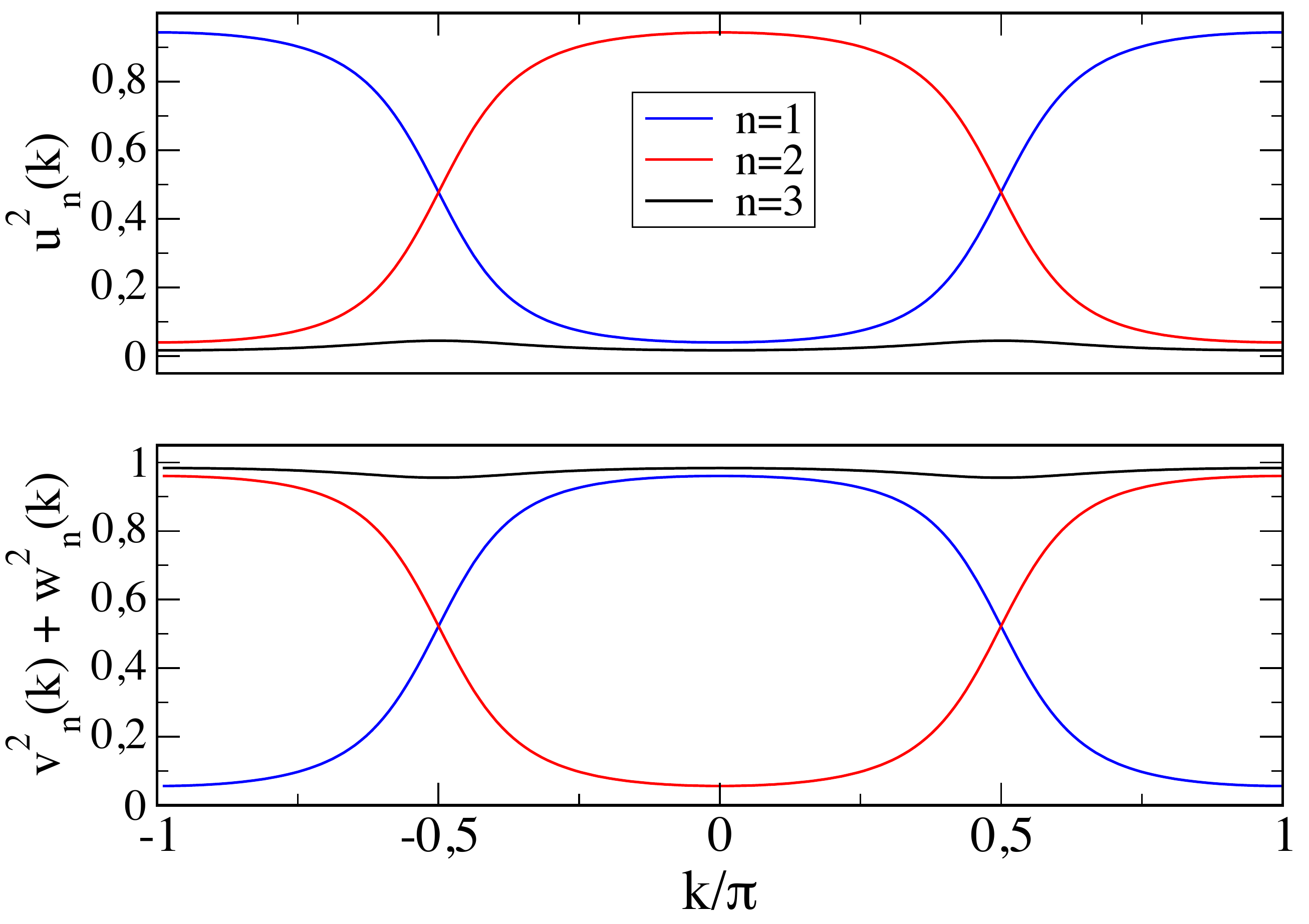}
\caption{Weights for c-electrons (top) and total f-electron weight (bottom) for hybridized quasiparticle bands along [11] $M-\Gamma -M$ direction (cf. Fig.~\ref{fig:dispersion}). For the central heavy band (n=3) conduction state weight $u_3^2(\bk)$ is small and total f-weight  $v_3^2(\bk)+ w_3^2(\bk)$ is close to one. For the upper (n=1) and lower (n=2) hybridized bands these weights alternate over the BZ. }
\label{fig:weights_cf}
\end{figure}
%%%%%%%%%%%%%%%%%%%%%%fig%%%%%%%%%%%%%%%%%%%%%%%%%%%%%%%
%
%
%\begin{widetext}
\begin{center}
\begin{table*}
    \centering
	\begin{tabular}[c]{lll}
	\hline\hline
	DOS cutoff \;\;\;\;\ \;\;\; & general $\bav,\de_s, \Delta$ \;\;\;\;\;\;\;\;\;\;\;\;\;  & \;\;\;\;\;special: $\Delta =0$ \;\;    \\
	\hline\\[-0.2cm]
	 $E_a=E_{20}$& $-D_c-\expa$ & \;\;\; $-D_c-\frac{\bav^2}{D_c}$ \\[0.1cm]
	 $E^-_b=E_{2\bQ}$ &$ -\fs \expa -\fs D_c\expb +\lam$ & \;\;\; $-\frac{\bav^2}{D_c}+\lam $ \\[0.1cm]
	$E^+_b=E_{30}$ & \;\;\;$ \fs \expa -\fs D_c\expb +\lam$   & \;\;\;\;\;  $ \lam$ \\[0.1cm]
	$E^-_c=E_{3\bQ}$ & $ -\fs \expa +\fs D_c\expb +\lam $ &\;\;\;\;\; $\lam$ \\[0.1cm]
        $E^+_c=E_{10}$ & \;\; $ \fs \expa +\fs D_c\expb +\lam    $  & \;\;\;\;\; $\frac{\bav^2}{D_c}+\lam$ \\[0.1cm]
        $E_d=E_{1\bQ}$& \;\; $D_c+\expa$ & \;\;\;\;\;  $D_c+\frac{\bav^2}{D_c}$ \\[0.4cm]        
         \hline\hline
	\end{tabular}
    \caption
    {Typical boundary cutoff values of the hybridized bands (Fig.~\ref{fig:dispersion}) and associated DOS functions
     in Fig.~\ref{fig:spec_sym}a.  Here $\bav^2=\bav^2_1+\bav^2_2, \de_s=\bav^2_1-\bav^2_2 $. 
     In the last column terms of order $\approx (\De^2/D_c)$ are suppressed. Furthermore $\lam \ll D_c$ is assumed.
      Particle-hole symmetry is preserved for $\de_s=0$ and $\lambda=0$.}
    \label{tbl:DOScutoff}
\end{table*}
\end{center}
%\end{widetext}
These relations generalize the simple one f-orbital hybridization formula mentioned in the introduction to the case of a CEF split
two-orbital system. Here the auxiliary functions $r_\bk,\phi_\bk$ are given by
\bea
r_\bk&=&\bigl(\frac{1}{3}[(\De^2+\bav^2)+\frac{1}{3}\epsilon^{\lam 2}_\bk]\bigr)^\frac{3}{2},
\\
\cos\phi_\bk&=&\frac{1}{2r_\bk}\bigl\{
\frac{1}{3}\epsilon^\lam_\bk[\frac{2}{9}\epsilon^{\lam 2}_\bk +(\De^2+\bav^2)]
-\De(\epsilon^\lam_\bk\De+\delta_s)
\bigr\},
\non
\label{eqn:auxiliary}
\eea
where we defined $\bav^2=\bav^2_1+\bav^2_2$  and  $\delta_s=\bav^2_1-\bav^2_2$.
The three quasiparticle bands $E_{n\bk}$ are shown in Fig.~\ref{fig:dispersion} for a special parameter set with $\bav_\tau$ evaluated selfconsitently as described in Sec.~\ref{sec:constraint}. For this figure (and the associated DOS in Fig.~\ref{fig:spec_sym}) we use parameters such that particle-hole symmetry is preserved.
There is a distinctive difference to the fully screened KL model (one orbital per conduction band): In the latter one 
has only an upper and a lower hybridized band similar to bands $n=1,2$ in Fig.~\ref{fig:dispersion} with $E_{3\bk}$ missing.
In the present underscreened model with two f-orbitals there is a third narrow band $E_{3\bk}$ within the large direct hybridization gap
$\Delta_{h1}^d$  in Fig.~\ref{fig:dispersion}. Thus the present model provides the existence of a heavy band extending throughout the Brillouin zone (BZ), contrary to the single f- orbital model where the heavy mass appears only on the zone center and boundary alternatively for the two bands.\\

In addition to the dispersion it is important to know the composition of Bloch states in terms of conduction and localized f-states for all wave vectors according to
\be
| \Psi^n_{\bk\si}\ra=u^n_\bk c^\dag_{\bk\si}|0\ra+v^n_\bk f^\dag_{1\bk\si}|0\ra+w^n_\bk f^\dag_{2\bk\si}|0\ra,
\label{eqn:blochstate}
\ee
where $n=1-3$ is the band index denoting band $E_{n\bk}$ and associated Bloch state $|\Psi^n_{\bk\si}\ra$. 
For these eigenstates of $\hh_\bk$ we obtain
\be
\bl
u^{n2}_\bk&=[\bav^2_1(E_{n\bk}-\eplay)+\bav^2_2(E_{n\bk}-\eplax)]^2/D^n_\bk,
\\
v^{n2}_\bk&=\bav^2_1(E_{n\bk}-\eplay)^2(E_{n\bk}-\epsilon_\bk)^2/D^n_\bk,
\\
w^{n2}_\bk&=\bav^2_2(E_{n\bk}-\eplax)^2(E_{n\bk}-\epsilon_\bk)^2/D^n_\bk ,  %\\[0.5cm]
\el
\ee
 and
 \be
 \bl
D^n_\bk=&(E_{n\bk}-\epsilon_\bk)^2[\bav^2_1(E_{n\bk}-\eplay)^2+\bav^2_2(E_{n\bk}-\eplax)^2]
\non\\
&+
[\bav^2_1(E_{n\bk}-\eplay)+\bav^2_2(E_{n\bk}-\eplax)]^2 .
\non
\label{eqn:blochcoeff}
\el
\ee
These coefficients fulfill the normalization condition $u^{n2}_\bk+v^{n2}_\bk+w^{n2}_\bk=1$. The c-electron weight $u^2_{n\bk}$ and total f- weight 
$v^{n2}_\bk+w^{n2}_\bk$ for the three bands is shown in Fig.~\ref{fig:weights_cf}. We note that the bands $n=1,2$ have partly c- or f- character which changes when $\bk$ moves across the BZ. (c.f. Fig.~\ref{fig:dispersion}). On the other hand the central narrow (heavy) band has predominantly f-electron character throughout the BZ. Its small dispersion is due to a small c-electron admixture.

\subsection{Band widths, hybridization gaps and effective masses}
\label{sec:widthgap}

Before we solve the central selfconsistency problem for the order parameters $\bav_\tau$ (or $V_\tau$)  we assume that 
they are already known and then  discuss certain characteristic features of the quasiparticle bands like widths and hybridization gaps. For  this purpose, to keep algebraic expressions simple  we restrict mostly to the symmetric case $\bav_1=\bav_2\equiv \bav$ with $\delta_s=0$ where the spectrum may show particle-hole symmetry for proper choice of $\mu$ such that $\lambda=0$. The conditon for its realization and also the asymmetric situation   $\bav_1\neq \bav_2$ will be discussed further in Sec.~\ref{sec:numerical} using the numerical results.\\

We first recall the two hybridized bands $E_{1,2\bk}$ in the one f-orbital KL model given in the Introduction. The essential 
nontrivial features of this model are: (i) a direct hybrization gap $\Delta_h^d=2\bav$ resulting from the anti-crossing of conduction band and renormalized f-level. (ii) a much smaller indirect hybridization gap $\Delta_h^{in}=2\frac{\bav^2}{D_c}\ll\Delta_h^d$  which defines a new low energy scale $T^*=\bav^2/D_c$ that is exponentially small compared to the conduction band width $2D_c$ (see also Appendix~\ref{sec:app2}). The fermionic quasiparticle picture of the KL is valid for temperatures much below the characteristric temperature $T^*$. The low energy scale is associated with very flat bands near the BZ center and boundaries. The corresponding Bloch states have high effective masses of the order $m^*/m_c\simeq D_c/T^* \gg 1$ ($m_c$ is the bare conduction band mass). The heavy mass quasiparticles and small (indirect) hybridization gap govern the low temperature $(T\ll T^*)$ thermodynamic, transport and dynamical physical properties of heavy fermion metals and Kondo semiconductors \cite{hewson:93,fazekas:99,thalmeier:05}.

In the underscreened two f-orbital model this simple picture has to be extended due to the appearance of an additional heavy (narrow) band within the main hybridization gap (Fig.~\ref{fig:dispersion}). The excitation spectrum may now be characterized by the following band widths $W_n$ and hybridization gaps $\De^d_h, \De^{in}_h$ which are energy differences of the three bands $E_{n\bk}$ at symmetry  $\bk-$ points, e.g. $\bk=0$, $\bQ'=(\pi/2,\pi/2)$ and $\bQ=(\pi,\pi)$. From Table~\ref{tbl:DOScutoff} we get $(\lam\ll D_c)$ the following 
\clearpage

\begin{figure}
%\vspace{0.5cm}
\includegraphics[width=0.92\columnwidth]{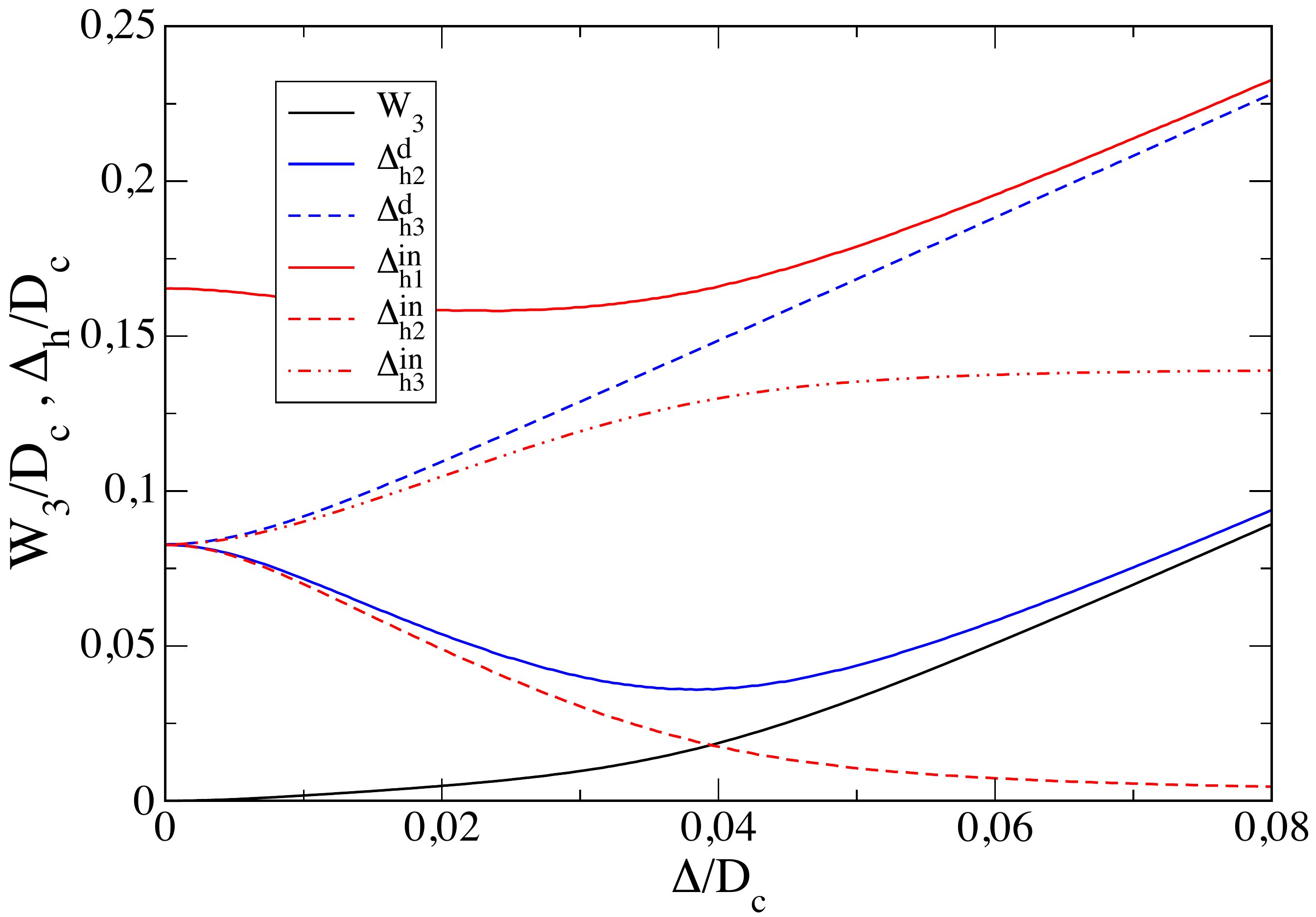}
\caption{Hybridization gaps and width of quasipartcle bands $E_{n\bk}$ $(n=1-3)$ (Fig.~\ref{fig:dispersion}) 
as function of quasi-quartet splitting $\De$. Exchange parameters are $\pJ_1=\pJ_2=0.7$ and $J_{12}=0.1$ with $\mu=-0.081$. Here $W_3=E_{3\bQ}-E_{30}$ is the central heavy band width
(full black). Direct hybridization gaps are $\De_{h2}^d=E_{10}-E_{30}$ (full blue) and $\De_{h3}^d=E_{3\bQ}-E_{2\bQ}$ (dashed blue). Indirect hybridization gaps are $\De_{h1}^{in}=E_{10}-E_{2\bQ}$, the overall indirect gap (full red) , $\De_{h2}^{in}=E_{10}-E_{3\bQ}$ (dashed red)  and  $\De_{h3}^{in}=E_{30}-E_{2\bQ}$ (dash-dotted, red). Some gaps behave non-monotonic as function of splitting $\De$.
}
\label{fig:widthgap}
\end{figure}

{\it Bandwidths:}\\
\bea
\bl
\non
W_{n=1,2}=&E_{n\bQ}-E_{n0}
\\
=&D_c+\fs\bigl(\frac{\bav^2}{D_c}+\frac{\De^2}{D_c}\bigr)
-\fs D_c\bigl[(\frac{\bav^2}{D_c^2})^2+\frac{\De_0^2}{D^2_c}\bigr]^\fs ,
\\
W_{3}=&E_{3\bQ}-E_{30}
\\
=&
-\bigl(\frac{\bav^2}{D_c}+\frac{\De^2}{D_c}\bigr)
+ D_c\bigl[(\frac{\bav^2}{D_c^2})^2+\frac{\De_0^2}{D^2_c}\bigr]^\fs .
\label{eqn:bandwidth}
\el
\eea
Here, for convenience we used both $\Delta_0$ and $\Delta=\frac{\De_0}{2}$. 
For orbital splitting $\Delta_0\rightarrow 0$ we have $W_{n=1,2}\rightarrow D_c$ and $W_3\rightarrow 0$.
The overall bandwith of  $E_{n=1,2\bk}$ bands is little affected by $\Delta$ (as is obvious from Fig.~\ref{fig:dispersion}), 
it is always of order $D_c$. On the other hand the width of the narrow central band  $E_{3\bk}$ depends sensitively on the orbital
splitting $\Delta_0$. For a finite dispersion of   $E_{3\bk}$ a finite splitting  $\Delta_0$ as well as a finite hybridization $\bav$  is required.
The $\Delta$-dependence of $W_3$ is shown  in Fig.~\ref{fig:widthgap} (full black line). This genuinely heavy band is a unique aspect of the quasi-quartet KL model.
In fact it directly characterizes the low energy scale because $W_3$ may be written as $(T^*=\bav^2/D_c)$
\bea
\bl
W_3=-T^*\!+(T^{*2}+\De_0^2)^\fs\simeq
\left\{
\begin{array}{ll}
\fs\De_0\bigl(\frac{\De_0}{T^*}\bigr)& \De_0 < T^* \\
\De_0\bigl[1-\bigl(\frac{T^*}{\De_0}\bigl)\bigr] &  \De_0 > T^*
\label{eqn:bandwidth3}
\end{array}
\right.
\non
\el
\\
\eea
where we neglected terms of order $(\De^2/D_c)$. As shown in Fig.~\ref{fig:widthgap} the orbital splitting $\De_0$ must be finite to get a  dispersive central band in Fig.~\ref{fig:dispersion}. For finite $\De_0$ when $T^*\rightarrow 0$ the central band degenerates into two flat subbands an energy $\De_0$ apart.\\

From Fig.~\ref{fig:dispersion} we also conclude that the bandstructure of the quasi-quartet model KL exhibits several hybridization gaps, in contrast to the single orbital model:
Three direct gaps $\De_{hi}^d$  and three indirect gaps  $\De_{hi}^{in}$  $(i=1-3)$ . For each class two are equivalent  for the symmetric case when $\bav_1=\bav_2$ and spectral symmetry holds. We then obtain:\\

%
%%%%%%%%%%%%%%%%%%%%%% figure %%%%%%%%%%%%%%%%%%%%%%%%%%%%
%%%%%%%%%%%%%%%%%%%%%%fig%%%%%%%%%%%%%%%%%%%%%%%%%%%%%%%
%
{\it Direct hybridization gaps:}
\bea
\De_{h1}^d&=&E_{1\bQ'}-E_{2\bQ'}=2(\bav^2+\De^2)^\fs,
\non\\
\De_{h3}^{d}&=&E_{3\bQ}-E_{2\bQ}=D_c\bigl[(\frac{\bav^2}{D_c^2})^2+\frac{\De_0^2}{D^2_c}\bigr]^\fs =\bigl(T^{*2}+\De^2_0\bigr)^\fs\non\\
&\equiv& E_{10}-E_{30}=\De_{h2}^d,
\label{eqn:hybgapdi}
\eea

{\it Indirect hybrization gaps:}
\bea
\De_{h1}^{in}&=&E_{10}-E_{2Q}=\bigl(\frac{\bav^2}{D_c}+\frac{\De^2}{D_c}\bigr)+D_c\bigl[(\frac{\bav^2}{D_c^2})^2+\frac{\De_0^2}{D^2_c}\bigr]^\fs\non\\
&&\simeq T^*+(T^{*2}+\De_0^2)^\fs,
\non\\
\De_{h3}^{in}&=&E_{30}-E_{2\bQ}=\bigl(\frac{\bav^2}{D_c}+\frac{\De^2}{D_c}\bigr)\simeq T^*\non\\
&\equiv& E_{10}-E_{3\bQ}=\De_{h2}^{in}
\label{eqn:hybgapin}
\eea
The first direct gap is also valid for general case $\bav_1\neq \bav_2$ $(\delta_s\neq 0)$.  For $\De\rightarrow 0$ $\De_{h1}^d\equiv 2\bav$ as in the single f-orbital KL and $\De_{h2,3}^{d}=T^*$. Thus in the two-orbital KL
there are also {\it direct} hybridization gaps  $\De_{h2,3}^{d}$ which are equal to the Kondo low energy scale $T^*$. 
 For the calculation of $T^*$ in terms of the microscopic model parameters the selfconsistency equation Eq.~(\ref{eqn:selfcons}) under the particle number constraint for $n_f$ has to be solved. We will do this for general $\bav_1,\bav_2$. Then, due to the lack of spectral symmetry all three direct as well as indirect hybridization gaps are inequivalent (Fig.~\ref{fig:widthgap}).\\

Corresponding to the flat parts of the dispersions we may also indroduce effective renormalized quasiparticle masses $m^*_n$ $(n=1-3)$ in relation to the underlying unhybridized tight binding model with a band mass $m_b=\frac{\hbar^2k_F}{D_c}$. We obtain 
\be
\frac{m_{1,2}^*}{m_b}=\bigl(\frac{D_c}{\bav}\bigr)^2=\frac{D_c}{T^*};\;\;\;
\frac{m^*_3}{m_b}\simeq\bigl(\frac{\bav}{\De}\bigr)^2=\frac{T^*D_c}{\De^2},
\ee
and therefore
\be
 \frac{m^*_3}{m^*_{1,2}}=\bigl(\frac{T^*}{\De}\bigr)^2,
\ee

Thus in the two-orbital model two types of heavy renormalized quasi-particle masses appear:
i) The $m^*_{1,2}$ effective mass of upper/lower hybridized partly heavy bands which are analogous to the one orbital model.
ii) The novel type $m^*_3$ effective mass of the central globally heavy band that appears only in the two-orbital model. 
The latter depends strongly on $\De$; it increases with decreasing CEF splitting and diverges for $\De=0$ leading to a flat unhybridized band. Any small residual quasiparticle interactions beyond the present mean field treatment will then localize these states into a twofold degenerate unscreened localized spin moment which is naturally expected for the present underscreened case. Thus the CEF splitting which completely suppresses the underscreening for $\De \gg T^*$ is necessary to stabilize the itinerancy of the central quasiparticle band. In fact its width and mass increase or decrease with increasing $\De$, respectively. We give a simple estimate for the localization due to renormalized quasiparticle interaction denoted $U^*$, assumed positive here. The scale of the latter is estimated from fluctuation expansion beyond mean-field solution for the one-orbital Anderson lattice model \cite{tesanovic:86}. In the present Kondo limit it is equivalent to
 $U^*=(I_{ex}/D_c)T^*$. The central narrow band will stay itinerant as long as $W_3\geq U^*$. This is approximately the case when $\Delta_0>(I_{ex}/D_c)T^*$. For much smaller $\Delta_0$, $W_3$ will shrink rapidly (Fig.~\ref{fig:widthgap}) and localization to a residual spin $S^*=\frac{1}{2}$ will occur. As mentioned in Sec.~\ref{sec:model} $S^*$ is weakly coupled to the remaining heavy band states and their coherent Fermi liquid behaviour is preserved. Eventually the effective intersite interactions in the lattice will lead to magnetic order of residual spins.
%
% %%%%%%%%%%%%%%%%%%%%% figure %%%%%%%%%%%%%%%%%%%%%%%%%%%%
%%%%%%%%%%%%%%%%%%%%%%fig%%%%%%%%%%%%%%%%%%%%%%%%%%%%%%%
%
%
% %%%%%%%%%%%%%%%%%%%%% figure %%%%%%%%%%%%%%%%%%%%%%%%%%%%
\begin{figure}
%\vspace{0.5cm}
{\centering
\includegraphics[width=0.9125\columnwidth]{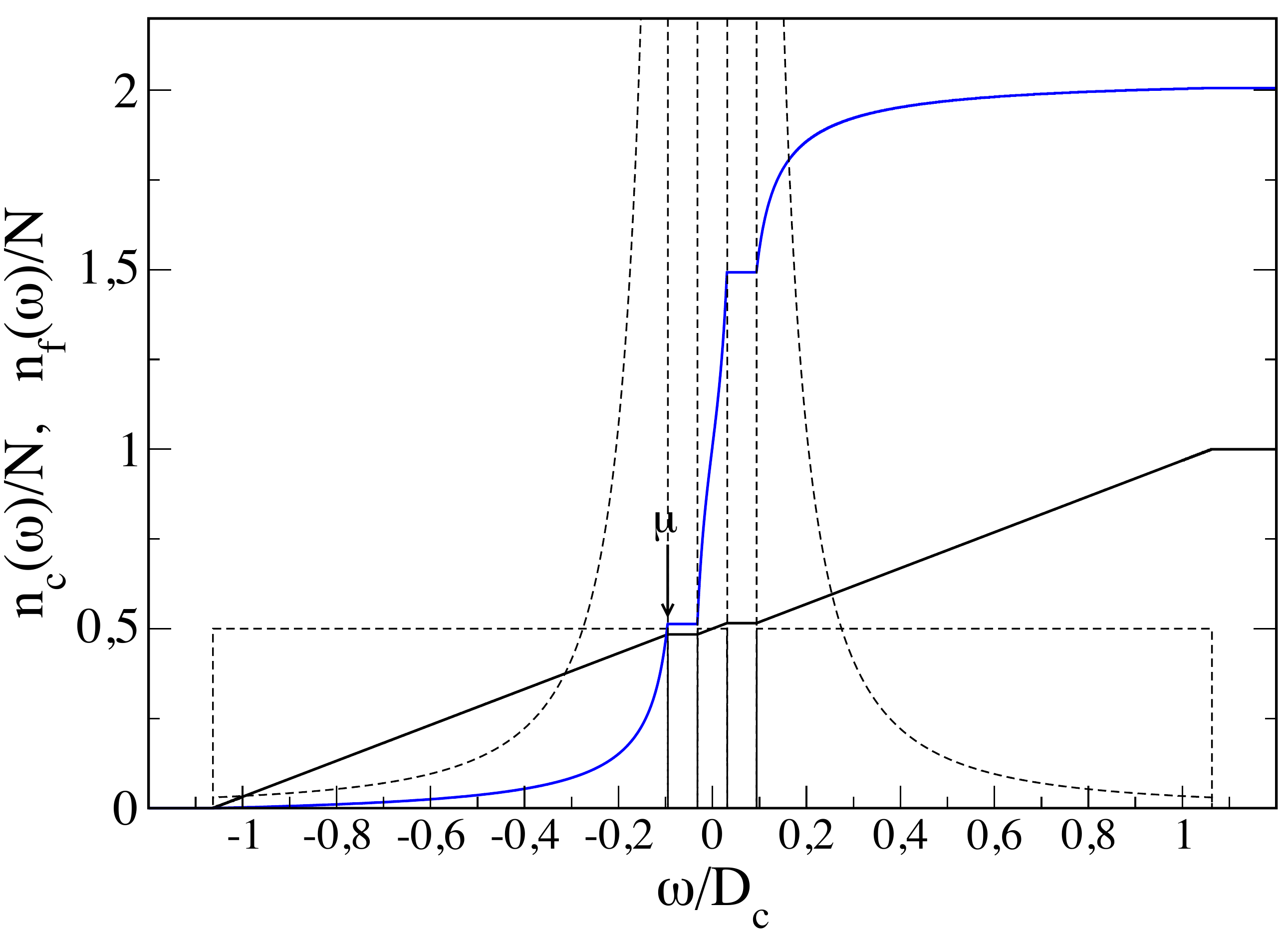}\\
\hspace{-0.2cm}
\includegraphics[width=0.95\columnwidth]{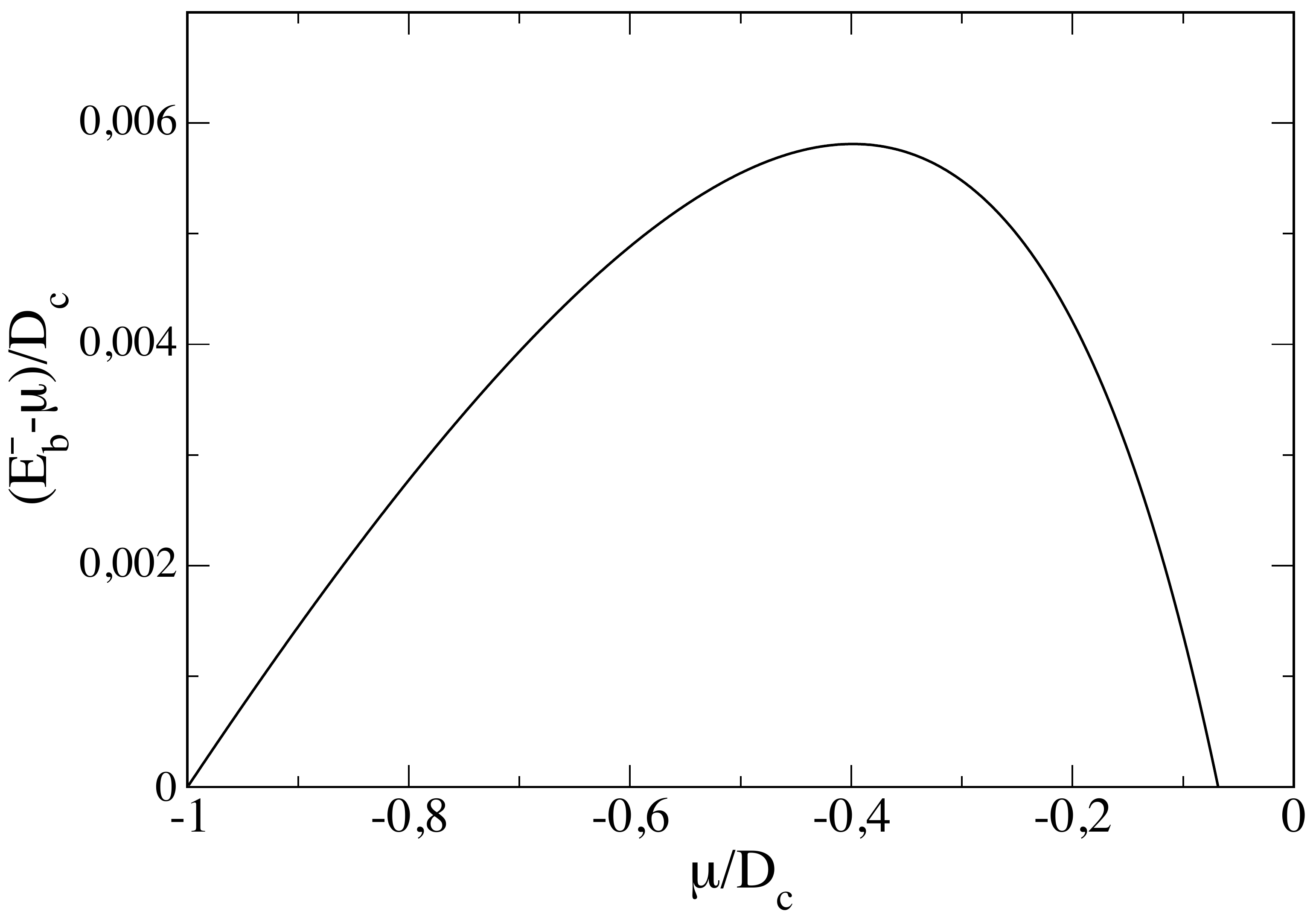}
}
\caption{(a) Integrated conduction and f-electron DOS $n_c(\omega)$ (full black) , $n_f(\omega)$ (full blue) for the symmetric case (parameters as in Fig.~\ref{fig:spec_sym}). By the  constraint the chemical potential $\mu=E^-_b$ lies near the upper edge (see (b)) of lower quasiparticle band such that $n_f(\mu)/N=\frac{1}{2}$. Here we are also below and close to half filling $n_c(\mu)/N=1/2$. When $\mu$ (or $n_c$) decreases, the DOS gap structure (dashed lines) is dragged along to lower energy to keep the f-occupation $n_f(\mu)/N=1/2$ fixed (see also Fig.\ref{fig:specsymm_chem}).
(b) Position of chemical potential with respect to upper edge $E_b^-$ of lower band. 
}
\label{fig:occ_cf}
\end{figure}
%%%%%%%%%%%%%%%%%%%%%%fig%%%%%%%%%%%%%%%%%%%%%%%%%%%%%%%
%
\section{Renormalized Green's functions and spectral functions}
\label{sec:Green}

The determination of Lagrange parameter $\lambda$ (the effective f-level position) from the selfconsistency equation and chemical potential $\mu$ from the particle number constraint is facilitated by the use of spectral functions. They may be computed from the Green's function matrix (in $c,f_1,f_2$ orbital space) of the model which is defined by 
\be
\hG_\bk(\om)=(\om-\hh_\bk)^{-1}=
\left(
 \begin{array}{ccc}
G_c& A_1& A_2 \\
A_1& G_{f1}& B \\
A_2& B& G_{f2}
\end{array}
\right)_{\bk,\om}
%\nonumber
\label{eqn:matGreen}
\ee
Each element may be given explicitly or in terms of its bare Green's function element (which is nonzero only
for the diagonal)  and the renormalized conduction electron Green's function $G_c(\om)$. Alternatively the elements may be presented in explicit form
as (using $d_\bk(\om)$ from Eq.~(\ref{eqn:determinant}))
\bea
\bl
G_c(\om)&=(\om-\eplax)(\om-\eplay)/d_\bk(\om)\\
G_{f1}(\om)&=\bigl[(\om-\epsilon_\bk)(\om-\eplay)-\bav^2_2\bigr]/d_\bk(\om)\non\\
G_{f2}(\om)&=\bigl[(\om-\epsilon_\bk)(\om-\eplax)-\bav^2_1\bigr]/d_\bk(\om)\non
\el
\\
\label{eqn:ExpGreen1}
\eea
 for the diagonal part and likewise for the off-diagonal terms:
\be
\bl
A_1(\om)&=\bav_1(\om-\eplay)/d_\bk(\om),
\\
A_2(\om)&=\bav_2(\om-\eplax)/d_\bk(\om),
\\
B(\om)&=\bav_1\bav_2/d_\bk(\om).
\label{eqn:ExpGreen2}
\el
\ee
The latter describe the mixing of orbital dynamics due to the Kondo interaction term. The pole structure of $d^{-1}_\bk(\om)$
is determined by the three quasiparticle energies (Eqs.~(\ref{eqn:determinant},\ref{eqn:qpbands})). The above form is therefore appropriate when,
e.g., one wants to calculate the dynamic magnetic susceptibility or optical conductivity in terms of quasiparticle excitation energies. For the moment we are only interested
in the orbitally projected density of states (DOS) functions. For this purpose the use of explicit quasiparticle bands may be circumvented, as was 
demonstrated already for the one-orbital KL model \cite{lacroix:79,newns:87}. In this case we can employ the following representations of diagonal  Green's function elements:
\bea
G_c(\om)&=&G_c^0(\om-\Sigma_c(\om))=
\frac{1}{[\om-\Sigma_c(\om)]-\epsilon_\bk}\non\\
G_{f\tau}(\om)&=&G^0_{f\tau}(\om)+\frac{\bav^2_\tau}{(\om-\epla)^2}G_c(\om)\non\\
\Sigma_c(\om)&=&\Sigma_{\tau}\frac{\bav^2_\tau}{\om-\epla};\;\;\;(\tau=1,2)
\label{eqn:ImpGreen1}
\eea
where $G_c^0(\om)=(\om-\epsilon_\bk)^{-1}$ and  $G_{f\tau}^0(\om)=(\om-\epla)^{-1}$ are the bare conduction electron and f-electron Green's functions and $\Sigma_c(\om)$ is the conduction electron self energy due to the Kondo interaction. The nondiagonal parts may 
be represented as
\be
\bl
A_\tau(\om)=&\frac{\bav_\tau}{(\om-\epla)}G_c(\om),
\\
B(\om)=&\frac{\bav_1\bav_2}{(\om-\eplax)(\om-\eplay)}G_c(\om). 
\label{eqn:ImpGreen2}
\el
\ee
%
%
% %%%%%%%%%%%%%%%%%%%%% figure %%%%%%%%%%%%%%%%%%%%%%%%%%%%%%%%%%%%%%%%%
\begin{figure}
%\vspace{0.5cm}
\includegraphics[width=0.95\columnwidth]{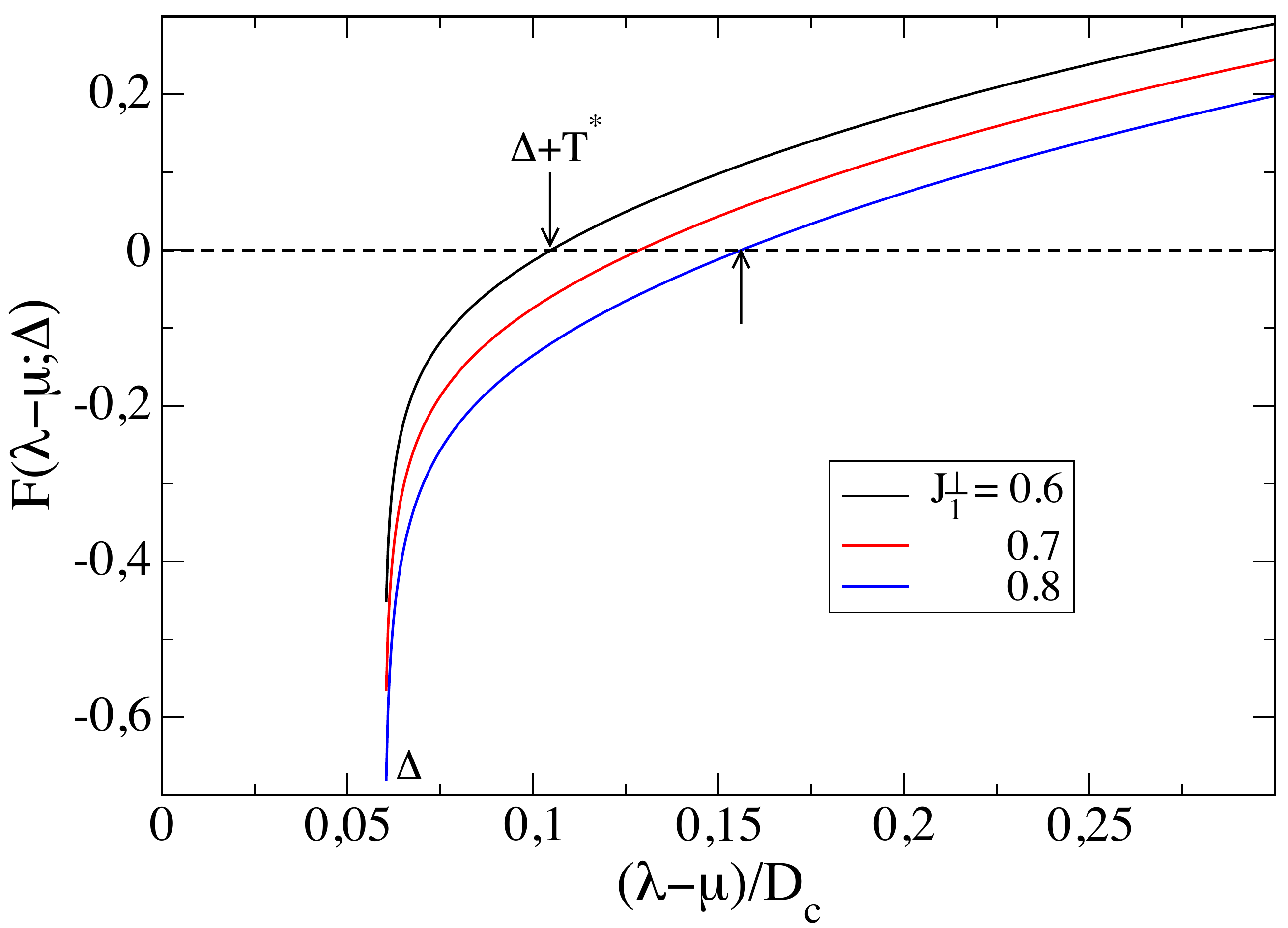}\hfill
\includegraphics[width=0.96\columnwidth]{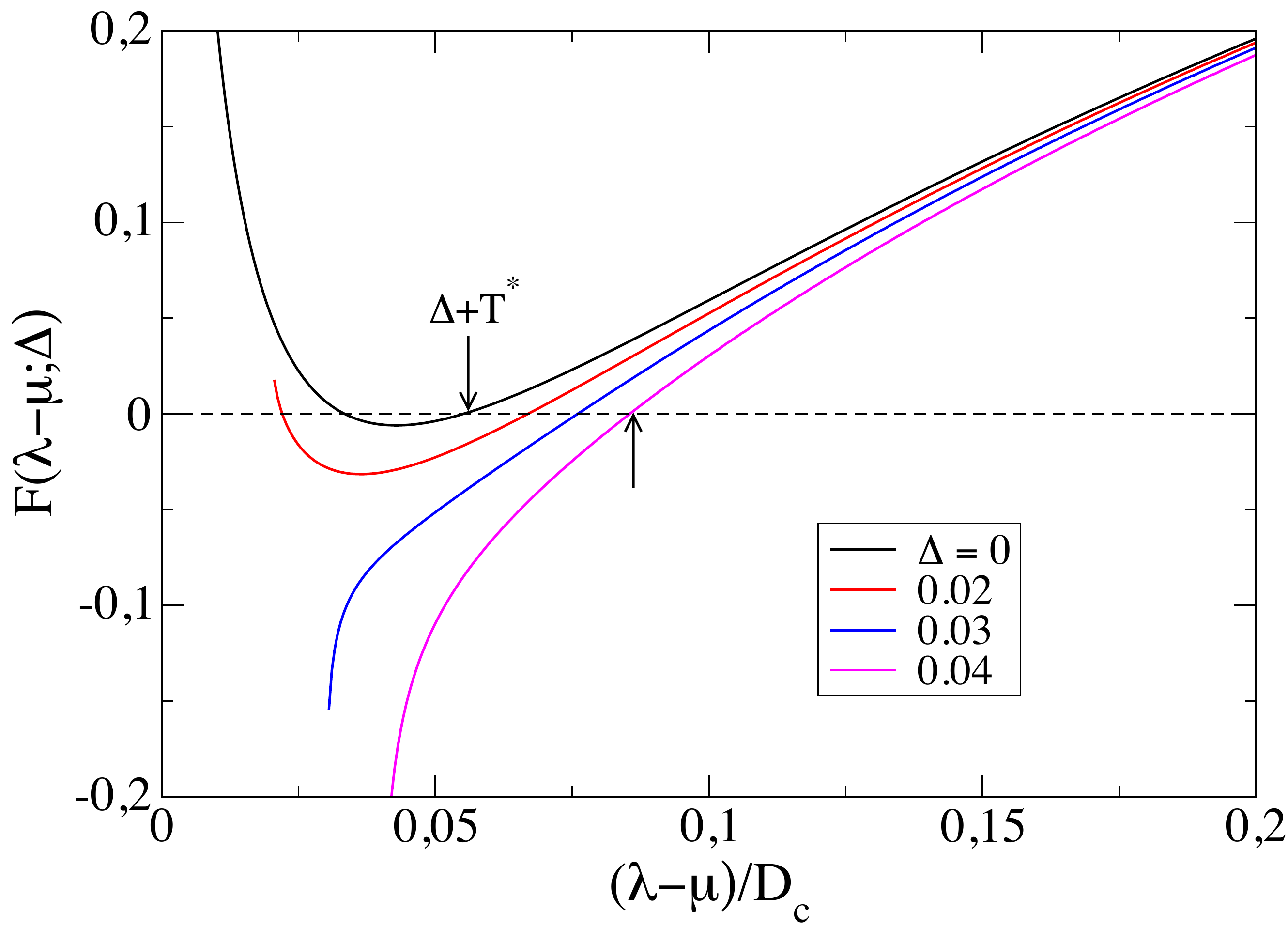}
\caption{Selfconsistency function $F(\lambda-\mu;\Delta)$. (a) for $\De=0.06$; 
$\pJ_2=0.6;\; J_{12}=0.1$ and various $J^\perp_1$. (b) for $\pJ_1=\pJ_2=0.6;\; J_{12}=0.1$ and various $\De$. The intercept with the dashed zero line give the solution for $\lambda-\mu=\De+T^*$. In (b) the lower intercept for small $\De$ is unphysical (it corresponds to higher ground state energy).}
\label{fig:faux}
\end{figure}
%%%%%%%%%%%%%%%%%%%%%%fig%%%%%%%%%%%%%%%%%%%%%%%%%%%%%%%%%%%%%%%%%%%%
%
Thus all Green's function elements of Eq.~(\ref{eqn:matGreen}) may be expressed by the renormalized conduction electron $G_c(\om)$ which contains the Kondo interaction effect via $\Sigma_c(\om)$. Then all spectral functions and generalized DOS functions can be calculated with 
\bea
\hat{\rho}(\om)=-\frac{1}{\pi}Im\frac{1}{N_s}\sum_\bk\hG(\bk\om)_{\om\rightarrow \omega +i\eta}.
\label{eqn:matDOS}
\eea
Using Eqs.~(\ref{eqn:ImpGreen1},\ref{eqn:ImpGreen2}) this can eventually be expressed via the bare conduction electron DOS $\rho_c^0(\omega)=(1/N_s)\Sigma_\bk\delta(\omega-\epsilon_\bk)$. There are two straightforward model expressions for this quantity:
\bea
\bl
\mbox{SQ-DOS:}\;& \rho_c^0(\omega)= \frac{1}{2D_c}\Theta_H(D_c-|\omega|),
\\
\mbox{TB-DOS:}\; &\rho_c^0(\omega)=\bigl(\frac{4}{\pi^2}\bigr)K\bigl(1-\frac{\omega^2}{D^2_c}\bigr)
 \frac{1}{2D_c}\Theta_H(D_c-|\omega|).
 \non
\label{eqn:DOS}
\el\\
\eea
%
%
%%%%%%%%%%%%%%%%%%%%%% figure %%%%%%%%%%%%%%%%%%%%%%%%%%%%
\begin{figure}
\includegraphics[width=0.90\columnwidth]{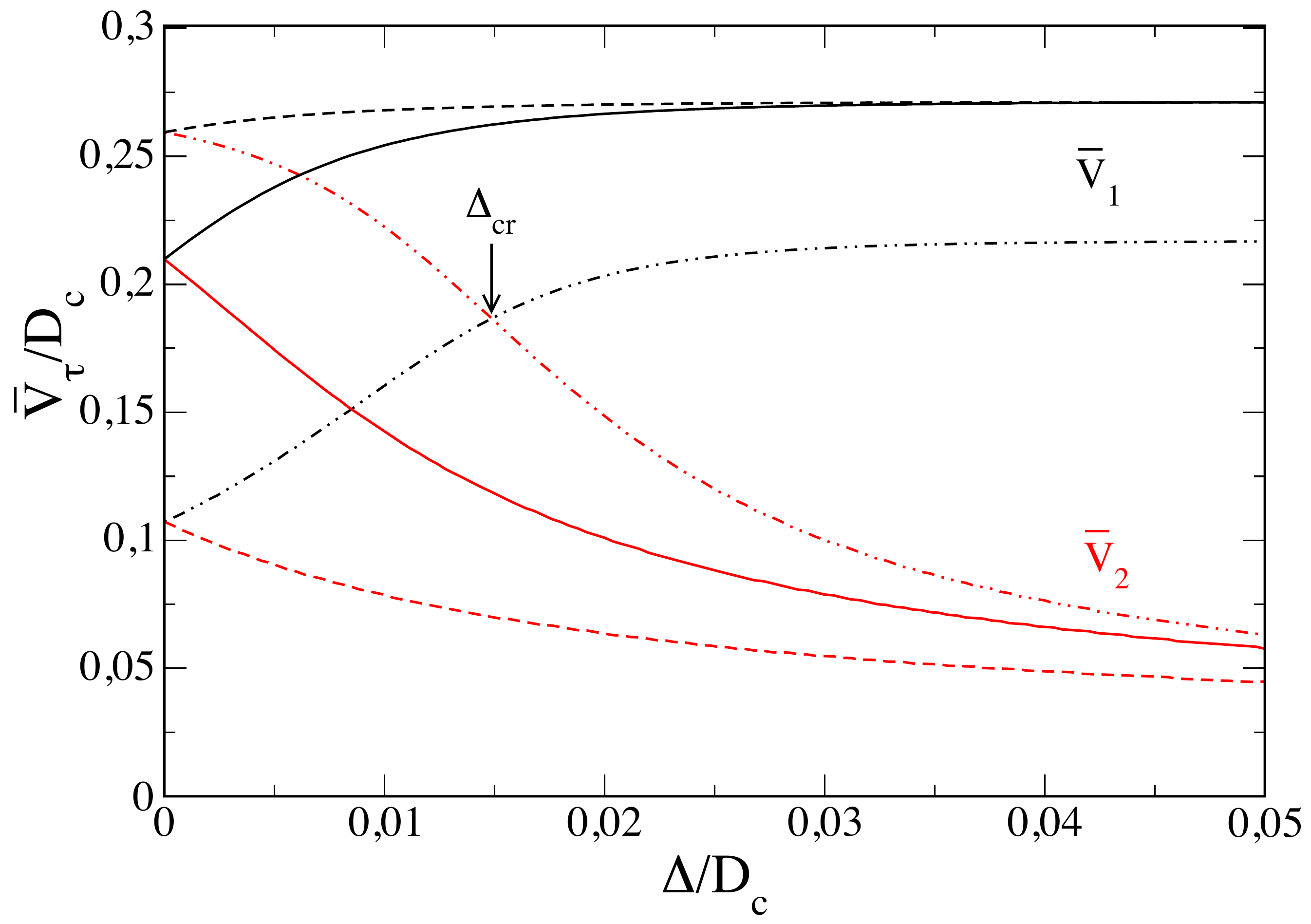}
\caption{Selfconsistent effective hybridizations $\bav_1$ (black), $\bav_2$ (red) as function of quasi-quartet splitting. In all cases $J_{12}=0.1$. $\pJ_1=\pJ_2=0.7$ (full); $\pJ_1=0.7,\pJ_2=0.6$ (dashed); $\pJ_1=0.6,\pJ_2 =0.7$ (dash-dotted). In the latter case the upper level has stronger exchange coupling $\pJ_2 > \pJ_1$, therefore there must be a crossing, of $\bav_1,\bav_2$ for a special finite $\Delta_{cr}$. For $\pJ_1=\pJ_2=0.7$ and $\De=0$ (symmetric $\bav_1=\bav_2$ case) we
note the agreement of $T^*=\bav^2/D_c\simeq 0.08$ with Fig.~\ref{fig:tstar}.}
\label{fig:vbar}
\end{figure}
%%%%%%%%%%%%%%%%%%%%%%fig%%%%%%%%%%%%%%%%%%%%%%%%%%%%%%%
%
%
%%%%%%%%%%%%%%%%%%%%%% figure %%%%%%%%%%%%%%%%%%%%%%%%%%%%
\begin{figure}
\vspace{-0.9cm}
\includegraphics[width=1.05\columnwidth]{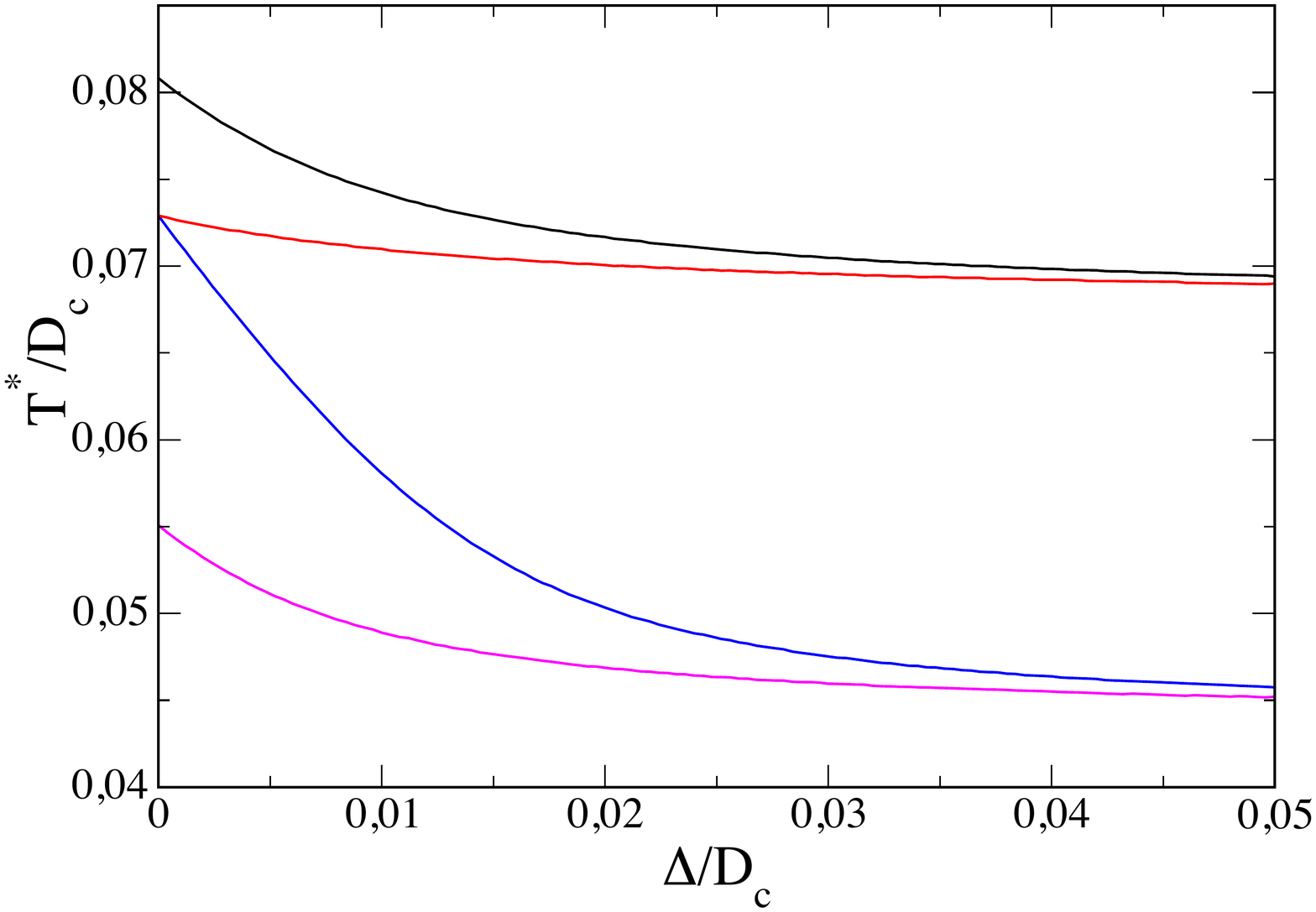}
\caption{Kondo energy scale $T^*$ as function of quasi-quartet splitting. In all cases $J_{12}=0.1$. (a) $\pJ_1=\pJ_2=0.7$ (black); $\pJ_1=\pJ_2=0.6$ (magenta); $\pJ_1=0.7,\pJ_2=0.6$ (red); $\pJ_1=0.6, \pJ_2=0.7$ (blue); the latter two  have to be degenerate for $\De=0$. }
\label{fig:tstar}
\end{figure}
%%%%%%%%%%%%%%%%%%%%%%fig%%%%%%%%%%%%%%%%%%%%%%%%%%%%%%%
%
%
The first is a constant square box DOS $\rho^0_c=1/2D_c$ within the interval $|\omega|\leq D_c$ ($2D_c=$ bare conduction band width).
The second possibility corresponds to the DOS of the 2D n.n. tight binding (TB) model also used for the dispersion in Fig.~\ref{fig:dispersion}, namely
$\epsilon_\bk=-(D_c/2)(\cos k_x+\cos k_y)$.
Due to the (complete) ellliptic function $K(x)$ (of the first kind) it has a logarithmic van Hove singularity at $\omega=0$ $(x=1)$.
Then, using Eqs.~(\ref{eqn:ImpGreen1},\ref{eqn:ImpGreen2}) all components of Eq.~(\ref{eqn:matDOS}) may be expressed by $\rho_c(\omega)$ 
 and $\Sigma_c(\omega)$. One obtains:
\bea
\rho_c(\omega)&=&\rho^0_c(\omega-\Sigma_c(\omega)),
\non\\
\rho_{f\tau}(\omega)&=&\frac{\bav^2_\tau}{(\omega-\epla)^2}\rho_c(\omega);\;\;\;  \rho_f(\omega)=\sum_\tau\rho_{f\tau}(\omega),
\non\\
\rho_{A\tau}(\omega)&=&\frac{\bav_\tau}{(\omega-\epla)}\rho_c(\omega).
\label{eqn:DOSrenorm}
\eea
 For special parameters (e.g. Figs.~\ref{fig:dispersion},\ref{fig:spec_sym}) the DOS functions may have the spectral symmetries $\rho_{c,f}(\omega)=\rho_{c,f}(-\omega)$ and  $\rho_{\tau}(\omega)=\rho_{\bar{\tau}}(-\omega)$, see Sec.~\ref{sec:numerical}.
The partial  f-DOS $\rho_{f\tau}$  and `hybridization DOS' $\rho_{A\tau}$ are derived from the renormalized  $\rho_c$  by multiplication  with (singular) prefactors.
Therefore we first discuss the renormalized conduction electron DOS itself.  For the square  DOS model of Eq.~(\ref{eqn:DOS}) it is shown in Fig.~\ref{fig:spec_sym}a. The two hybridization gaps and the central band are clearly visible. The cutoff energies of the bands (DOS and hybridization gap boundaries)  are designated in analogy to the single f-orbital KL model \cite{lacroix:79}. They are summarized in Table~\ref{tbl:DOScutoff} (see also  Fig.~\ref{fig:dispersion}).
The limits in the last column are for $\Delta\rightarrow 0$, i.e. the degenerate f- quartet. In this case $E^+_b=E^-_c=\lam$ are degenerate because the central $E_{3\bk}$ quasiparticle band is dispersionless in agreement with Eq.~(\ref{eqn:bandwidth3}) ($W_3=0)$.\\

The more physically relevant (total) f-DOS is shown in Fig.~\ref{fig:spec_sym}(b,c), obtained by using Eq.~(\ref{eqn:DOSrenorm}) with square-DOS (b) or TB-DOS (c) models, respectively. The typical DOS singularities at the main hybridization gap boundaries $E^-_b,E^+_c$ are visible  for $\bav_1=\bav_2=\bav$. They result from the flat portions of the lower/upper hybricized bands  $E_{n=1,2}(\bk)$ in Fig.~\ref{fig:dispersion}. Furthermore the overall flat central band $E_3(\bk)$  produces a strong 
narrow f- DOS peak around  $\lam$ whose width is given by Eq.~(\ref{eqn:bandwidth3}). It has mostly contributions from $f_1,f_2$- states (Fig.~\ref{fig:weights_cf}) which are shown as dashed lines in Fig.~\ref{fig:spec_sym}b,c.  The according DOS for the alternative TB model (which really corresponds to Figs.~\ref{fig:dispersion},\ref{fig:weights_cf}) is shown in Fig.~\ref{fig:spec_sym}c. Because of the van Hove singularity the DOS of the central band $E_3(\bk)$ shows additional structure., but essentially the qualitative band widths and hybridization gaps are unchanged.

\section{Selfconsistency relation and constraints}
\label{sec:constraint}

The DOS functions of Sec~\ref{sec:Green} may be used to express particle number constraints and selfconsistency condition as
\be
\bl
\label{eqn:implicit}
N\int_{-\infty}^\mu\rho_c(\omega)d\omega &= n_c, 
\\
N\int_{-\infty}^\mu\rho_f(\omega)d\omega &= n_f\equiv 1,
 \\
N\int_{-\infty}^\mu\rho_{A\tau}(\omega)d\omega &= V_\tau .
\el
\ee
Where $N=2$ is the doublet pseudospin degeneracy. These are four equations which determine $\mu,\lambda, \bav_1,\bav_2$ 
(or  $V_1,V_2$). To gain some insight it is useful to investigate analytical approximate solutions of these implicit equations. For simplicity we use the square DOS model in for $\rho_c^0(\omega)$ in Eq.~(\ref{eqn:DOS}). It has the advantage that the selfconsistency Eq.~(\ref{eqn:selfcons}) can be expressed as an algebraic equation and the Kondo scale $T^*$ depends smoothly on $\mu$.  We restrict to the symmetric case ($\bav_1=\bav_2$, $\lam=0$) with $n_c < 1$. \\

From the first of  Eq.~(\ref{eqn:implicit}) one obtains the conduction electron number $n_c(\omega)$. It is shown further below in Fig.~\ref{fig:occ_cf}a where it has linear behaviour in $\omega$ aside from the two small plateaux caused by the hybridization gap $\Delta^{d}_{h2,3}$ of Eq.~(\ref{eqn:hybgapdi}). Up to the plateau energy we have ($n_c=n_c(\mu)$):
\be
\label{eqn:Kondo1}
\mu=(n_c-1)D_c-\bigl(\frac{\bav^2}{D_c}+\frac{\De^2}{D_c}\bigr).
\ee
For the symmetric case according to Fig.~\ref{fig:spec_sym}a  $n_c=1-W_3/(2D_c)$ holds. For this value of $n_c$ when the conduction states of the lower band are filled Eq.~(\ref{eqn:Kondo1}) leads indeed, together with Eq.~(\ref{eqn:bandwidth}) to $\mu=E^-_b$. i.e. the chemical potential is pinned to the upper  edge $E^-_b$ of the $E_{2\bk}$ band where $n_f(\mu)=1$  (Fig.~\ref{fig:occ_cf}a,b). Assuming $\De\ll T^*$ we may approximate $n_c=1-\De^2/(T^*D_c)$. For smaller $n_c$ the chemical potential drops  below the edge $(E_b^- -\mu >0)$ (Fig.~\ref{fig:occ_cf}a,b).\\

Now we describe the essential procedure how the value of  $\lambda-\mu$ may be approximately obtained from the selfconsistency equation (last in Eq.~(\ref{eqn:implicit})). It may be written as
\be
V_\tau=\frac{N\bav_\tau}{2D_c}\ln\frac{\mu-\lambda-(-1)^\tau\De}{E_a-\lambda-(-1)^\tau\De}
\equiv \frac{N\bav_\tau}{2D_c}F_\tau .
\ee
Here we defined the auxiliary function
\bea
F_\tau(\mu,\lam)=\int^\mu_{-\infty}\frac{\hrho_c(\omega)}{(\omega-\epla)}
&&\simeq \ln\frac{\lambda-\mu+(-1)^\tau\De}{(D_c+\mu)+\lambda-\mu+(-1)^\tau\De}\non\\
&&\simeq
\left\{
\begin{array}{ll}
\ln\frac{\lam -\mu-\De}{Dc} \;\;(\tau=1)\\
\ln\frac{\lam -\mu+\De}{Dc} \;\;(\tau=2)
\label{eqn:ffunc}
\end{array}
\right.
\eea
where in the first approximation we used $E_a\simeq-D_c$,  the second approximation
holds for $\lambda, |\mu|,\De \ll D_c$. In  $F_1,F_2$ we must have $\lam-\mu >\De$ by definition.
Using the transformation given by Eq.~(\ref{eqn:hybtrans}) we get the matrix selfconsistency equation
\be
\left(
 \begin{array}{cc}
(N\rho_0)F_1-\frac{a_1}{A}& -\frac{b}{A} \\
 -\frac{b}{A}& (N\rho_0)F_2-\frac{a_2}{A}\\
\end{array}
\right)
\left(
 \begin{array}{c}
\bav_1 \\
\bav_2 
\end{array}
\right)
=0.
\label{eqn:matselfcons}
\ee
It implies a relation between the two effective hybridizations given by
\be
\bav_1^2=R\bav_2^2;\;\;\;R=\frac{(N\rho_0)F_2A-a_1}{(N\rho_0)F_1A-a_2},
\label{eqn:hybratio}
\ee
where from now on we use the definitions
\be
\bl
a_\tau&=-\fs(J^\perp_\tau+\fs J_{12}), \;\;\; b=-\frac{1}{4}J_{12} ,
\\
A&=a_1a_2-b^2=\frac{1}{4}[J^\perp_1J^\perp_2+\fs J_{12}(J^\perp_1+J^\perp_2)],
\\
B&=\fs(a_1+a_2)=-\fs\bigl[\fs(J^\perp_1+J^\perp_2)+\fs J_{12}\bigl].
\label{eqn:excon}
\el
\ee
The solution for $\lambda-\mu$ is then defined by the (secular) selfconsistency equation, i.e., the vanishing of the matrix determinant function $F(\lam-\mu;\De)$ given by
\be
\bl
F(\lam-\mu,\De)=&(N\rho_0)^2(a_1a_2-b^2)F_1F_2
\\
&-(N\rho_0)(a_1F_1+a_2F_2)+1=0.
\label{eqn:numselfcons}
\el
\ee
This is the fundamental equation for the problem which determines $\lambda-\mu$, the effective f-level position above the Fermi energy. It adjusts itself such that the $n_f=1$ constraint is respected.
 We may obtain the explicit solution in the simplest special case with $\Delta=0$;  $J^\perp_1=J^\perp_2=J_\perp$  and $J_{12}=0$ where we get the approximate $(|\mu|\ll D_c)$ result
\be
\bl
T^*(0):&=T^*_0=D_c\exp\bigl(-\frac{2}{g}\bigr); \;\;\; g=(N\rho_0)J_\perp
.
\label{eqn:barekondo}
\el
\ee
Here $T^*_0$ is the Kondo temperature of each individual doublet since we assumed they are decoupled ( $J_{12}=0$) . More general cases are
treated in Appendix \ref{sec:app2}. In the most general situation the selfconsistency equation can be solved for $\lambda-\mu$ only numerically. 
For large $\De$ is has a unique solution (Fig~\ref{fig:faux}), for smaller $\Delta$ there are two solutions. The larger is the physical one because it
corresponds to lower KL ground state energy and also connects adiabatically to the unique solution for larger $\De$ (Fig.~\ref{fig:faux}b).
To extract the Kondo scale $T^*(\De)$ for the general asymmetric case  from the numerical solution  of  Eq.~(\ref{eqn:numselfcons}) for $\lam-\mu$ we subtract the effect of $\Delta$ on $\lambda-\mu$ (Fig.~\ref{fig:faux}b) according to (see also Appendix~\ref{sec:app2})
\be
\lam-\mu=\De+T^*(\De).
\label{eqn:delkondo}
\ee
%
%
% %%%%%%%%%%%%%%%%%%%%% figure %%%%%%%%%%%%%%%%%%%%%%%%%%%%
\begin{figure}
%\vspace{0.5cm}
\includegraphics[width=0.95\columnwidth]{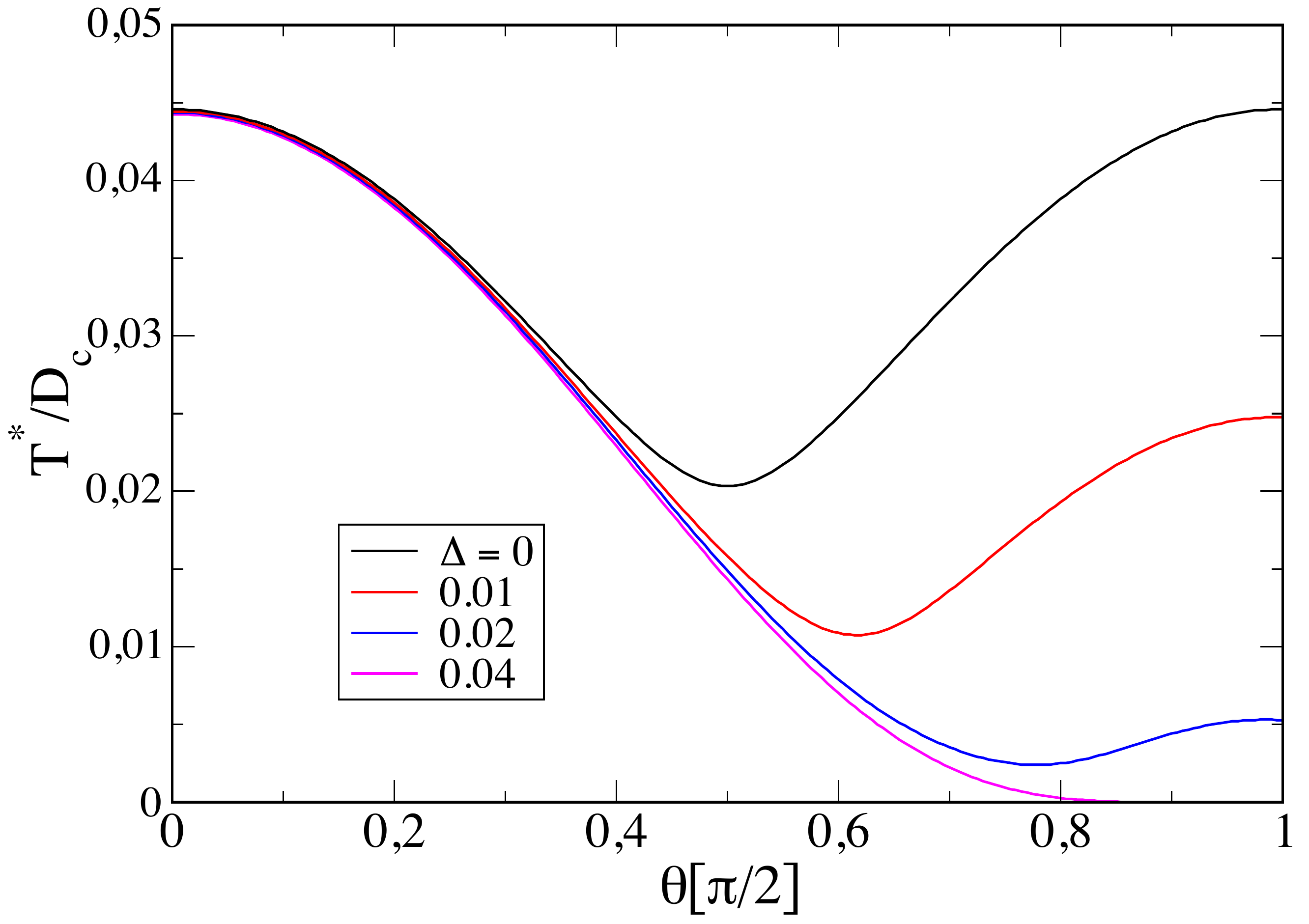}\hfill
\caption{(a)Variation of $T^*$ with $\theta=\tan^{-1}(\pJ_2/\pJ_1)$ for various CEF energies $\De$
and $\pJ=0.6, J_{12}=0.1$ ($\pJ_1=J_\perp\cos\theta,  \pJ_2=J_\perp\sin\theta)$. For $\De=0$ due to degenerate upper/lower level the curve is symmetric around $\theta =\frac{\pi}{4}$ or $\pJ_2=\pJ_1$. For increasing $\De$ the influence of the upper level on the
Kondo effect is progressively reduced, Therefore $T^*\rightarrow 0$ for $\theta\rightarrow \frac{\pi}{2}$, 
$(\pJ_1\rightarrow 0)$.
}
\label{fig:tstar_theta}
\end{figure}
%%%%%%%%%%%%%%%%%%%%%%fig%%%%%%%%%%%%%%%%%%%%%%%%%%%%%%%
%
Likewise we may then also compute the effective hybridisations $\bav_\tau$. They are obtained by using Eq.~(\ref{eqn:hybratio}) $\bav_1^2=R\bav^2_2$ together with the constraint $n_f=1$ (Eq.~(\ref{eqn:DOSrenorm})) for the f-electron occupation. The latter may be expressed as
\be
n_f=\hG_1\bav^2_1+\hG_2\bav^2_2=1.
\ee
From the two relations we finally obtain
\be
\bl
\bav^2_1&=\frac{R}{\hG_2+R\hG_1}; \;\;\;\;
\bav^2_2=\frac{1}{\hG_2+R\hG_1},\;\;\;
\label{eqn:bav}
\el
\ee
with $R$ given by Eq.~(\ref{eqn:hybratio}). In these expressions we defined $\hG_\tau=(N\rho_0)G_\tau$  with $G_\tau$ obtained from
\bea
&&G_\tau(\mu,\lam)=\int^\mu_{-\infty}\frac{\hrho_c(\omega)}{(\omega-\epla)^2}d\omega \non\\
&&=\frac{D_c+\mu}{[\lam-\mu+(-1)^\tau\De][(D_c+\mu)+\lam-\mu+(-1)^\tau\De]} \non\\
&&\simeq
\left\{
\begin{array}{ll}
\frac{1}{\lam -\mu-\De} \;\;(\tau=1)\\
\frac{1}{\lam -\mu+\De} \;\;(\tau=2)
\label{eqn:gfunc}
\end{array}
\right.
\eea
where again the approximation holds for  $\lambda, |\mu|,\De \ll D_c$.
Without splitting $(\De=0)$ and equal couplings $\pJ_1=\pJ_2$ this leads to $R=1$ and then the symmetric case $\bav^2_1=\bav^2_2$ is realized. In general, for a ratio of $\pJ_2/\pJ_1> 1$ there is always a value of $\De_{cr}>0$ for which the symmetric case occurs (Fig.~\ref{fig:vbar}). 
The original hybridization order parameters $V_1,V_2$ of Eq.~(\ref{eqn:selfcons}) may be obtained from the 
relation $V_\tau=(N\bav_\tau/2D_c)F_\tau$. With $\lambda-\mu$ and $\bav_\tau$ determined the spectral functions  of Eq.~(\ref{eqn:DOSrenorm}) with the proper selfconsistent energy scales can be determined. 
As a last step $n_c(\mu)$  may be obtained by numerical integration of the first equation in  Eq.~(\ref{eqn:implicit}). In order to satisfy the $n_f=1$ constraint the effectice f-level position adjusts itself such that the chemical potential $\mu$ aways lies close the upper edge $E_b^-$of the lower (n=2) band  for $n_c <1$ as shown in Fig.~\ref{fig:occ_cf}a,b.
%The upper $(n=1,3)$ bands must remain unoccupied due to this constraint.
\\
%
% %%%%%%%%%%%%%%%%%%%%% figure %%%%%%%%%%%%%%%%%%%%%%%%%%%%
\begin{figure}
%\vspace{0.5cm}
\includegraphics[width=0.95\columnwidth]{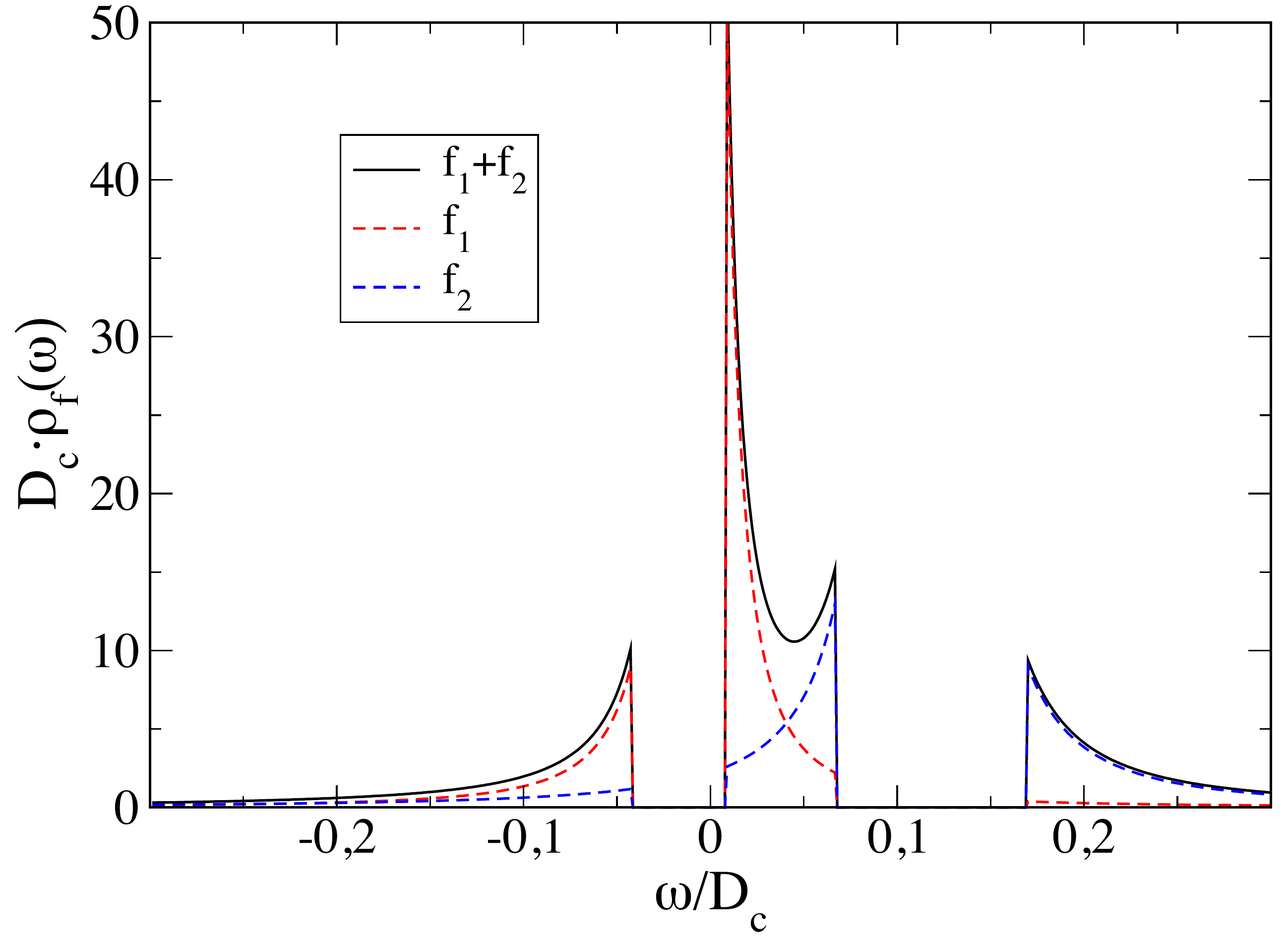}
\includegraphics[width=0.95\columnwidth]{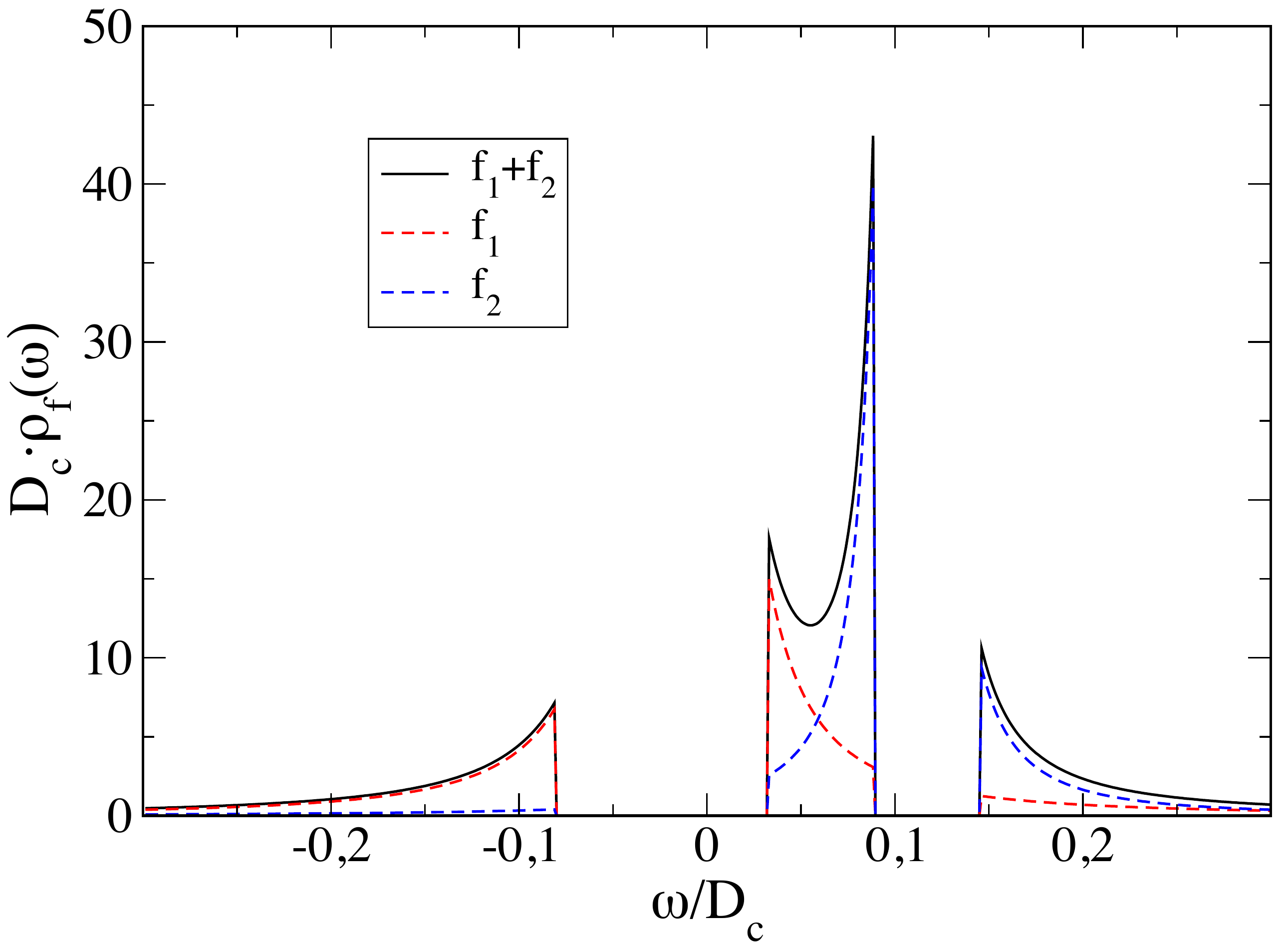}
\caption{Renormalized f-DOS of quasiparticles for various asymmetric cases without particle-hole symmetry.
(a) $\pJ_1=0.4, \pJ_2=0.8, J_{12}=0.2, \De=0.06, \mu =-0.04$  (b) $\pJ_1=0.6, \pJ_2=0.8, J_{12}=0.2, \De=0.06, \mu =-0.08$.
%(c)  $\pJ_1=0.7, \pJ_2=0.6, J_{12}=0.2, \De=0.07, \mu =-0.04$. 
When the ratio $\pJ_2/\pJ_1$ decreases and $\Delta$ grows the central band DOS shifts to larger energies, increasing the left ($\De^{in}_{h3}$) and decreasing the right ($\De^{in}_{h2}$) hybridisation gap. The width of the central band also increases with $\De$.}
\label{fig:spec_asym}
\end{figure}
%%%%%%%%%%%%%%%%%%%%%%fig%%%%%%%%%%%%%%%%%%%%%%%%%%%%%%%
%

Finally we give a closed expression for the ground state energy in terms of  the selfconsistently determined $\lam$, order parameters $\bav_\tau$ and auxiliary quantities discussed above. Using Eq.~(\ref{eqn:gsenergy}) we can write $E^\lam_{gs}$ as
\bea
&&E_{gs}^\lam/N_s=(N\rho_0)K_b-\De(N\rho_0)\bigl[\bav^2_1G_1-\bav^2_2G_2\bigr]\non\\
&&-\fs(N\rho_0)^2\bigl[(J^\perp_1+\fs J _{12})F_1^2\bav_1^2+(J^\perp_2+\fs J_{12})F_2^2\bav_2^2\bigr] \non\\
&&-\fs(N\rho_0)^2J_{12}F_1F_2\bav_1\bav_2 
\label{eqn:gsenergy_expl}
\eea
In the first term we introduced another auxiliary function for the total enery of the renormalized conduction band 
as given by
\be
K_b(\mu)=\int^\mu_{-\infty}\hrho_c(\omega)\omega d\omega =\fs(\mu^2-E_a^2).
\ee
The ground state energy of the uncoupled system without exchange terms is obtained as
\be
E_{gs0}/N_s=\fs\bigr[(N\rho^c_0)(\mu^2-D_c^2)-\De\bigl],
\label{eqn:gsenergy0}
\ee
and serves as a reference for the (negative) energy gain 
$\delta E_{gs}(\De ;\mu)/N_s=(E_{gs}^\lam-E_{gs0})/N_s$ due to the KL quasiparticle formation. For the most simple case  with only the ground state active $(\pJ_2=J_{12}=0)$ we obtain, using $E_a\simeq-(D_c+T^*)$ the result  $\delta E_{gs}/N_s=-T^*[1+\ln(D_c/T^*)]$, therefore the Kondo energy gain is of order $T^*$.

\section{Numerical solution for the selfconsistency relation and spectral functions}
\label{sec:numerical}

The general selfconsistency Eq.~(\ref{eqn:numselfcons}) has no closed solution for the central quantity $\lam-\mu$, except in the simplest case $\De=0$ (see Appendix \ref{sec:app2}). Since the dependence of Kondo lattice properties on the quasi-quartet splitting is an imortant issue of this work we now have to solve it numerically. This means we are looking for the root $\lam(\De)-\mu$ of $F(\lam-\mu,\Delta)=0$ (Eq.~\ref{eqn:numselfcons}). Alternativeliy one may solve directly for $T^*(\De)$ via the iterative equation Eq.~(\ref{eqn:tkiterate}).
The function  $F(\lam-\mu,\Delta)$ is shown in Fig,~\ref{fig:faux} for sets of exchange ($\pJ_1,\pJ_2,J_{12}$) and fixed $\De$ (a) and for fixed exchange set and various $\De$ (b). The $(\lambda-\mu)$- position of the zero gives the Kondo scale (Eq.~(\ref{eqn:delkondo}))  and determines the selfconsistent solution for $\bav_\tau$ using Eq.~(\ref{eqn:bav}).\\ 

The resulting Kondo low energy scale $T^*(\De)$ is shown in Fig.~\ref{fig:tstar} for various exchange parameter sets as function of $\De$. As expected it decreases with increasing quasi-quartet splitting because the spin-flip processes (elastic within excited $\Ga_7$ level and inelastic between $\Ga_6\leftrightarrow\Ga_7$ will be suppressed. On the other hand for the real quartet case $\De =0$ the Kondo scale $T^*$ becomes symmetric with respect to $\pJ_1,\pJ_2$ (red and blue curve in (a)).

When $T^*$ or likewise $\lambda-\mu$ is known the effective hybridizations $\bav_\tau$ which determine quasiparticle bands and spectrum can be calculated. It is shown in Fig.~\ref{fig:vbar} for essentially the same parameters as in Fig.~\ref{fig:tstar}. Their relative size of $\bav_1$ (black) and $\bav_2$ (red)  depends crucially on the order of $\pJ_1,\pJ_2$. For $\pJ_1>\pJ_2$ we naturally  have $\bav_1>\bav_2$ already at $\De=0$ and their difference increases slightly with increasing $\De$ (dashed lines). For $\pJ_1=\pJ_2$ degeneracy at $\De=0$ must occur. But in both cases for finite $\De$ we have always $\bav_1>\bav_2$ which will lead to  an asymmetric spectrum as discussed below. The most interesting case is $\pJ_2>\pJ_1$, i.e. when the excited $\Ga_7$ has a larger exchange coupling than the ground state $\Ga_6$. Therefore when $\De=0$ we must also have $\bav_2 > \bav_1$. Because the Kondo effect for the excited state decreases rapidly for increasing $\De$ so must $\bav_2$. Then necessarily a crossing of both curves where $\bav_1=\bav_2$ at a special value $\De_{cr}$ that depends on the set $(\pJ_1,\pJ_2,J_{12})$. At this value the spectrum may be symmetric for a proper choice of the chemical potential (Fig.~\ref{fig:spec_sym}).\\

Using these selfconsistent values of $\lambda, \bav_\tau$ we may calculate the quasiparticle bands from Eq.~(\ref{eqn:qpbands}), with a special particle-hole symmetric example given in Fig.~\ref{fig:dispersion}. Its main heavy-band features are characterized by the band widths and hybridization gaps given in Sec.~\ref{sec:widthgap}. For the general band structure  the most important ones are depicted in Fig.~\ref{fig:widthgap} as function of $\Delta$. For this case $\pJ_1=\pJ_2$ and the band structure will be asymmetric for $\De>0$.  We can clearly see that the band width $W_3$ of the central $E_{3\bk}$ heavy band first increases quadratically and then linearly with $\Delta$ compatible with by Eq.~(\ref{eqn:bandwidth3}) (full black line). The evolution of the direct gaps is shown by the red curves and the indirect gaps by blue curves. We observe that one direct and indirect gap show non-monotonic behaviour which is even more pronounced when $\pJ_1\neq \pJ_2$.  Note that $\Delta^d_{h2,3}$ and  $\Delta^{in}_{h2,3}$  are never equal because the particle-hole symmetry is absent for all $\De$, however for $\pJ_2 >\pJ_1$ they may show a crossing similar to $\bav_1,\bav_2$ in Fig.~\ref{fig:vbar}.\\

The exchange parameters $(\pJ_1\pJ_2,J_{12})$ of the model are an independent set (Appendix~\ref{sec:app1}) that correspond to the CEF parameters. Therefore one should also know how the low energy Kondo scale $T^*$ changes with e.g. the ratio of upper/lower level exchange $\pJ_2/\pJ_1$ (Fig.~\ref{fig:tstar_theta}). For this purpose we use the polar parametrization $J^\perp_1=J_\perp\cos\theta$,   $J^\perp_1=J_\perp\sin\theta$ discussed in the Appendix~\ref{sec:app1}. Here $\theta$ varies in the interval $[0,\frac{\pi}{2}]$ corresponding to the change from $\pJ_1=\pJ,\pJ_2=0$ to $\pJ_1=0,\pJ_2=\pJ$.
For $\De=0$  upper/lower level are degenerate and therefore $T^*(\theta)$ must be symmetric
around $\theta =\frac{\pi}{4}$ or $\pJ_2=\pJ_1$. When $\De$ increases the upper level contributes
progressively  less to the effective hybridization and therefore, together with the complete decoupling
of the lower level for $\theta\rightarrow \frac{\pi}{2}$,  $(\pJ_1\rightarrow 0)$ we observe $T^*\rightarrow 0$ in 
Fig.~\ref{fig:tstar_theta}. For increasing $J_{12}$ the minimum in $T^*(\theta)$ becomes less pronounced and
essentially vanishes for the isotropic case $J_{12}=J_\perp$.\\

We discussed already basic features of the spectral function in the case that particle-hole symmetry of the original TB band or square DOS model is preserved. This requires special conditions for the selfconsistent solution: Firstly we must have $\bav_1=\bav_2$, i.e. equal effective hybridization strength. This can only be achieved either for $\Delta=0$ or by fine tuning the CEF energy $\De$
to a special value $\Delta_{cr}$ depending on the exchange parameter set $\pJ_1,\pJ_2,J_{12}$. For this value the curves of $\bav_1(\Delta)$ and  $\bav_2(\Delta)$ cross (Fig.~\ref{fig:vbar}) which is only possible if $\pJ_2>\pJ_1$. Furthermore since the hybridization of bands happens around the effective f-level position $\lambda$ we must tune $\lambda=0$ by setting the chemical potential to $\mu=-(\De+T^*)$ to achieve the full particle hole-symmetry of quasiparticle bands as depicted in Figs.~\ref{fig:dispersion},\ref{fig:spec_sym}. \\

\begin{figure}
%\vspace{0.5cm}
\includegraphics[width=0.95\columnwidth]{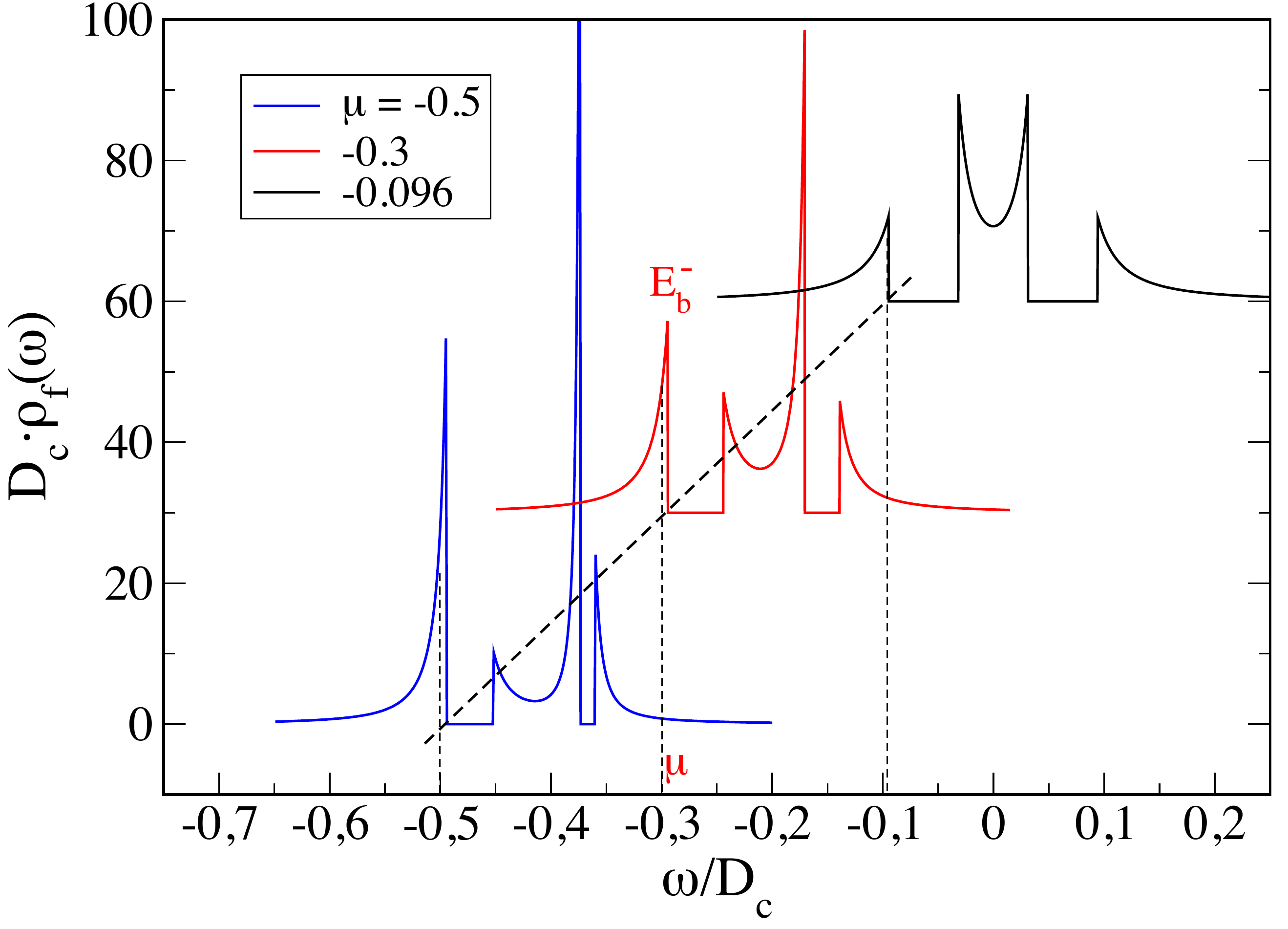}
\caption{Evolution of total f-quasiparticle DOS with chemical potential $\mu$ starting from the symmetric case
$\mu=-0.096$ ($n_c\simeq 1$) via $\mu=-0.3$ to $\mu=-0.5$ ($n_c\simeq 0.5$). Exchange and CEF parameters like in Figs.~\ref{fig:spec_sym},\ref{fig:occ_cf}. As $\mu$ moves to the left the quasiparticle DOS is dragged along to fulfill the $n_f=1$ constraint. Therefore $\mu$ is always pinned in the lower f-DOS peak just below $E^-_b$ (connected by thick dashed line, cf. Fig.~\ref{fig:occ_cf}b). The red and black spectrum have an ordinate offset of thirty units for clarity.}
\label{fig:specsymm_chem}
\end{figure}

Therefore the symmetric case requires rather special conditions to be realized. In the case of general size of exchange constants and CEF splitting the total f-spectrum will be quite asymmetric and also the interchange symmetry $f_1\leftrightarrow f_2$ visible in  Fig.~\ref{fig:spec_sym} will be lost. We present a few characteristic examples for the general case in Fig.~\ref{fig:spec_asym} (and also Fig.~\ref{fig:specsymm_chem})
for the underlying square-DOS conduction electron model.
In (a) we have $\pJ_2>\pJ_1$ and $\De$ still sufficiently small such that $\bav_2>\bav_1$. Therefore the upper hybridization gap will be larger as compared to the lower gap. While the lower and upper bands are still roughly symmetric the central heavy band is now quite asymmetric because it has shifted out of the center of the overall hybridisation gap which would correspond to $\De^{in}_{h1}=E_{10}-E_{2\bQ}$ in the TB model.
%
% %%%%%%%%%%%%%%%%%%%%% figure %%%%%%%%%%%%%%%%%%%%%%%%%%%%
%%%%%%%%%%%%%%%%%%%%%%fig%%%%%%%%%%%%%%%%%%%%%%%%%%%%%%%
%
In (b) we still have the case $\pJ_2>\pJ_1$ but now $\De$ is sufficiently large to achieve already the inverse relation $\bav_1>\bav_2$. Therefore the lower hybridization gap has now become larger than the upper one because the central band is shifted towards the upper band. As a consequence the skewing of $f_1,f_2$ distribution is now opposite to that in (a). Thus by varying
$\De$ or the ratio $\pJ_2/\pJ_1$ (i.e. $\theta$) one may shift the central heavy band more or less continuously through the overall hybridzation gap and also change its width $W_3$. For the case $\pJ_1 >\pJ_2$ we will have $\bav_1\gg\bav_2$ and thererfore the lower hybridization gap dominates while the upper one becomes very narrow. Roughly the upper and central bands have merged into one band separated by a large gap from the lower one. This resembles now the one-orbital case except that the lower and combined upper DOS parts are very asymmetric.

We mostly considered the case for slightly less than half filling $n_c<1$. For such case the chemical potential is pinned very close to the upper
edge $E^-_b$ of the lowest band to satisfy the constraint $n_f=1$ (Fig.~\ref{fig:occ_cf}a). When $\mu$ is decreased one always stays in a metallic situation with  the distance $E^-_b-\mu$  behaving nonmonotonic (Fig.~\ref{fig:occ_cf}b).. 
The systematic change of the quasiparticle f-DOS with $\mu$, starting from the symmetric case (close to half filling $n_c\approx 1$) is presented in Fig.~\ref{fig:specsymm_chem}. It demonstrates that shifting the chemical potential to lower values, i.e. reducing the conduction band filling $n_c$ the hybridized quasiparticle spectrum is dragged along with $\mu$ to lower energies such that the chemical potential remains pinned in the DOS peak of the lowest band in accordance with Fig.~\ref{fig:occ_cf}b . In this way the f-constraint $n_f=1$ is respected at each band filling $n_c$.

\section{Conclusion and Outlook}
\label{sec:conclusion}

In this work we gave a detailed investigation of the quasiparticle spectrum of the underscreened quasi-quartet Kondo lattice and the related heavy band structure and associated effective hybridzation gaps.
A two-orbital model representing Kramers doublets slightly split by a CEF as is frequently found in tetragonal Ce- or Yb- compounds has been used. We started from an exchange model where f-charge fluctuations are already completely suppressed and single f-(electron- or hole-) occupancy is realized. A fermionic representation of the f-conduction electron exchange is employed  and  the interacting model is treated within a constrained  mean field theory that ensures the single f-occupancy.\\ 

We derived the effective heavy quasiparticle bands and their partial DOS functions. The two orbital KL model with $N=2$ degeneracy of conduction band and $2N$ localized states has much richer features than the single orbital KL model. Firstly  there is in addition a characteristic central heavy band which lies inside the main hybridization gap. It consists mainly of a superpositon of the two f-states with a small admixture of conduction states that is responsible for the overall dispersion of the central band. The width of this band is itself ot the order $W_3=(T^{*2}+\Delta_0^2)^\fs-T^*$ making it a genuinely heavy band of mostly f-character. This is in contrast to the upper and lower band which change their character from  light conduction states to heavy hybridzed f-states and vice versa when traversing the Brillouin zone. Furthermore, due the central heavy band there are now in general three different direct and indirect hybridization gaps, as opposed to just one of each in the single-orbital model. In particular the two-orbital model has now also a {\it direct} hybridization gap of the order of the Kondo scale $T^*$.
A rather surprising result is the dependence of the central heavy band width on the CEF splitting $\De$. Although the dispersion of the latter is due to the hybridization with c-states, it will be nonzero only for finite CEF splitting $\De$ . In other words for $\De=0$  the central band will collapse to a flat band which will be localized into a  'left-over' spin by residual quasiparticle interactions. This is an inevitable and natural consequence of the underscreend model.\\

The dependence of $T^*$, the effective hybridizations $\bav_\tau$ and the associated quasiparticle band width and hybridization gaps on the CEF splitting has been investigated for various parameters of the exchange model. The latter is characterized by two (orbital-) diagonal and one off-diagonal exchange constants which are derived from the CEF states. As expected the increase of $\De$ decreases $T^*$ due to the reduction of the effective hybridization of the upper doublet when $\De$ increases. This also explains the reduction of $T^*$ when the ratio $\pJ_2/\pJ_1=\tan\theta$ is tuned from small to large values.

The novel central heavy band in the underscreened KL model is also evident in the various partial DOS spectra. In general the spectrum is asymmetric with respect to the effective f level position at $\lambda$, i.e. the central band is placed off-center in the overall hybridization gap between the lower and upper bands. Also the distribution of $f_1,f_2$ orbital weights is asymmetric. This situation prevails for any $\De$ in the case that $\pJ_2<\pJ_1$. In the opposite case $\pJ_2>\pJ_1$ there exists a critical $\Delta_{cr}$ where the effective hybridizations become equal. Then, with suitable choice of chemical potential (such that $\lambda=0$) or $n_c$ the quasiparticle spectrum may become symmetric in energy. In any case the upper band edge of the lower quasiparticle band must stay pinned to the chemical potential to ensure the single (electron or hole) f-occupancy.  Then the central and upper band must stay unoccupied. This is due to the total suppression of charge fluctuations in the present 2-orbital KL model. An analogous Anderson lattice- type model where only the total $n_f+n_c$ occupancy must be preserved allows for c-f charge fluctuations. In this case, with suitable tuning of parameters one could also shift the Fermi level more easily into the central genuinely heavy band. This would also apply to the two-orbital KL model with magnetic polarization.\\

The interest in this model stems mainly from possible applications to magnetic order in Kondo lattice compounds. Our detailed investigation of quasiparticle structure lays the foundation for considering this question more realistically than in the canonical but oversimplified 1-orbital KL model. For the magnetism the influence of excited CEF levels in practice always has to be taken into account. We may  speculate how to approach it in the framework of the present model theory: In the one-orbital KL model the true ground state may 
exhibit ferromagnetic or antiferromagnetic order, depending on conduction band filling \cite{lacroix:79,bernhard:15,li:10}. In this case the magnetism appears simply due to the rigid band splitting of hybridized bands into (pseudo-) spin up- and down bands with an ensuing polarization of bands that generates the moment. In the present two-orbital KL model there is an interesting alternative possibility.
Because we have two CEF split f-Kramers doublets with a nondiagonal exchange ($J_{12}$) with conduction electrons it is possible to develop excitonic (induced) magnetism involving the two orbitals. This may be investigated by either directly breaking the (pseudo-spin) symmetry and minimize the ground state energy with respect to the possible magnetic order parameters, as in the one-orbital KL model. Or one may investigate the magnetic response in the paramagnetic phase to find channels
for the instability. Furthermore one may study the dynamic magnetic response to see whether the magnetism appears via the softening of a mixed CEF/Kondo spin-exciton mode. The detailed investigation of the excitation spectrum in this work provides the solid foundation for such analysis.\\

\section*{ACKNOWLEDGMENTS}
A.A. acknowledges support through  National Research Foundation  (NRF) funded by the Ministry of Science of Korea (2015R1C1A1A01052411) and (2017R1D1A1B03033465), and by  Max Planck POSTECH / KOREA Research Initiative (No. 2011-0031558)
programs through NRF funded by the Ministry of Science of Korea. 
%

%\newpage

\appendix
\section{Fermionic representation of the quasi-quartet CEF states and Kondo exchange model}
\label{sec:app1}

In this appendix we give the schematic derivation  of  the fermionic representation of Hamiltonian in Eq.~(\ref{eqn:modelHam})
from the original Kondo exchange Hamiltonian in Eq.~(\ref{eqn:cfHam}). In the latter the 4f-charge fluctuations present in the underlying Anderson- type 
model are already eliminated by a
Schrieffer-Wolff transformation \cite{fazekas:99} for the limit $U_{ff}\gg D_c$ where $U_{ff}$ is the Coulomb repulsion of
localized 4f states and $2D_c$ the conduction band width. Then the degrees of freedom are the conduction electrons and CEF split 4f states resulting from the spin-orbit ground state multiplet $(J=\frac{5}{2})$ for $Ce^{3+}(4f^1)$ and  $(J=\frac{7}{2})$ for $Yb^{3+}(4f^{13})$ single electron and hole-type cases, respectively. In a tetragonal environment for e.g. frequently realized  122- structure of heavy fermion compounds, the action of the CEF with $D_{4h}$ symmetry splits the $(2J+1)$- fold degenerate 4f multiplets into a series of three (Ce) or four (Yb) Kramers doublets  belonging to either $\Gamma_6$ or $\Gamma_7$ representation of $D_{4h}$ (naturally this means that mixed multiple representations must occur). It may happen, in particular in Yb- compounds like, e.g.  YbRu$_2$Ge$_2$ \cite{jeevan:06,takimoto:08}  when the two lowest doublets of   $\Gamma_6$ and $\Gamma_7$ symmetry are close in energy compared to the overall CEF splitting thus
forming a 'quasi-quartet' state (in cubic symmetry $O_h$ they would combine to a true $\Gamma_8$ quartet as for YbB$_{12}$ \cite{alekseev:04,akbari:09} or CeB$_{6}$ \cite{ohkawa:85,shiina:97}). We focus here on the Yb- case when both $\Gamma_6$, $\Gamma_7$ appear twofold, therefore their wave functions are linear superpositions of free 4f-in states $|JM\rangle$ $ (|M|\leq J)$ whose coefficients depend explicitly on the CEF parameters due to the mixing of each pair of representations.

As discussed in Ref.~ \onlinecite{takimoto:08} the quasi-quartet $\Gamma_6$-$\Gamma_7$ pair can be represented as
\be
\bl
|\Ga_6\pm\ra&=\alpha_{11}|\pm\frac{7}{2}\ket+\alpha_{12}|\mp\fs\ket,
\\
|\Ga_7\pm\ra&=\beta_{11}|\mp\frac{5}{2}\ket+\beta_{12}|\pm\frac{3}{2}\ket,
\el
\label{eqn:CEFstate}
\ee
where the pseudo-spin $\sigma\equiv\sigma_z=\pm 1 (\equiv \pm)$ represents the twofold Kramers degeneracy due to time reversal invariance.
We assume without loss of generality that $\Ga_6$ is the lower and $\Ga_7$ the upper doublet split by an energy $\De_0$. The CEF energies may then be defined symmetrically as $E_6=-\frac{\De_0}{2}$ and $E_7=+\frac{\De_0}{2}$.
Because the CEF states are those of single 4f electrons $(Ce^{3+}, 4f^{1})$ or holes   $(Yb^{3+}, 4f^{13})$ they may be represented as 
\bea
|\Ga_{6\si}\ket=f^\dag_{1\si}|0\ket ;\;\;\;  |\Ga_{7\si}\ket=f^\dag_{2\si}|0\ket ,
\label{eqn:fmapping}
\eea
where the 'orbital' index $\tau=1,2$ corresponds to $\Ga_6,\Ga_7$, respectively. Here $|0\ket= |f^0,J=0\ket$ or  $ |f^{14},J=0\ket$ is the vacuum or reference state corresponding to the empty or full 4f-shell from which $f^\dag_{\tau\si}$ creates an electron or hole, respectively. The fermionic representation is then given by
\bea
H_{CEF}=-\frac{\De_0}{2}\sum_{i\si}\bigl(f^\dag_{i1\si}f_{i1\si}-f^\dag_{i2\si}f_{i2\si}\bigr) .
\label{eqn:CEFHam}
\eea
To express the exchange interactions of the Kondo term in fermionic variables it is also  helpful to introduce the pseudo  spins of the CEF Kramers doublets via the relation \cite{takimoto:08}
\bea
S^\alpha_{\tau\tau'}=\fs\sum_{\si\si'}f^\dag_{\tau\si}\hat{\si}^\alpha_{\si\si'}f_{\tau'\si'}^{},
\label{eqn:pseudo}
\eea
where $\hat{\si}^\alpha$ $(\alpha=x,y,z \; \mbox{or}\; \pm ,z)$ are the Pauli matrices for the $S=\fs$ Kramers pseudo spins for both orbitals $\tau=1,2$. Explicitly this translates into
 \bea
 S^z_{\tau\tau}&=&\fs(f^\dag_{\tau\ua}f_{\tau\ua}-f^\dag_{\tau\da}f_{\tau\da}); \;\;\;
 S^+_{\tau\tau}=f^\dag_{\tau\ua}f_{\tau\da}; \;\;  S^-_{\tau\tau}= f^\dag_{\tau\da}f_{\tau\ua}\non\\
 S^z_{\tau\bar{\tau}}&=&\fs(f^\dag_{\tau\ua}f_{\bar{\tau}\ua}-f^\dag_{\tau\da}f_{\bar{\tau}\da}); \;\;\;
 S^+_{\tau\bar{\tau}}=f^\dag_{\tau\ua}f_{\bar{\tau}\da}; \;\;  S^-_{\tau\bar{\tau}}= f^\dag_{\tau\da}f_{\bar{\tau}\ua}\non\\
 \label{eqn:pseudo2}
 \eea
 Here the pairs $(\tau\bar{\tau})$ are defined as either $(1,2)$ or $(2,1)$.\\
 
The original Kondo Hamiltonian resulting from the Schrieffer-Wolff transformation comprises all $(2J+1)$ states of the relevant 4f multiplet of total angular momentum J. It can be written as \cite{jensen:91}
%\begin{widetext}
\bea
\bl
\hspace{-0.61cm}
H_{ex}=
&\;
(g_J-1)I_{ex}\sum_i\bs_i\cdot \bJ_i=
\fs(g_J-1)I_{ex} \; \times
\\&\!\!
\sum_i
\bigl[J^+_ic^\dag_{i\da}c_{i\ua}+J^-_ic^\dag_{i\ua}c_{i\da}+
J_i^z(c^\dag_{i\ua}c_{i\ua}-c^\dag_{i\da}c_{i\da})\bigr],
\label{eqn:sfHam}
\el
\eea
%\end{widetext}
%
where $c^\dag_{I\si}$  creates a conduction electron at lattice site $i$, $J^\alpha_i$ are the components of the 4f total angular momentum operator. Furthermore $I_{ex}$ is the bare (physical) spin exchange constant (assuming the convention $I_{ex}>0$ for AF exchange) and $g_J$ the Land\'e factor to project it to the J-multiplet. Since we want to restrict to the lowest two doublets we can express the $J_\alpha$-operators in this subspace by the pseudospin operators Eq.~(\ref{eqn:pseudo}) of these doublets according to
\be
\bl
J^z=&c^z_{11}S^z_{11}+c^z_{22}S^z_{22},
\\
J^\pm=&c_{11}S^\pm_{11}+c_{22}S^\pm_{22}+c_{12}\frac{1}{\sqrt{2}}(S^\pm_{12}+S^\pm_{21}).
\label{eqn:JSequiv}
\el
\ee
The nondiagonal terms $\sim c_{12}$ are important as they lead to inelastic transitions between the split doublets thus coupling the Kondo screening of both pseudospins. The linear transformation coefficients may be directly obtained from the CEF wave functions of the original $\Ga_6$,$\Ga_7$ states by the relations
\be
\bl
c^z_{11}&=7-8\alpha^2_{12}; \;\;\;  c^z_{22}=-5+8\beta^2_{12},
\\
c_{11}&=4\alpha_{12}^2; \;\;\; c_{22}=4\sqrt{3}\beta_{12}\sqrt{1-\beta^2_{12}},
\\
c_{12}&=\sqrt{7}\sqrt{(1-\alpha^2_{12})(1-\beta^2_{12})} +\sqrt{30}\alpha_{12}\beta_{12},
\label{eqn:Jcoeff}
\el
\ee
where the normalization conditons $\alpha^2_{11}+\alpha^2_{12}=1$ and $\beta^2_{11}+\beta^2_{12}=1$ have been applied.
Using the equivalence of Eq.~(\ref{eqn:JSequiv}) and Eqs.~(\ref{eqn:pseudo},\ref{eqn:pseudo2}) in the Hamiltonian  of Eq.~(\ref{eqn:sfHam})  we finally arrive at the effective
quasi-quartet Kondo interaction $H_{ex}=H_{cf}+H^{cf}_{12}$ in Eq.~(\ref{eqn:modelHam}). Furthermore adding the CEF potential of Eq.~(\ref{eqn:CEFHam}) and the bare conduction electron part leads to the total model Hamiltonian $H$ of Eq.~(\ref{eqn:modelHam}).
The interaction constants in this model can then be expressed \cite{takimoto:08} by the bare sf- exchange constant and  the coefficients in Eq.~(\ref{eqn:Jcoeff}) according to $(J_0=(g_J-1)I_{ex})$:
\be
\bl
J^\perp_1&=c_{11}J_0; \;\;\; J^\perp_2=c_{22}J_0; \;\;\; J_{12}=\frac{1}{\sqrt{2}}c_{12}J_0
\\
J^z_1&=c^z_{11}J_0; \;\;\; J^z_2=c^z_{22}J_0; \;\;\; J^z_{12}=0
\el
\ee
The vanishing of $J^z_{12}$ is a symmetry property independent of CEF wave function coefficients. It is obvious from Eq.~(\ref{eqn:CEFstate}) since the $\Gamma_6,\Gamma_7$ doublets do not contain $|JM\ket$ states with equal $M$. The overall sign of these constants (positive for AF and negative for FM) depends on the sign of the bare spin exchange $I_{ex}$ the size of $g_J$ and the sign of CEF derived coefficients in Eq.~(\ref{eqn:Jcoeff}). In the main text we will restrict to the case of only positive (antiferromagnetic) effective exchange constants and sometimes use the polar parametrization $J^\perp_1=J_\perp\cos\theta$,   $J^\perp_1=J_\perp\sin\theta$
with 
\bea
\bl
J_\perp&=J_0(c_{11}^2+c_{22}^2)^\fs; \;\;\; 
\\
\tan\theta&=\frac{c_{22}}{c_{11}}=\sqrt{3}\frac{\beta_{12}}{\alpha_{12}^2}
\sqrt{1-\beta_{12}^2}.
\el
\eea
The above relations map the microscopic independent parameters $(J_0,\alpha_{12},\beta_{12})$ to the independent in-plane exchange model sets $(J^\perp_1, J^\perp_2,J_{12})$ or $(J_\perp,\theta ,J_{12})$. The out-of-plane exchange parameter sets
$(J^z_1,J^z_2)$ are then fixed by the $\alpha_{12},\beta_{12}$ values corresponding to the in-plane set.

The fermionic representation for the exchange model of Eq.~(\ref{eqn:modelHam}) has been derived starting
from the mapping in Eq.~(\ref{eqn:fmapping}). The latter is possible only for reference (vacuum) states $|0\rangle$ which are fully symmetric such that total angular momentum $J=0$. This is always the  case for $Ce^{3+}$ or $Yb^{3+}$ with reference states corresponding to the empty $(4f^0)$ or full $(4f^{14})$ shell, respectively. Furthermore it can be used for the nearly half filled case of $Eu^{2+}$ and $Sm^{3+}$ which also have reference states $(4f^6)$ with $J=0$ \cite{hanzawa:98,takimoto:11}. However for arbitrary occupation of the f-shell the representation of CEF states with  Fermi operators is not possible and more general Hubbard or 'standard basis' operators \cite{figueira:94} with more complicated commutation relations have to be employed.

\section{Approximate expression for the low energy scale T$^*(\De)$}
\label{sec:app2}

The low energy scale $T^*(\De)$ (Eq.~(\ref{eqn:delkondo})) is determined by the solution of the selfconsistency equation Eq.~(\ref{eqn:numselfcons}) which can generally be solved only numerically. In the simplest case $(\De=0, J_1=J_2,J_{12}=0))$ $T^*_0=T^*(0)$ reduces to the expression of the  Kondo temperature in Eq.~(\ref{eqn:barekondo}). We can also give a closed expression of $T^*_0$ for the general exchange model.
For $\De =0$ we have $F_1=F_2\equiv F_0$ with $F_0=\ln[(\lam-\mu)/D_c]$ and then Eq.~(\ref{eqn:numselfcons}) reduces to 
\be
(N\rho_0)^2(a_1a_2-b^2)F^2_0-(N\rho_0)(a_1+a_2)F_0+1=0. 
\label{eqn:numselfcons0}
\ee
Using Eq.~(\ref{eqn:excon}) this selfconsistency equation  has two possible closed solutions which are given by $(|\mu|\ll D_c)$
\be
\bl
F_0&=\ln[(\lam-\mu)/D_c]
\\
&=-\frac{1}{(N\rho^c_0)|\frac{A}{B}|}
\bigl[1\mp(1-\frac{A}{B^2})^\fs\bigr]
\equiv -\frac{2}{\tg_\perp},
\\
T_0^*&=\lam-\mu=D_c\exp\bigl(-\frac{2}{\tg^0_\perp}\bigr),
\label{eqn:tkondo_app}
\el
\ee
where the effective Kondo coupling strength is then given by
\bea
\hspace{-0.64cm}
\bl
\tg^0_\perp&=(N\rho^c_0)\frac{2|\frac{A}{B}|}{1-(1-\frac{A}{B^2})^\fs}\\
&=(N\rho^c_0)\frac{J^{\perp 2}_{av}+\bar{J}_\perp J_{12}}
{(\bar{J}_\perp +\fs J_{12})-\fs\bigl[(J^\perp_1-J^\perp_2)^2+J^2_{12}\bigr]^\fs}
\label{eqn:coup0_app}
\el
\eea
Here we defined the two types of orbital-averaged exchange as $J^\perp_{av}=(J^\perp_1J^\perp_2)^\fs$ and
$\bar{J}_\perp=\fs(J^\perp_1+J^\perp_2)$. Note that of the above two solutions we use only  $(-)$ because it 
has the larger effective coupling $\tg_\perp$ and hence the largest Kondo energy scale and therefore the lowest
ground state energy. The (-) solution corresponds also to the largest value of $\lam-\mu$ in the graphical solution 
plot of Fig.~\ref{fig:faux}b. In the special case of $J_{12}=0$ of two decoupled doublets one can show that 
$\tg_\perp=max(J^\perp_1,J^\perp_2)$ and furthermore if both are equal then we recover Eq.~(\ref{eqn:barekondo}).
In the true quartet case ($\pJ_1=\pJ_2=J_{12}\equiv J_\perp$) with $SU(4)$ symmetry we obtain 
$\tg_\perp=(2N\rho^c_0)J_\perp$ and the corresponding larger Kondo scale $T^*_0$ due to $2N=4$ degeneracy.

It is also possible to give an approximate closed expression for $T^*(\De)$. Using Eq.~(\ref{eqn:delkondo}) we may,
after some algebra, reformulate the selfconsistency in Eq.~(\ref{eqn:numselfcons}) as 
\be
\bl
T^*(\De)&=D_c\exp\bigl(-\frac{2}{\tg_\perp(T^*,\De)}\bigr)
\\
-\frac{2}{\tg_\perp(T^*,\De)}&=
\frac{(N\rho_0)a_2\ln\frac{D_c}{\De_0+T^*}+1}
{(N\rho_0)^2A\ln\frac{D_c}{\De_0+T^*}+(N\rho_0)a_1}
\label{eqn:tkiterate}
\el
\ee
This is still equivalent to  Eq.~(\ref{eqn:numselfcons}). It has the form appropriate for iterative solution for $T^*(\De)$. 
If we stop after the first iteration step, i.e., replacing $T^*(\De) \rightarrow T_0^*$ at the r.h.s we get an approximate
closed expression 
\be
T^*(\De)=D_c\exp\Bigl[\frac{(N\rho_0)a_2\ln\frac{D_c}{\De_0+T_0^*}+1}
{(N\rho_0)^2A\ln\frac{D_c}{\De_0+T_0^*}+(N\rho_0)a_1}\Bigr],
\label{eqn:tkdel}
\ee
where $T^*_0$ is given by Eqs.~(\ref{eqn:tkondo_app},\ref{eqn:coup0_app}) and the exchange parameters $a_1,a_2,A$
in Eq.~(\ref{eqn:excon}). This approximation formula works well when $T^*(\De)$ dependence is not too rapid such that the
first iteration is sufficient. This is the case for $\pJ_1\geq\pJ_2$ when the Kondo effect is dominated by the lower doublet and the upper one has moderate influence. For the opposite case $\pJ_2\geq\pJ_1$ ,
$T^*(\De)$ decays rapidly with $\De$ (see Fig.~\ref{fig:tstar}) the approximate formula gives a too rapid decrease with $\De$ as compared to the numerical solution of Eq.~(\ref{eqn:numselfcons}) or Eq.(\ref{eqn:tkiterate}).\\

The fundamental quantity in the constrained mean field theory is $\lambda$ ( or $\lam-\mu$), the  position of the effective f-level which is adjusted to constrain (on the average) to the Hilbert space with occupation $n_f=1$. For $\Delta=0$ the value of $\lambda-\mu$ corresponds directly to the Kondo energy scale $T^*$. For nonzero $\Delta$ the definition of the latter is not unambiguous. One way is to subtract directly the CEF energy according to Eq.~(\ref{eqn:delkondo}), another way is to define it as $T^*=\bav^2/D_c$ via the hybridization gaps in the symmetric case as discussed in Sec.~\ref{sec:widthgap}. Here we discuss the connection between the two definitions. Using Eqs.~(\ref{eqn:bav},\ref{eqn:gfunc}) and assuming $T^*$ from Eq.~(\ref{eqn:delkondo})  we obtain the relation
\bea
\frac{\bav^2}{D_c}=T^*\Bigl(\frac{1+\frac{\De_0}{T^*}}{1+\frac{R}{1+R}\frac{\De_0}{T^*}}\Bigr).
\label{eqn:tstarcomp}
\eea
In particular then for $\De=0$ always $\frac{\bav^2}{D_c}=T_0^*$ as given by Eqs.~(\ref{eqn:tkondo_app},\ref{eqn:coup0_app}). For the general symmetric case when $\De=\De_{cr}$ we have $\bav_1=\bav_2$ or $R=1$. When $\De_{cr}/T^*\ll 1$ the above equation then leads to
\bea
\frac{\bav^2}{D_c}=T^*+\De = \lambda-\mu
\eea
consistent with  the relation in  Eq.~(\ref{eqn:delkondo}). In the opposite limit   $\De_{cr}/T^*\gg 1$ the upper level influence on $T^*$ is negligible and indeed from Eq.~(\ref{eqn:tstarcomp}) we get  $T^*=\frac{\bav_1^2}{D_c}$ . Therefore both definitions of $T^*$ are consistent in the cases where they can be applied simultaneously.

%%%%%%%%%%%%%%%%%%%%%%%%%      References        %%%%%%%%%%%%%%%%%%%%
%\newpage
%\bibliographystyle{prsty}
\bibliography{References}

%merlin.mbs apsrev4-1.bst 2010-07-25 4.21a (PWD, AO, DPC) hacked
%Control: key (0)
%Control: author (8) initials jnrlst
%Control: editor formatted (1) identically to author
%Control: production of article title (-1) disabled
%Control: page (0) single
%Control: year (1) truncated
%Control: production of eprint (0) enabled
\begin{thebibliography}{40}%
\makeatletter
\providecommand \@ifxundefined [1]{%
 \@ifx{#1\undefined}
}%
\providecommand \@ifnum [1]{%
 \ifnum #1\expandafter \@firstoftwo
 \else \expandafter \@secondoftwo
 \fi
}%
\providecommand \@ifx [1]{%
 \ifx #1\expandafter \@firstoftwo
 \else \expandafter \@secondoftwo
 \fi
}%
\providecommand \natexlab [1]{#1}%
\providecommand \enquote  [1]{``#1''}%
\providecommand \bibnamefont  [1]{#1}%
\providecommand \bibfnamefont [1]{#1}%
\providecommand \citenamefont [1]{#1}%
\providecommand \href@noop [0]{\@secondoftwo}%
\providecommand \href [0]{\begingroup \@sanitize@url \@href}%
\providecommand \@href[1]{\@@startlink{#1}\@@href}%
\providecommand \@@href[1]{\endgroup#1\@@endlink}%
\providecommand \@sanitize@url [0]{\catcode `\\12\catcode `\$12\catcode
  `\&12\catcode `\#12\catcode `\^12\catcode `\_12\catcode `\%12\relax}%
\providecommand \@@startlink[1]{}%
\providecommand \@@endlink[0]{}%
\providecommand \url  [0]{\begingroup\@sanitize@url \@url }%
\providecommand \@url [1]{\endgroup\@href {#1}{\urlprefix }}%
\providecommand \urlprefix  [0]{URL }%
\providecommand \Eprint [0]{\href }%
\providecommand \doibase [0]{http://dx.doi.org/}%
\providecommand \selectlanguage [0]{\@gobble}%
\providecommand \bibinfo  [0]{\@secondoftwo}%
\providecommand \bibfield  [0]{\@secondoftwo}%
\providecommand \translation [1]{[#1]}%
\providecommand \BibitemOpen [0]{}%
\providecommand \bibitemStop [0]{}%
\providecommand \bibitemNoStop [0]{.\EOS\space}%
\providecommand \EOS [0]{\spacefactor3000\relax}%
\providecommand \BibitemShut  [1]{\csname bibitem#1\endcsname}%
\let\auto@bib@innerbib\@empty
%</preamble>
\bibitem [{\citenamefont {Newns}\ and\ \citenamefont {Read}(1987)}]{newns:87}%
  \BibitemOpen
  \bibfield  {author} {\bibinfo {author} {\bibfnamefont {D.~M.}\ \bibnamefont
  {Newns}}\ and\ \bibinfo {author} {\bibfnamefont {N.}~\bibnamefont {Read}},\
  }\href@noop {} {\bibfield  {journal} {\bibinfo  {journal} {Advances in
  Physics}\ }\textbf {\bibinfo {volume} {36}},\ \bibinfo {pages} {799}
  (\bibinfo {year} {1987})}\BibitemShut {NoStop}%
\bibitem [{\citenamefont {Hewson}(1993)}]{hewson:93}%
  \BibitemOpen
  \bibfield  {author} {\bibinfo {author} {\bibfnamefont {A.~C.}\ \bibnamefont
  {Hewson}},\ }\href@noop {} {\emph {\bibinfo {title} {The Kondo problem to
  heavy fermions}}}\ (\bibinfo  {publisher} {Cambridge University Press},\
  \bibinfo {address} {Cambridge},\ \bibinfo {year} {1993})\BibitemShut
  {NoStop}%
\bibitem [{\citenamefont {Tsunetsugu}\ \emph {et~al.}(1997)\citenamefont
  {Tsunetsugu}, \citenamefont {Sigrist},\ and\ \citenamefont
  {Ueda}}]{tsunetsugu:97}%
  \BibitemOpen
  \bibfield  {author} {\bibinfo {author} {\bibfnamefont {H.}~\bibnamefont
  {Tsunetsugu}}, \bibinfo {author} {\bibfnamefont {M.}~\bibnamefont {Sigrist}},
  \ and\ \bibinfo {author} {\bibfnamefont {K.}~\bibnamefont {Ueda}},\
  }\href@noop {} {\bibfield  {journal} {\bibinfo  {journal} {Review of Modern
  Physics}\ }\textbf {\bibinfo {volume} {69}},\ \bibinfo {pages} {809}
  (\bibinfo {year} {1997})}\BibitemShut {NoStop}%
\bibitem [{\citenamefont {Kuramoto}(2000)}]{kuramoto:00}%
  \BibitemOpen
  \bibfield  {author} {\bibinfo {author} {\bibfnamefont {Y.}~\bibnamefont
  {Kuramoto}},\ }\href@noop {} {\emph {\bibinfo {title} {Dynamics of Heavy
  Electrons}}}\ (\bibinfo  {publisher} {Clarendon Press},\ \bibinfo {address}
  {Oxford},\ \bibinfo {year} {2000})\BibitemShut {NoStop}%
\bibitem [{\citenamefont {Thalmeier}\ and\ \citenamefont
  {Zwicknagl}(2005)}]{thalmeier:05}%
  \BibitemOpen
  \bibfield  {author} {\bibinfo {author} {\bibfnamefont {P.}~\bibnamefont
  {Thalmeier}}\ and\ \bibinfo {author} {\bibfnamefont {G.}~\bibnamefont
  {Zwicknagl}},\ }\enquote {\bibinfo {title} {Handbook on the {P}hysics and
  {C}hemistry of {R}are {E}arths},}\ \ (\bibinfo  {publisher} {Elsevier},\
  \bibinfo {address} {Amsterdam},\ \bibinfo {year} {2005})\ Chap.\ \bibinfo
  {chapter} {219}, pp.\ \bibinfo {pages} {135--287}\BibitemShut {NoStop}%
\bibitem [{\citenamefont {Ohkawa}(1985)}]{ohkawa:85}%
  \BibitemOpen
  \bibfield  {author} {\bibinfo {author} {\bibfnamefont {F.~J.}\ \bibnamefont
  {Ohkawa}},\ }\href@noop {} {\bibfield  {journal} {\bibinfo  {journal} {J.
  Phys. Soc. Jpn.}\ }\textbf {\bibinfo {volume} {54}},\ \bibinfo {pages} {3909}
  (\bibinfo {year} {1985})}\BibitemShut {NoStop}%
\bibitem [{\citenamefont {Shiina}\ \emph {et~al.}(1997)\citenamefont {Shiina},
  \citenamefont {Shiba},\ and\ \citenamefont {Thalmeier}}]{shiina:97}%
  \BibitemOpen
  \bibfield  {author} {\bibinfo {author} {\bibfnamefont {R.}~\bibnamefont
  {Shiina}}, \bibinfo {author} {\bibfnamefont {H.}~\bibnamefont {Shiba}}, \
  and\ \bibinfo {author} {\bibfnamefont {P.}~\bibnamefont {Thalmeier}},\
  }\href@noop {} {\bibfield  {journal} {\bibinfo  {journal} {J. Phys. Soc.
  Jpn.}\ }\textbf {\bibinfo {volume} {66}},\ \bibinfo {pages} {1741} (\bibinfo
  {year} {1997})}\BibitemShut {NoStop}%
\bibitem [{\citenamefont {Sundermann}\ \emph {et~al.}(2018)\citenamefont
  {Sundermann}, \citenamefont {Yavas}, \citenamefont {Chen}, \citenamefont
  {Kim}, \citenamefont {Fisk}, \citenamefont {Kasinathan}, \citenamefont
  {Haverkort}, \citenamefont {Thalmeier}, \citenamefont {Severing},\ and\
  \citenamefont {Tjeng}}]{sundermann:18}%
  \BibitemOpen
  \bibfield  {author} {\bibinfo {author} {\bibfnamefont {M.}~\bibnamefont
  {Sundermann}}, \bibinfo {author} {\bibfnamefont {H.}~\bibnamefont {Yavas}},
  \bibinfo {author} {\bibfnamefont {K.}~\bibnamefont {Chen}}, \bibinfo {author}
  {\bibfnamefont {D.~J.}\ \bibnamefont {Kim}}, \bibinfo {author} {\bibfnamefont
  {Z.}~\bibnamefont {Fisk}}, \bibinfo {author} {\bibfnamefont {D.}~\bibnamefont
  {Kasinathan}}, \bibinfo {author} {\bibfnamefont {M.~W.}\ \bibnamefont
  {Haverkort}}, \bibinfo {author} {\bibfnamefont {P.}~\bibnamefont
  {Thalmeier}}, \bibinfo {author} {\bibfnamefont {A.}~\bibnamefont {Severing}},
  \ and\ \bibinfo {author} {\bibfnamefont {L.~H.}\ \bibnamefont {Tjeng}},\
  }\href@noop {} {\bibfield  {journal} {\bibinfo  {journal} {Phys. Rev. Lett.}\
  }\textbf {\bibinfo {volume} {120}},\ \bibinfo {pages} {016402} (\bibinfo
  {year} {2018})}\BibitemShut {NoStop}%
\bibitem [{\citenamefont {Lacroix}\ and\ \citenamefont
  {Cyrot}(1979)}]{lacroix:79}%
  \BibitemOpen
  \bibfield  {author} {\bibinfo {author} {\bibfnamefont {C.}~\bibnamefont
  {Lacroix}}\ and\ \bibinfo {author} {\bibfnamefont {M.}~\bibnamefont
  {Cyrot}},\ }\href@noop {} {\bibfield  {journal} {\bibinfo  {journal} {Phys.
  Rev. B}\ }\textbf {\bibinfo {volume} {20}},\ \bibinfo {pages} {1969}
  (\bibinfo {year} {1979})}\BibitemShut {NoStop}%
\bibitem [{\citenamefont {Ikeda}\ and\ \citenamefont
  {Miyake}(1996)}]{ikeda:96}%
  \BibitemOpen
  \bibfield  {author} {\bibinfo {author} {\bibfnamefont {H.}~\bibnamefont
  {Ikeda}}\ and\ \bibinfo {author} {\bibfnamefont {K.}~\bibnamefont {Miyake}},\
  }\href@noop {} {\bibfield  {journal} {\bibinfo  {journal} {J. Phys. Soc.
  Jpn.}\ }\textbf {\bibinfo {volume} {65}},\ \bibinfo {pages} {1769} (\bibinfo
  {year} {1996})}\BibitemShut {NoStop}%
\bibitem [{\citenamefont {Hanzawa}(2002)}]{hanzawa:02}%
  \BibitemOpen
  \bibfield  {author} {\bibinfo {author} {\bibfnamefont {K.}~\bibnamefont
  {Hanzawa}},\ }\href@noop {} {\bibfield  {journal} {\bibinfo  {journal} {J.
  Phys. Soc. Jpn.}\ }\textbf {\bibinfo {volume} {71}},\ \bibinfo {pages} {1481}
  (\bibinfo {year} {2002})}\BibitemShut {NoStop}%
\bibitem [{\citenamefont {Irkhin}(2017)}]{irkhin:17}%
  \BibitemOpen
  \bibfield  {author} {\bibinfo {author} {\bibfnamefont {V.~Y.}\ \bibnamefont
  {Irkhin}},\ }\href@noop {} {\bibfield  {journal} {\bibinfo  {journal}
  {Physics-Uspekhi}\ }\textbf {\bibinfo {volume} {60}},\ \bibinfo {pages} {747}
  (\bibinfo {year} {2017})}\BibitemShut {NoStop}%
\bibitem [{\citenamefont {Zhang}\ and\ \citenamefont {Yu}(2000)}]{zhang:00}%
  \BibitemOpen
  \bibfield  {author} {\bibinfo {author} {\bibfnamefont {G.-M.}\ \bibnamefont
  {Zhang}}\ and\ \bibinfo {author} {\bibfnamefont {L.}~\bibnamefont {Yu}},\
  }\href@noop {} {\bibfield  {journal} {\bibinfo  {journal} {Phys. Rev. B}\
  }\textbf {\bibinfo {volume} {62}},\ \bibinfo {pages} {76} (\bibinfo {year}
  {2000})}\BibitemShut {NoStop}%
\bibitem [{\citenamefont {Li}\ \emph {et~al.}(2010)\citenamefont {Li},
  \citenamefont {Zhang},\ and\ \citenamefont {Yu}}]{li:10}%
  \BibitemOpen
  \bibfield  {author} {\bibinfo {author} {\bibfnamefont {G.-B.}\ \bibnamefont
  {Li}}, \bibinfo {author} {\bibfnamefont {G.-M.}\ \bibnamefont {Zhang}}, \
  and\ \bibinfo {author} {\bibfnamefont {L.}~\bibnamefont {Yu}},\ }\href@noop
  {} {\bibfield  {journal} {\bibinfo  {journal} {Phys. Rev. B}\ }\textbf
  {\bibinfo {volume} {81}},\ \bibinfo {pages} {094420} (\bibinfo {year}
  {2010})}\BibitemShut {NoStop}%
\bibitem [{\citenamefont {Liu}\ \emph {et~al.}(2013)\citenamefont {Liu},
  \citenamefont {Zhang},\ and\ \citenamefont {Yu}}]{liu:13}%
  \BibitemOpen
  \bibfield  {author} {\bibinfo {author} {\bibfnamefont {Y.}~\bibnamefont
  {Liu}}, \bibinfo {author} {\bibfnamefont {G.-M.}\ \bibnamefont {Zhang}}, \
  and\ \bibinfo {author} {\bibfnamefont {L.}~\bibnamefont {Yu}},\ }\href@noop
  {} {\bibfield  {journal} {\bibinfo  {journal} {Phys. Rev. B}\ }\textbf
  {\bibinfo {volume} {87}},\ \bibinfo {pages} {134409} (\bibinfo {year}
  {2013})}\BibitemShut {NoStop}%
\bibitem [{\citenamefont {Li}\ \emph {et~al.}(2015)\citenamefont {Li},
  \citenamefont {Liu}, \citenamefont {Zhang},\ and\ \citenamefont
  {Yu}}]{li:15}%
  \BibitemOpen
  \bibfield  {author} {\bibinfo {author} {\bibfnamefont {H.}~\bibnamefont
  {Li}}, \bibinfo {author} {\bibfnamefont {Y.}~\bibnamefont {Liu}}, \bibinfo
  {author} {\bibfnamefont {G.-M.}\ \bibnamefont {Zhang}}, \ and\ \bibinfo
  {author} {\bibfnamefont {L.}~\bibnamefont {Yu}},\ }\href@noop {} {\bibfield
  {journal} {\bibinfo  {journal} {J. Phys. Condens. Matter}\ }\textbf {\bibinfo
  {volume} {27}},\ \bibinfo {pages} {425601} (\bibinfo {year}
  {2015})}\BibitemShut {NoStop}%
\bibitem [{\citenamefont {Beach}\ and\ \citenamefont
  {Assaad}(2008)}]{beach:08}%
  \BibitemOpen
  \bibfield  {author} {\bibinfo {author} {\bibfnamefont {K.~S.~D.}\
  \bibnamefont {Beach}}\ and\ \bibinfo {author} {\bibfnamefont {F.~F.}\
  \bibnamefont {Assaad}},\ }\href@noop {} {\bibfield  {journal} {\bibinfo
  {journal} {Phys. Rev. B}\ }\textbf {\bibinfo {volume} {77}},\ \bibinfo
  {pages} {205123} (\bibinfo {year} {2008})}\BibitemShut {NoStop}%
\bibitem [{\citenamefont {Hoshino}\ and\ \citenamefont
  {Kuramoto}(2013)}]{hoshino:13}%
  \BibitemOpen
  \bibfield  {author} {\bibinfo {author} {\bibfnamefont {S.}~\bibnamefont
  {Hoshino}}\ and\ \bibinfo {author} {\bibfnamefont {Y.}~\bibnamefont
  {Kuramoto}},\ }\href@noop {} {\bibfield  {journal} {\bibinfo  {journal}
  {Phys. Rev. Lett.}\ }\textbf {\bibinfo {volume} {111}},\ \bibinfo {pages}
  {026401} (\bibinfo {year} {2013})}\BibitemShut {NoStop}%
\bibitem [{\citenamefont {Akbari}\ \emph {et~al.}(2009)\citenamefont {Akbari},
  \citenamefont {Thalmeier},\ and\ \citenamefont {Fulde}}]{akbari:09}%
  \BibitemOpen
  \bibfield  {author} {\bibinfo {author} {\bibfnamefont {A.}~\bibnamefont
  {Akbari}}, \bibinfo {author} {\bibfnamefont {P.}~\bibnamefont {Thalmeier}}, \
  and\ \bibinfo {author} {\bibfnamefont {P.}~\bibnamefont {Fulde}},\
  }\href@noop {} {\bibfield  {journal} {\bibinfo  {journal} {Phys. Rev. Lett.}\
  }\textbf {\bibinfo {volume} {102}},\ \bibinfo {pages} {106402} (\bibinfo
  {year} {2009})}\BibitemShut {NoStop}%
\bibitem [{\citenamefont {Akbari}\ and\ \citenamefont
  {Thalmeier}(2012)}]{akbari:12}%
  \BibitemOpen
  \bibfield  {author} {\bibinfo {author} {\bibfnamefont {A.}~\bibnamefont
  {Akbari}}\ and\ \bibinfo {author} {\bibfnamefont {P.}~\bibnamefont
  {Thalmeier}},\ }\href@noop {} {\bibfield  {journal} {\bibinfo  {journal}
  {Phys. Rev. Lett.}\ }\textbf {\bibinfo {volume} {108}},\ \bibinfo {pages}
  {146403} (\bibinfo {year} {2012})}\BibitemShut {NoStop}%
\bibitem [{\citenamefont {Jeevan}\ \emph {et~al.}(2006)\citenamefont {Jeevan},
  \citenamefont {Geibel},\ and\ \citenamefont {Hossain}}]{jeevan:06}%
  \BibitemOpen
  \bibfield  {author} {\bibinfo {author} {\bibfnamefont {H.~S.}\ \bibnamefont
  {Jeevan}}, \bibinfo {author} {\bibfnamefont {C.}~\bibnamefont {Geibel}}, \
  and\ \bibinfo {author} {\bibfnamefont {Z.}~\bibnamefont {Hossain}},\
  }\href@noop {} {\bibfield  {journal} {\bibinfo  {journal} {Phys. Rev. B}\
  }\textbf {\bibinfo {volume} {73}},\ \bibinfo {pages} {020407} (\bibinfo
  {year} {2006})}\BibitemShut {NoStop}%
\bibitem [{\citenamefont {Takimoto}\ and\ \citenamefont
  {Thalmeier}(2008)}]{takimoto:08}%
  \BibitemOpen
  \bibfield  {author} {\bibinfo {author} {\bibfnamefont {T.}~\bibnamefont
  {Takimoto}}\ and\ \bibinfo {author} {\bibfnamefont {P.}~\bibnamefont
  {Thalmeier}},\ }\href@noop {} {\bibfield  {journal} {\bibinfo  {journal}
  {Phys. Rev. B}\ }\textbf {\bibinfo {volume} {77}},\ \bibinfo {pages} {045105}
  (\bibinfo {year} {2008})}\BibitemShut {NoStop}%
\bibitem [{\citenamefont {Jeevan}\ \emph {et~al.}(2011)\citenamefont {Jeevan},
  \citenamefont {Adroja}, \citenamefont {Hillier}, \citenamefont {Hossain},
  \citenamefont {Ritter},\ and\ \citenamefont {Geibel}}]{jeevan:11}%
  \BibitemOpen
  \bibfield  {author} {\bibinfo {author} {\bibfnamefont {H.~S.}\ \bibnamefont
  {Jeevan}}, \bibinfo {author} {\bibfnamefont {D.~T.}\ \bibnamefont {Adroja}},
  \bibinfo {author} {\bibfnamefont {A.~D.}\ \bibnamefont {Hillier}}, \bibinfo
  {author} {\bibfnamefont {Z.}~\bibnamefont {Hossain}}, \bibinfo {author}
  {\bibfnamefont {C.}~\bibnamefont {Ritter}}, \ and\ \bibinfo {author}
  {\bibfnamefont {C.}~\bibnamefont {Geibel}},\ }\href@noop {} {\bibfield
  {journal} {\bibinfo  {journal} {Phys. Rev. B}\ }\textbf {\bibinfo {volume}
  {84}},\ \bibinfo {pages} {184405} (\bibinfo {year} {2011})}\BibitemShut
  {NoStop}%
\bibitem [{\citenamefont {Huesges}\ \emph {et~al.}(2018)\citenamefont
  {Huesges}, \citenamefont {Kliemt}, \citenamefont {Krellner}, \citenamefont
  {Sarkar}, \citenamefont {Klau{\ss}}, \citenamefont {Geibel}, \citenamefont
  {Rotter}, \citenamefont {Novak}, \citenamefont {Kunes},\ and\ \citenamefont
  {Stockert}}]{huesges:18}%
  \BibitemOpen
  \bibfield  {author} {\bibinfo {author} {\bibfnamefont {Z.}~\bibnamefont
  {Huesges}}, \bibinfo {author} {\bibfnamefont {K.}~\bibnamefont {Kliemt}},
  \bibinfo {author} {\bibfnamefont {C.}~\bibnamefont {Krellner}}, \bibinfo
  {author} {\bibfnamefont {R.}~\bibnamefont {Sarkar}}, \bibinfo {author}
  {\bibfnamefont {H.-H.}\ \bibnamefont {Klau{\ss}}}, \bibinfo {author}
  {\bibfnamefont {C.}~\bibnamefont {Geibel}}, \bibinfo {author} {\bibfnamefont
  {M.}~\bibnamefont {Rotter}}, \bibinfo {author} {\bibfnamefont
  {P.}~\bibnamefont {Novak}}, \bibinfo {author} {\bibfnamefont
  {J.}~\bibnamefont {Kunes}}, \ and\ \bibinfo {author} {\bibfnamefont
  {O.}~\bibnamefont {Stockert}},\ }\href@noop {} {\bibfield  {journal}
  {\bibinfo  {journal} {New J. Phys.}\ }\textbf {\bibinfo {volume} {20}},\
  \bibinfo {pages} {073021} (\bibinfo {year} {2018})}\BibitemShut {NoStop}%
\bibitem [{\citenamefont {Hafner}\ \emph {et~al.}(2018)\citenamefont {Hafner},
  \citenamefont {Rai}, \citenamefont {Banda}, \citenamefont {Klient},
  \citenamefont {Krellner}, \citenamefont {Sichelschmidt}, \citenamefont
  {Morosan}, \citenamefont {Geibel},\ and\ \citenamefont {Brando}}]{hafner:18}%
  \BibitemOpen
  \bibfield  {author} {\bibinfo {author} {\bibfnamefont {D.}~\bibnamefont
  {Hafner}}, \bibinfo {author} {\bibfnamefont {B.}~\bibnamefont {Rai}},
  \bibinfo {author} {\bibfnamefont {J.}~\bibnamefont {Banda}}, \bibinfo
  {author} {\bibfnamefont {K.}~\bibnamefont {Klient}}, \bibinfo {author}
  {\bibfnamefont {C.}~\bibnamefont {Krellner}}, \bibinfo {author}
  {\bibfnamefont {J.}~\bibnamefont {Sichelschmidt}}, \bibinfo {author}
  {\bibfnamefont {E.}~\bibnamefont {Morosan}}, \bibinfo {author} {\bibfnamefont
  {C.}~\bibnamefont {Geibel}}, \ and\ \bibinfo {author} {\bibfnamefont
  {M.}~\bibnamefont {Brando}},\ }\href@noop {} {\bibfield  {journal} {\bibinfo
  {journal} {preprint}\ } (\bibinfo {year} {2018})}\BibitemShut {NoStop}%
\bibitem [{\citenamefont {Saso}\ and\ \citenamefont {Harima}(2003)}]{saso:03}%
  \BibitemOpen
  \bibfield  {author} {\bibinfo {author} {\bibfnamefont {T.}~\bibnamefont
  {Saso}}\ and\ \bibinfo {author} {\bibfnamefont {H.}~\bibnamefont {Harima}},\
  }\href@noop {} {\bibfield  {journal} {\bibinfo  {journal} {J. Phys. Soc.
  Jpn.}\ }\textbf {\bibinfo {volume} {72}},\ \bibinfo {pages} {1131} (\bibinfo
  {year} {2003})}\BibitemShut {NoStop}%
\bibitem [{\citenamefont {Cornut}\ and\ \citenamefont
  {Coqblin}(1972)}]{cornut:72}%
  \BibitemOpen
  \bibfield  {author} {\bibinfo {author} {\bibfnamefont {B.}~\bibnamefont
  {Cornut}}\ and\ \bibinfo {author} {\bibfnamefont {B.}~\bibnamefont
  {Coqblin}},\ }\href@noop {} {\bibfield  {journal} {\bibinfo  {journal} {Phys.
  Rev. B}\ }\textbf {\bibinfo {volume} {5}},\ \bibinfo {pages} {4541} (\bibinfo
  {year} {1972})}\BibitemShut {NoStop}%
\bibitem [{\citenamefont {Perkins}\ \emph {et~al.}(2007)\citenamefont
  {Perkins}, \citenamefont {Nunez-Regueiro}, \citenamefont {Coqblin},\ and\
  \citenamefont {Iglesias}}]{perkins:07}%
  \BibitemOpen
  \bibfield  {author} {\bibinfo {author} {\bibfnamefont {N.~B.}\ \bibnamefont
  {Perkins}}, \bibinfo {author} {\bibfnamefont {M.~D.}\ \bibnamefont
  {Nunez-Regueiro}}, \bibinfo {author} {\bibfnamefont {B.}~\bibnamefont
  {Coqblin}}, \ and\ \bibinfo {author} {\bibfnamefont {J.~R.}\ \bibnamefont
  {Iglesias}},\ }\href@noop {} {\bibfield  {journal} {\bibinfo  {journal}
  {Phys. Rev. B}\ }\textbf {\bibinfo {volume} {76}},\ \bibinfo {pages} {125101}
  (\bibinfo {year} {2007})}\BibitemShut {NoStop}%
\bibitem [{\citenamefont {Thomas}\ \emph {et~al.}(2011)\citenamefont {Thomas},
  \citenamefont {da~Rosa~Sim{$\tilde{o}$}es}, \citenamefont {Iglesias},
  \citenamefont {Lacroix}, \citenamefont {Perkins},\ and\ \citenamefont
  {Coqblin}}]{thomas:11}%
  \BibitemOpen
  \bibfield  {author} {\bibinfo {author} {\bibfnamefont {C.}~\bibnamefont
  {Thomas}}, \bibinfo {author} {\bibfnamefont {A.~S.}\ \bibnamefont
  {da~Rosa~Sim{$\tilde{o}$}es}}, \bibinfo {author} {\bibfnamefont {J.~R.}\
  \bibnamefont {Iglesias}}, \bibinfo {author} {\bibfnamefont {C.}~\bibnamefont
  {Lacroix}}, \bibinfo {author} {\bibfnamefont {N.~B.}\ \bibnamefont
  {Perkins}}, \ and\ \bibinfo {author} {\bibfnamefont {B.}~\bibnamefont
  {Coqblin}},\ }\href@noop {} {\bibfield  {journal} {\bibinfo  {journal} {Phys.
  Rev. B}\ }\textbf {\bibinfo {volume} {83}},\ \bibinfo {pages} {014415}
  (\bibinfo {year} {2011})}\BibitemShut {NoStop}%
\bibitem [{\citenamefont {Thomas}\ \emph {et~al.}(2014)\citenamefont {Thomas},
  \citenamefont {da~Rosa~Sim{$\tilde{o}$}es}, \citenamefont {Lacroix},
  \citenamefont {Iglesias},\ and\ \citenamefont {Coqblin}}]{thomas:14}%
  \BibitemOpen
  \bibfield  {author} {\bibinfo {author} {\bibfnamefont {C.}~\bibnamefont
  {Thomas}}, \bibinfo {author} {\bibfnamefont {A.~S.}\ \bibnamefont
  {da~Rosa~Sim{$\tilde{o}$}es}}, \bibinfo {author} {\bibfnamefont
  {C.}~\bibnamefont {Lacroix}}, \bibinfo {author} {\bibfnamefont {J.~R.}\
  \bibnamefont {Iglesias}}, \ and\ \bibinfo {author} {\bibfnamefont
  {B.}~\bibnamefont {Coqblin}},\ }\href@noop {} {\bibfield  {journal} {\bibinfo
   {journal} {Journal of Magnetism and Magnetic Materials}\ }\textbf {\bibinfo
  {volume} {372}},\ \bibinfo {pages} {247} (\bibinfo {year}
  {2014})}\BibitemShut {NoStop}%
\bibitem [{\citenamefont {Fazekas}(1999)}]{fazekas:99}%
  \BibitemOpen
  \bibfield  {author} {\bibinfo {author} {\bibfnamefont {P.}~\bibnamefont
  {Fazekas}},\ }\href@noop {} {\emph {\bibinfo {title} {Lecture notes on
  electron correlation and magnetism}}}\ (\bibinfo  {publisher} {World
  Scientific},\ \bibinfo {address} {Singapore},\ \bibinfo {year}
  {1999})\BibitemShut {NoStop}%
\bibitem [{\citenamefont {Nozieres}\ and\ \citenamefont
  {Blandin}(1980)}]{nozieres:80}%
  \BibitemOpen
  \bibfield  {author} {\bibinfo {author} {\bibfnamefont {P.}~\bibnamefont
  {Nozieres}}\ and\ \bibinfo {author} {\bibfnamefont {A.}~\bibnamefont
  {Blandin}},\ }\href@noop {} {\bibfield  {journal} {\bibinfo  {journal} {J.
  Physique}\ }\textbf {\bibinfo {volume} {41}},\ \bibinfo {pages} {193}
  (\bibinfo {year} {1980})}\BibitemShut {NoStop}%
\bibitem [{\citenamefont {Coleman}(2015)}]{coleman:15}%
  \BibitemOpen
  \bibfield  {author} {\bibinfo {author} {\bibfnamefont {P.}~\bibnamefont
  {Coleman}},\ }\href@noop {} {\emph {\bibinfo {title} {Introduction to
  many-body pysics}}}\ (\bibinfo  {publisher} {Cambridge University Press},\
  \bibinfo {address} {Cambridge},\ \bibinfo {year} {2015})\BibitemShut
  {NoStop}%
\bibitem [{\citenamefont {Tesanovic}\ and\ \citenamefont
  {Valls}(1986)}]{tesanovic:86}%
  \BibitemOpen
  \bibfield  {author} {\bibinfo {author} {\bibfnamefont {Z.}~\bibnamefont
  {Tesanovic}}\ and\ \bibinfo {author} {\bibfnamefont {O.~T.}\ \bibnamefont
  {Valls}},\ }\href@noop {} {\bibfield  {journal} {\bibinfo  {journal}
  {Physical Review B}\ }\textbf {\bibinfo {volume} {34}},\ \bibinfo {pages}
  {1918} (\bibinfo {year} {1986})}\BibitemShut {NoStop}%
\bibitem [{\citenamefont {Bernhard}\ and\ \citenamefont
  {Lacroix}(2015)}]{bernhard:15}%
  \BibitemOpen
  \bibfield  {author} {\bibinfo {author} {\bibfnamefont {B.~H.}\ \bibnamefont
  {Bernhard}}\ and\ \bibinfo {author} {\bibfnamefont {C.}~\bibnamefont
  {Lacroix}},\ }\href@noop {} {\bibfield  {journal} {\bibinfo  {journal} {Phys.
  Rev. B}\ }\textbf {\bibinfo {volume} {92}},\ \bibinfo {pages} {094401}
  (\bibinfo {year} {2015})}\BibitemShut {NoStop}%
\bibitem [{\citenamefont {Alekseev}\ \emph {et~al.}(2004)\citenamefont
  {Alekseev}, \citenamefont {Mignot}, \citenamefont {S.}, \citenamefont
  {Nefeodova}, \citenamefont {Shitsevalova}, \citenamefont {Paderno},
  \citenamefont {Bewley}, \citenamefont {Eccleston}, \citenamefont
  {Clementyev}, \citenamefont {Lazukov}, \citenamefont {Sadikov},\ and\
  \citenamefont {Tiden}}]{alekseev:04}%
  \BibitemOpen
  \bibfield  {author} {\bibinfo {author} {\bibfnamefont {P.~A.}\ \bibnamefont
  {Alekseev}}, \bibinfo {author} {\bibfnamefont {J.-M.}\ \bibnamefont
  {Mignot}}, \bibinfo {author} {\bibfnamefont {K.}~\bibnamefont {S.}}, \bibinfo
  {author} {\bibfnamefont {E.~V.}\ \bibnamefont {Nefeodova}}, \bibinfo {author}
  {\bibfnamefont {N.~Y.}\ \bibnamefont {Shitsevalova}}, \bibinfo {author}
  {\bibfnamefont {Y.~P.}\ \bibnamefont {Paderno}}, \bibinfo {author}
  {\bibfnamefont {R.~I.}\ \bibnamefont {Bewley}}, \bibinfo {author}
  {\bibfnamefont {R.~S.}\ \bibnamefont {Eccleston}}, \bibinfo {author}
  {\bibfnamefont {E.~S.}\ \bibnamefont {Clementyev}}, \bibinfo {author}
  {\bibfnamefont {V.~N.}\ \bibnamefont {Lazukov}}, \bibinfo {author}
  {\bibfnamefont {I.~P.}\ \bibnamefont {Sadikov}}, \ and\ \bibinfo {author}
  {\bibfnamefont {N.~N.}\ \bibnamefont {Tiden}},\ }\href@noop {} {\bibfield
  {journal} {\bibinfo  {journal} {J. Phys. Condens. Matter}\ ,\ \bibinfo
  {pages} {2631}} (\bibinfo {year} {2004})}\BibitemShut {NoStop}%
\bibitem [{\citenamefont {Jensen}\ and\ \citenamefont
  {Mackintosh}(1991)}]{jensen:91}%
  \BibitemOpen
  \bibfield  {author} {\bibinfo {author} {\bibfnamefont {J.}~\bibnamefont
  {Jensen}}\ and\ \bibinfo {author} {\bibfnamefont {A.~R.}\ \bibnamefont
  {Mackintosh}},\ }\href@noop {} {\emph {\bibinfo {title} {Rare Earth Magnetism
  Structures and Excitations}}}\ (\bibinfo  {publisher} {Clarendon Press},\
  \bibinfo {address} {Oxford},\ \bibinfo {year} {1991})\BibitemShut {NoStop}%
\bibitem [{\citenamefont {Hanzawa}(1998)}]{hanzawa:98}%
  \BibitemOpen
  \bibfield  {author} {\bibinfo {author} {\bibfnamefont {K.}~\bibnamefont
  {Hanzawa}},\ }\href@noop {} {\bibfield  {journal} {\bibinfo  {journal} {J.
  Phys. Soc. Jpn.}\ }\textbf {\bibinfo {volume} {67}},\ \bibinfo {pages} {3151}
  (\bibinfo {year} {1998})}\BibitemShut {NoStop}%
\bibitem [{\citenamefont {Takimoto}(2011)}]{takimoto:11}%
  \BibitemOpen
  \bibfield  {author} {\bibinfo {author} {\bibfnamefont {T.}~\bibnamefont
  {Takimoto}},\ }\href@noop {} {\bibfield  {journal} {\bibinfo  {journal} {J.
  Phys. Soc. Jpn.}\ }\textbf {\bibinfo {volume} {80}},\ \bibinfo {pages}
  {123710} (\bibinfo {year} {2011})}\BibitemShut {NoStop}%
\bibitem [{\citenamefont {Figueira}\ \emph {et~al.}(1994)\citenamefont
  {Figueira}, \citenamefont {Foglio},\ and\ \citenamefont
  {Martinez}}]{figueira:94}%
  \BibitemOpen
  \bibfield  {author} {\bibinfo {author} {\bibfnamefont {M.~S.}\ \bibnamefont
  {Figueira}}, \bibinfo {author} {\bibfnamefont {M.~E.}\ \bibnamefont
  {Foglio}}, \ and\ \bibinfo {author} {\bibfnamefont {G.~G.}\ \bibnamefont
  {Martinez}},\ }\href@noop {} {\bibfield  {journal} {\bibinfo  {journal}
  {Phys. Rev. B}\ }\textbf {\bibinfo {volume} {50}},\ \bibinfo {pages} {17933}
  (\bibinfo {year} {1994})}\BibitemShut {NoStop}%
\end{thebibliography}%

\end{document}